\documentclass[aps,pra,reprint]{revtex4-1}
\usepackage{blindtext}
\usepackage{graphicx}
\usepackage{amsmath}
\usepackage{amssymb}
\usepackage[english]{babel}
\usepackage{xcolor}
\usepackage{siunitx}
\usepackage[colorlinks=true, pdfstartview=FitV, linkcolor=blue,
citecolor=blue, urlcolor=blue]{hyperref}
\usepackage[capitalise]{cleveref}
\usepackage{float}

\makeatletter
\let\saved@includegraphics\includegraphics
\AtBeginDocument{\let\includegraphics\saved@includegraphics}
\renewenvironment*{figure}{\@float{figure}}{\end@float}
\renewenvironment*{table}{\@float{table}}{\end@float}
\makeatother

\makeatletter
\newcommand\footnoteref[1]{\protected@xdef\@thefnmark{\ref{#1}}\@footnotemark}
\makeatother

\begin{document}

\title{Hole spin qubits in Si FinFETs with fully tunable spin-orbit coupling\\
and sweet spots for charge noise}

\author{Stefano Bosco}
\author{Bence Het\'enyi}
\author{Daniel Loss}

\affiliation{Department of Physics, University of Basel, Klingelbergstrasse 82, 4056 Basel, Switzerland}

\begin{abstract}
The strong spin-orbit coupling in hole spin qubits enables fast and electrically tunable gates, but at the same time enhances the susceptibility of the qubit to charge noise. Suppressing this noise is a significant challenge in semiconductor quantum computing.
Here, we show theoretically that hole Si FinFETs are not only very compatible with modern CMOS technology, but they present operational sweet spots where the charge noise is completely removed.
The presence of these sweet spots is a result of the interplay between the anisotropy of the material and the triangular shape of the FinFET cross-section, and  it does not require an extreme fine-tuning of the electrostatics of the device. 
We present how the sweet spots appear in FinFETs grown along different crystallographic axes and we study in detail how the behaviour of these devices change when the cross-section area and aspect ratio are varied. We identify designs that maximize the qubit performance and could pave the way towards a scalable spin-based quantum computer. 
\end{abstract}

\maketitle

\section{Introduction}

Strong spin-orbit coupling \cite{WinklerSpinOrbitCoupling2003} is a desirable ingredient to build a scalable spin-based quantum computer \cite{PhysRevA.57.120, RevModPhys.79.1217}, enabling fast and fully electrical manipulations of quantum bits \cite{nadj2010spin,Nowack1430,hendrickx2020fast}.
Promising platforms to reach large values of spin-orbit interactions are $p$-doped semicondutor nanowires, where the charge carriers are holes rather than electrons \cite{PhysRevLett.95.076805,PhysRevLett.98.097202,doi:10.1146/annurev-conmatphys-030212-184248,marx2020spin,scappucci2020germanium,maurand2016cmos,LiPauliSpinBlockade2015,Hendrickxsingleholespinqubit2019}. 
When holes are strongly confined in two directions, an externally tunable electric field generates a large effective spin-orbit field \cite{DRkloeffel1,DRkloeffel3} that results in ultrafast Rabi frequencies, larger than $400$~MHz \cite{froning2020ultrafast,wang2020ultrafast}, and in spin-orbit lengths of tens of nanometers \cite{de2018spin,PhysRevLett.112.216806,PhysRevB.93.121408,doi:10.1002/adma.201906523,wang2017electrical,froning2020strong,PhysRevB.87.161305}, shorter than typical interdot distances.
The regime of strong coupling between spins and photons in microwave resonators~\cite{landig2018coherent,mi2018coherent} has been predicted in these systems \cite{DRkloeffel2}, which could enable  long-range coupling between distant qubits.
Because of the large spin-orbit interaction, hole-superconductor heterostructures have attracted also much interest as platforms to detect and manipulate Majorana bound states \cite{sun2018helical,PhysRevLett.105.177002,PhysRevB.90.195421}.

On the other hand, large interactions between spin and charge degrees of freedom render the system strongly susceptible to charge noise, reducing the qubit lifetime \cite{Yonedaquantumdotspinqubit2018,froning2020ultrafast,culcer2009dephasing,bermeister2014charge}.
Efforts to find operational sweet spots where charge noise is reduced have been focusing on planar Ge qubits \cite{wang2019suppressing} or considering single atoms \cite{PhysRevLett.116.246801}, as well as artificial spin-orbit fields~\cite{PhysRevB.100.125430}. The appearance of sweet spots depending on the direction of the applied magnetic field has also been analyzed \cite{TanttuControllingspinorbitinteractions2018,VenitucciElectricalmanipulationsemiconductor2018}. 
At the working points identified in these studies, the spin-orbit interaction is not susceptible to small fluctuations of the electric field, but remains finite.
However, in electrostatically defined quantum dots in hole nanowires, there are additional noise channels that are not suppressed at these working points.
For example, because of the large value of the spin-orbit coupling, the fluctuations of the size of the dot strongly couple to the spin and lead to decoherence.
To remove charge noise in these systems, one needs the ability to on-demand fully switch \textit{ON} and \textit{OFF} the spin-orbit interactions  depending on whether the qubit is operational or idle.

We find that such a spin-orbit switch naturally occurs in $p$-doped Silicon Fin Field Effect Transistors (FinFETs)  \cite{6330653,kuhlmann2018ambipolar,geyer2020silicon,siliconFin}, thus making these devices ideal candidates to reliably store quantum information.
Silicon  is highly compatible with modern semiconductor industry and is one of the frontrunner materials for scalable large scale quantum computers. State-of-the-art electron Si qubits can operate reliably at temperatures higher than 1~K \cite{yang2020operation,petit2020high}, and high fidelity two-qubit gates~\cite{Veldhorsttwoqubitlogicgate2015,Watsonprogrammabletwoqubitquantum2018,ZajacResonantlydrivenCNOT2018,PhysRevX.9.021011,huang2019fidelity}, as well as singlet-triplet qubit operations~\cite{TakedaResonantlyDrivenSingletTriplet2020}, control over higher spin states~\cite{PhysRevX.10.041010} and scalable readout schemes \cite{ZhengRapidgatebasedspin2019,WestGatebasedsingleshotreadout2019,yoneda2020quantum,PhysRevX.10.021006,SeedhouseParityreadoutsilicon2020} have been demonstrated in Si.
In addition, Si offers the unique possibility to drastically reduce the hyperfine noise \cite{PhysRevLett.88.186802,PhysRevB.70.195340,fischer2009spin,PhysRevLett.105.266603,PhysRevB.78.155329,PhysRevB.85.195323,PhysRevB.101.115302} by isotopic purification \cite{Takedafaulttolerantaddressablespin2016,ZhaoSinglespinqubitsisotopically2019}, resulting in spin qubits whose performances are essentially limited only by the charge noise~\cite{Yonedaquantumdotspinqubit2018}.

The tunability of the spin-orbit coupling by varying the electric field is a well-known feature of semiconductor nanowires \cite{scappucci2020germanium,doi:10.1146/annurev-conmatphys-030212-184248}, but in many typical geometries, such as wires with rectangular  \cite{DRkloeffel3, voisin2016electrical,PhysRevLett.120.137702} or circular cross-sections \cite{DRkloeffel1,DRkloeffel2,froning2020ultrafast,froning2020strong}, the spin-orbit interaction is only fully removed when there is no external electric field.
While working without a DC electric field can be possible for etched \cite{reed1986spatial} or self-assembled \cite{doi:10.1080/00107510210123845} quantum dots, in electrostatically defined nanostructures, an external gate potential is required to delimit the dot. Consequently, the electric field cannot be easily set to zero, resulting in a residual spin-orbit interaction, which degrades the qubit performance.
In contrast, a crucial feature of the Si FinFETs studied here is their nearly triangular cross-section, which results in sweet spots where the spin-orbit coupling can be switched off at finite values of the electric field, thus removing the charge noise.
In fact, we show that holes confined in triangular wires present a large spin-orbit coupling even without electric fields and, depending on the design of the fin, an external gate potential can suppress this intrinsic coupling.

The present paper is organized as follows. 
In Sec.~\ref{Sec:Model}, we introduce different state-of-the-art FinFET designs \cite{6330653,kuhlmann2018ambipolar,geyer2020silicon,siliconFin}, including Silicon-on-Insulator (SOI) FinFETs and bulk Si FinFETs, and we discuss the theoretical model used in our analysis.

In Sec.~\ref{Sec:Equilateral}, we consider an ideal fin with an equilateral triangular cross-section and study the long wavelength dynamics of the holes confined there. Because Si is an anisotropic semiconductor \cite{WinklerSpinOrbitCoupling2003}, we pay particular attention to how different growth directions affect the effective spin-orbit interactions \cite{PhysRevB.99.115317, DRkloeffel3, xiong2020strong, PhysRevResearch.2.033036}.
By using a simple theoretical model, which only includes heavy and light holes, and by considering realistic inhomogeneous electric field profiles, we identify qualitatively distinct mechanisms that remove the spin-orbit coupling and that are suitable for different device designs.
Here, we also comment on the the effect of a possible moderate strain on the spin-orbit switch.

In Sec.~\ref{Sec:SOHS}, we extend the theoretical model of the FinFET by including the spin-orbit split-off hole band (SOHs).  This band is energetically separated from heavy and light holes by the bulk Si spin-orbit gap \cite{WinklerSpinOrbitCoupling2003}, but it strongly influences the hole behaviour in small wires and we find that it can even remove the spin-orbit sweet spot in wires with a triangular cross-section of side shorter than $35$ nm.
Fortunately, we find that a more careful device design can counteract the action of the SOHs and we discuss a possible way of recovering the sweet spot.

Finally, in Sec.~\ref{Sec:Charge-noise}, we study the charge noise of spin $1/2$ qubits \cite{PhysRevA.57.120} in Si FinFETs. We find that working close to the spin-orbit sweet spot drastically suppresses the influence of charge noise on the qubit lifetime, and strongly improves the dephasing time. By including in our analysis the fluctuations of the $g$-factor as a function of the electric field, a charge noise mechanism that is not directly related to the effective spin-orbit coupling of the wire, the exact position of the sweet spot is slightly shifted, but the charge noise can still be exactly cancelled, resulting in a system fully resilient against small charge fluctuations.

\section{Theoretical Model}
\label{Sec:Model}

\begin{figure}[t]
	 \includegraphics[width=0.45\textwidth]{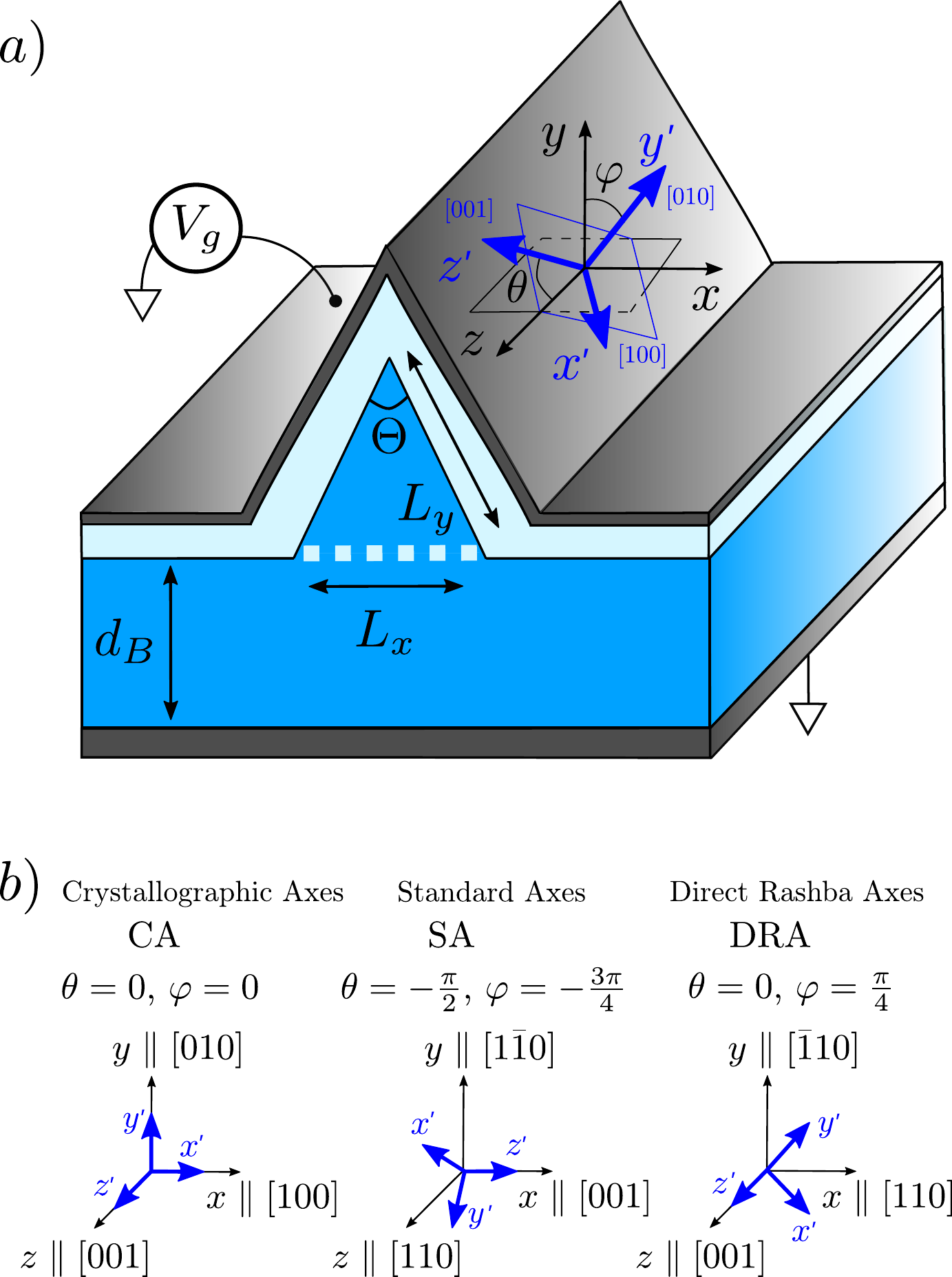}
	\caption{Sketch of a Si FinFET. In a), the blue and lightblue areas represent the semiconductor and the dielectric, respectively, while the gray areas are the metallic gates.  The orientation of the Si FinFET with respect to the crystallographic axes (blue axes) depends on the angles $\theta$ and $\varphi$. The wire extends along the $z$-direction and the fin has an isosceles triangular shape with base $L_x$, equal sides $L_y$ and height parallel to the $y$-direction. 
	The apex angle of the triangle is $\Theta$. We call wide and narrow FinFETs the $\Theta>\frac{\pi}{3}$ and $\Theta<\frac{\pi}{3}$ devices, respectively; the FinFET is equilateral when $\Theta=\frac{\pi}{3}$. 
The dashed line at the bottom of the fin indicates the lower boundary of the fin. In a SOI FinFET there is a clear physical separation between the Si substrate and the wire, which is provided by a thick dielectric layer at the position of the dashed line. In contrast, in bulk Si FinFET, there is no physical separation between substrate and the fin. In this case, the holes are localized in the fin by negative values of the gate potential $V_g$ applied with respect to a grounded back gate at a distance $d_B$ from the bottom of the triangle. 
In b), we show the main orientations of the axes of confinement with respect to the crystallographic axes.
\label{fig:Geometry-finfet}}
\end{figure}

In this paper, we analyze the hole Si FinFET sketched in Fig.~\ref{fig:Geometry-finfet}. 
The fin extends in the $z$-direction and it defines a nanowire with an isosceles triangular cross-section with equal sides $L_y$ and base $L_x$. 
We study two different FinFET designs: Silicon-on-Insulator and bulk FinFETs. In SOI FinFETs, the triangular fin lies on top of a dielectric material, while in bulk FinFETs, it lies on top of a Si substrate. The apex of the  fin is covered by a dielectric with an ideal metallic gate placed on top. The top gate is fixed at a potential $V_g$ measured with respect to a back-gate at a distance $d_B$ from the bottom of the wire.
In bulk FinFET,  a negative gate potential $V_g$ is required to localize the hole wavefunction inside the fin, while in SOI FinFETs, the holes are confined in the wire by the dielectric  and $V_g$ can attain positive values, too. 

The dynamics of this system is accurately described by the Hamiltonian
\begin{equation}
\label{eq:total-H}
H=H_{\text{LK}}+V_{\text{HW}}(x,y)+ V_E(x,y)  \ ,
\end{equation}
which comprises the hole kinetic energy $H_{\text{LK}}$ and two distinct potential energies $V_{\text{HW}}$ and $V_{E}$.

The potential $V_{\text{HW}}$ captures the abrupt interfaces between the semiconductor and the dielectric and because of the large energy gap between the materials, we model it by requiring the wavefunction to vanish at the edges of the system (hard-wall boundary conditions). In contrast, $V_{E}$ describes the smoother and externally tunable electrostatic potential generated by $V_g$. In the cross-section of the wire, this term is well-approximated by the multipole expansion
\begin{equation}
\label{eq:Multipole-E}
V_E(x,y)= -e \textbf{E}\cdot \textbf{r}- \frac{e}{2} \textbf{r} \cdot \underline{\delta\!E}  \cdot \textbf{r} \ ,
\end{equation}
that includes a homogeneous electric field vector $\textbf{E}=(E_x,E_y)$ and a tensor modelling the inhomogeneous component of the electric field $(\underline{\delta\!E})_{ij}$.
The coordinate system $\textbf{r}=({x,y})$ is centred in the center of mass of the wire. Importantly, both the homogeneous and inhomogeneous components of the electric field depend linearly on the external gate potential $V_g$ and their strength can be tuned by the gate design. More details on this approximation and a thorough discussion on the values of $E_i$ and $\delta\!E_{ij}$ in the FinFETs analyzed in this paper can be found in  App.~\ref{app:E-field}.

A precise description of the kinetic energy of the holes in the valence bands of semiconductors is provided by the $4\times 4$ Luttinger-Kohn (LK) Hamiltonian  \cite{LuttingerModel, WinklerSpinOrbitCoupling2003}
\begin{equation}
\label{eq:LK_Hamiltonian_4}
H_\text{LK}'=\left(\!\gamma_1+\frac{5}{2}\gamma_2\!\right)\!\frac{p'^2}{2m}-\frac{\gamma_2}{m} \textbf{p}'^2\cdot \textbf{J}'^2-\frac{2\gamma_3}{m}  p_i'p_j' \left\lbrace J_{i}',J_{j}' \right\rbrace+\text{cp}\ ,
\end{equation}
which describes the mixing of heavy holes (HHs) and light holes (LHs) with spin $3/2$ and $1/2$, respectively.
Here, $m$ is the bare electron mass, we use the anticommutator $\{ A,B \}=(AB+BA)/2$, and $\text{cp}$ stands for cyclic permutations.
Also, we defined $p'^2=p_{x}'^2+p_{y}'^2+p_{z}'^2$ and the vectors $\textbf{p}'^2=(p_{x}'^2,p_{y}'^2,p_{z}'^2)$ and $\textbf{J}'^2=(J_{x}'^2,J_{y}'^2,J_{z}'^2)$, where $p_i'=-i\hbar \partial_{i'}$ are canonical momenta and the four-dimensional matrices $J_{i}'$ are  spin 3/2 matrices. The primed coordinate system is aligned to the main crystallographic axes, i.e. $x'\parallel [100]$, $y'\parallel[010]$ and $z'\parallel [001]$.

The LK Hamiltonian is parametrized by three material-dependent dimensionless quantities: $\gamma_{1,2,3}$. Here, we use the values of $\gamma_i$ given in Ref.~\cite{WinklerSpinOrbitCoupling2003}.
 Si is an anisotropic semiconductor because the parameters $\gamma_2$ and $\gamma_3$ are quite different. As a consequence, the low-energy description of the system strongly depends on the orientation of the nanowire with respect to the crystallographic axes \cite{PhysRevB.99.115317, DRkloeffel3, xiong2020strong, PhysRevResearch.2.033036}. 
In our convention, the nanowire always extends along the $z$-direction, and to conveniently account for different growth directions,  we transform the LK Hamiltonian as $H_\text{LK}'\rightarrow H_\text{LK}$ by performing a rotation of an angle $\theta$ around the $y'\parallel[010]$ crystallographic axis and a subsequent rotation of an angle $\varphi$ around the rotated $z'$ axis, see Fig.~\ref{fig:Geometry-finfet}. This transformation aligns the coordinate system and the spin-matrices to the axes $({x,y,z})$ in the figure; the Hamiltonian $H_\text{LK}$ and the relation between the primed and unprimed coordinates are given explicitly in Eqs.~\eqref{eq:Eq:LK_rotated} and~\eqref{eq:coordinate_system_rotated}, respectively.

The most relevant orientations of the wire for this paper  are summarized in Fig.~\ref{fig:Geometry-finfet}b). In particular, it is instructive to study the behaviour of wires where the coordinate system is aligned to the crystallographic axes, and of wires grown along the $[110]$ direction, as standardly done in experiments \cite{kuhlmann2018ambipolar,geyer2020silicon,siliconFin,doi:10.1002/adma.201906523,voisin2016electrical}. We also consider the growth direction that maximizes the direct Rashba spin-orbit interactions in inversion symmetric Silicon nanowires \cite{DRkloeffel3} and quantum dots \cite{PhysRevB.99.115317}.
We will refer to these orientations as to Crystallographic Axes (CA), Standard Axes (SA), and Direct Rashba Axes (DRA), respectively.

The $4\times 4$  LK Hamiltonian in Eq.~\eqref{eq:LK_Hamiltonian_4} captures accurately the physics of Si nanowires with large cross-sections, however, to describe smaller wires one needs to include the contribution of the spin $1/2$ spin-orbit split-off  holes (SOHs), that are gapped from the HHs and LHs by a material-dependent spin-orbit energy $\Delta_0$. 
The SOHs are negligible only when $\Delta_0$ is much larger than the confinement energy, such that the HH-LH subspace is well-separated in energy from the subspace of the SOHs. 
In Silicon, the gap $\Delta_0\approx 44.1$~meV \cite{WinklerSpinOrbitCoupling2003}, is comparable to the confinement energy in narrow wires, leading to a strong influence of the SOHs on the response, especially in the presence of an electric field. To take the SOHs fully into account, we compare our results obtained with the $4\times 4$ Hamiltonian~\eqref{eq:LK_Hamiltonian_4} to more accurate results obtained with the $6\times 6$ extension of the LK Hamiltonian, the complete form of which can be found for example in Appendix C of Ref.~\cite{WinklerSpinOrbitCoupling2003}. 
The conduction electrons are separated by a much larger gap from the holes, and so their contribution is always neglected in the present analysis.\\

Without magnetic field, the eigenvalues of the Hamiltonian in Eq.~\eqref{eq:total-H} are doubly-degenerate Kramers partners.
When the wavelength of the hole wavefunction along the nanowire is larger than the confinement length in the cross-section, the low-energy physics of the system is well described by an effective nanowire Hamiltonian $H_\text{NW}$ that only acts on the lowest pair of eigenstates. 
To second order in the momentum $p_z$ along the wire, we obtain
\begin{equation}
\label{eq:wire-Ham}
H_\text{NW}=\frac{p_z^2}{2m^*}+ \pmb{v}\cdot\pmb{\sigma} p_z \ ,
\end{equation}
where $\pmb{\sigma}$ is a vector of Pauli matrices acting on the subspace of the lowest Kramers partners.
The nanowire Hamiltonian is parametrized by an effective mass $m^*$ and a spin-orbit velocity vector $\pmb{v}$  that can be found in perturbation theory.
In particular, one can decompose $H$ into powers of $p_z$ as $H=H_0+H_1 p_z+H_2 p_z^2$, with $H_0=H_\text{LK}(p_z=0)+V_{\text{HW}}(x,y)+ V_E(x,y)$, and introduce the unitary matrix $M_E$ that diagonalizes $H_0$, i.e. $\Big(M_E^\dagger H_0 M_E\Big)_{nm}=\epsilon_E^n \delta_{nm}$. 
To obtain accurate results, we compute the eigenvectors $M_E$ and the eigenenergies $\epsilon_E$ numerically by discretizing the Hamiltonian $H_0$.
By standard perturbation theory, it is straightforward to find 
\begin{equation}
\label{eq:SO-Exact}
\Big(\pmb{v}\cdot\pmb{\sigma}\Big)_{ij}= \Big(M_E^\dagger H_1 M_E\Big)_{ij} \ , 
\end{equation}
 where $i,j$ only act on the lowest pair of Kramers partners; a similar perturbative expression for the effective mass is given in Eq.~\eqref{eq:mass-Exact}. 
The subscript $E$ in the eigensystem $M_E$ and $\epsilon_E$ emphasizes the dependence on the electric field. For this reason, the effective parameters of the wire Hamiltonian are externally tunable by the gate potential $V_g$, which controls the electrostatic potential $V_E(x,y)$.
We also introduce the spin-orbit length
\begin{equation}
\label{eq:SO-length}
l_{so}=\frac{\hbar}{m^*|\pmb{v}|} \ ,
\end{equation}
that characterizes the spin-orbit interactions relative to the inertia of the particle.\\

To define a spin qubit, we  include an external, homogeneous magnetic field $\textbf{B}$.
For weak values of the magnetic field, typically below one Tesla, we can safely neglect the orbital contribution of the magnetic field and only focus on the coupling of the magnetic field to the spin-degree of freedom via the Zeeman energy, which in the $4\times 4$ Luttinger-Kohn Hamiltonian is $H_Z=2\textbf{B}\cdot(\kappa \textbf{J}+q\textbf{J}^3)$. Here, $\kappa$ and $q$ are material-dependent parameters for the magnetic interactions. The magnetic interactions when SOHs are included, as well as the precise value of $\kappa$ and $q$ for Si,  can be found in Ref.~\citep{WinklerSpinOrbitCoupling2003}.
Projecting the Zeeman Hamiltonian onto the groundstate of the wire, to linear order in $B$, the effective Hamiltonian in Eq.~\eqref{eq:wire-Ham} acquires the correction
\begin{equation}
\label{eq:Zeeman-energy-NW}
H_\text{NW}^Z= \frac{1}{2}\pmb{\Delta} \cdot\pmb{\sigma} \ , 
\end{equation}
where we introduce the vector $\pmb{\Delta}= \mu_B \underline{g}\cdot \textbf{B}$. Here,  $\mu_B$ is the Bohr magneton and $\underline{g}$ is a dimensionless $3\times 3$  matrix of $g$-factors. From perturbation theory, we obtain the electric field dependent Zeeman interactions
\begin{equation}
\label{eq:Delta_Zeeman}
\Big(\pmb{\Delta} \cdot\pmb{\sigma}\Big)_{ij}=2 \Big(M_E^\dagger H_Z M_E\Big)_{ij} \ .
\end{equation}

\section{Equilateral FinFETs}
\label{Sec:Equilateral}

To have a simple model of FinFETs, we consider first a nanowire with a triangular cross-section and we require that the hole wavefunction vanishes at the boundaries of the triangle, see Fig.~\ref{fig:Geometry-finfet}. 
This model provides an accurate description of SOI FinFETs, but it is questionable in bulk Si FinFETs, where there is no sharp interface at the bottom of the fin and the wavefunction can leak into the bulk.
In this case, however, the hard-wall approximation still provides a good qualitative understanding of the system, especially when the hole wavefunction is strongly confined inside the fin by a large negative gate potential $V_g$. 
The effect of the substrate in a bulk Si FinFET is discussed in App.~\ref{Sec:Deviation}. 

The choice of a triangular fin is crucial in our analysis. In fact, compared to rectangular or circular nanowires, a triangular cross-section lacks  inversion symmetry in the $(x,y)$ plane, i.e.  $V_\text{HW}(x,y)\neq V_\text{HW}(-x,-y)$, and consequently the triangular nanowire can present large intrinsic spin-orbit interactions without external homogeneous electric fields, i.e. $\pmb{v}_0\equiv \pmb{v}(\textbf{E}=0) \neq 0$ \footnote{
If $\textbf{E}=0$ and the cross-section is inversion symmetric, then $V(x,y)\equiv V_\text{HW}(x,y)+V_E(x,y)=\mathcal{I} V(x,y)\mathcal{I}^\dagger= V(-x,-y)$, where $\mathcal{I}=e^{-i\pi F_z}$ is the inversion operator in the $(x,y)$ plane; $F_z$ is the component parallel to the wire of the total angular momentum. It follows that the ground subspace of $H_\text{LK}(p_z=0)+V(x,y)$ can be labelled by $| i, \uparrow \downarrow \rangle$ and is composed of a degenerate Kramers doublet $|\uparrow\downarrow\rangle$ that is also an eigenstate of $\mathcal{I}$ to eigenvalue $i$. Let us consider the correction to the LK Hamiltonian $H_1 p_z$ to linear order in $p_z$, such that the states  $| i, \uparrow\downarrow \rangle$ are still approximate eigenstates. Because $H_1$ anticommutes with $\mathcal{I}$, i.e. $\mathcal{I} H_1 \mathcal{I}^\dagger=-H_1$, $H_1$ is an block off-diagonal matrix in the basis of the eigenstates of $\mathcal{I}$.  Consequently, in the groundstate subspace there cannot be spin-orbit interactions linear in $p_z$ and because time-reversal symmetry prohibits different masses for the Kramers partners, the first possible spin-orbit interactions are $\propto p_z^3$.
}.

In this section, we examine how the spin-orbit interaction varies as a function of the gate potential and of the growth direction in an ideal case, where the cross-section is an equilateral triangle of side $L$.  
In Si FinFETs, the triangular cross-section can be made rather equilateral, see e.g. \cite{geyer2020silicon,siliconFin}, however it is often the case that the fin is a more narrow \cite{kuhlmann2018ambipolar} or wide \cite{doi:10.1002/adma.201906523} isosceles triangle. The spin-orbit coupling in isosceles triangles with different aspect ratios is analyzed in Sec.~\ref{Sec:SOHS}.

A convenient orthonormal basis to describe this system comprises the eigenfunctions of the two-dimensional Laplace operator $p_x^2+p_y^2$ vanishing at the boundary of an equilateral triangle. Because of the highly symmetric geometry, the eigenfunctions can be expressed in terms of trigonometric functions, see App.~\ref{app:Laplace} [in particular Eq.~\eqref{eq:sol-LaplEq}] and Ref.~\cite{mccartin2011laplacian} for more details.
A natural energy scale for this problem is the confinement energy
\begin{equation}
\label{eq:Confinement_energy}
\epsilon_c=\frac{16 \hbar^2 \pi^2 }{3m L^2}\gamma_1 \ ,
\end{equation}
which characterizes the energy gap between different orbital states, see Eq.~\eqref{eq:Laplace-eigenvalues}.
This energy is quite large, for example in a Silicon wire with side $L=35$~nm, the quantization energy is $\epsilon_c\approx 14 $~meV, approximately 30$\%$ of the gap to the split-off band $\Delta_0\approx 44.1$~meV \cite{WinklerSpinOrbitCoupling2003}.
While for  $L\gtrsim 35$~nm, the $4\times 4$ LK Hamiltonian in Eq.~\eqref{eq:LK_Hamiltonian_4} is valid, in smaller wires such a strong quantization results in a large contribution of the SOHs, which demands a more detailed analysis that fully includes these states.
In the  following, we will refer to small (large) wires when the side $L$ is smaller (larger) than $35$~nm.
To gain a qualitative understanding of the system, we begin our analysis by studying large nanowires by using the $4\times 4$ LK Hamiltonian; a detailed analysis of the effect of the SOHs is postponed to Sec.~\ref{Sec:SOHS}.

We first compute the intrinsic spin-orbit velocity $\pmb{v}_0$ in the absence of electric fields, focusing on its dependence on the growth direction. Then, we separately describe the effect of homogeneous and inhomogeneous electric fields.
We discuss when the electric field-induced spin-orbit coupling compensates for the intrinsic spin-orbit interactions, yielding convenient operational sweet spots where spin-orbit effects vanish.

\subsection{Intrinsic spin-orbit velocity}
\label{Sec:Intrinsic-SO}

Without external fields, a simple yet satisfactory description of the system is provided by a reduced $12$ dimensional Hamiltonian $H_{12}$ that includes only the lowest three orbital states in Eq.~\eqref{eq:sol-LaplEq}.
$H_{12}$ parametrically depends on the growth orientation via the angles $\varphi$ and $\theta$, see Fig.~\ref{fig:Geometry-finfet}, and its general expression can be obtained by combining Eqs.~\eqref{eq:momenta-3x3} and~\eqref{eq:H12-general}; when $\theta=0$, $H_{12}$ is explicitly given in Eq.~\eqref{Eq:12times12E}.
By using a second order Schrieffer-Wolff transformation \cite{WinklerSpinOrbitCoupling2003,bravyi2011schrieffer} on $H_{12}$, the intrinsic spin-orbit velocity $\pmb{v}_0$ can be written as 
\begin{equation}
\label{eq:SO-Vector-Equilateral}
\pmb{v}_0=\frac{\hbar}{m L} (\gamma_3-\gamma_2)\ \pmb{\alpha}_0(\theta,\varphi) \ ,
\end{equation}
where $\pmb{\alpha}_0(\theta,\varphi)$ is a dimensionless three-dimensional vector that characterizes strength and direction of the spin-orbit field and depends on the growth direction and on the Luttinger parameters. 
Importantly, for an equilateral triangle, the intrinsic spin-orbit interaction is a result of the anisotropy of the semiconductor and it vanishes when $\gamma_2=\gamma_3$.
For this reason, materials such as Silicon, where the anisotropy is large, are a convenient choice to study this effect.

The general dependence of these quantities on the growth directions in Silicon is discussed in App.~\ref{app:Effective Parameters-GD}, see, in particular, Fig.~\ref{fig:vel_GD}.
From the analysis, we observe that the maximal value of $|\pmb{v}_0|\approx 2.91\hbar/(mL)$ is reached when the wire extends along one of the crystallographic axes, i.e. when $z\parallel [100]$, $[010]$, or $[001]$. In this case, there is no spin-orbit coupling in the direction of the wire, i.e. $(\pmb{v}_0)_z=0$, and we define the complex quantity $\alpha_0(\theta)=(\pmb{\alpha}_0)_x(\theta,\varphi)+i(\pmb{\alpha}_0)_y(\theta,\varphi)$, where to simplify the notation we suppress the explicit dependence of $\alpha_0(\theta)$ on $\varphi$.
In particular, we find
\begin{subequations}
\label{eq:alpha_pt}
\begin{align}
\label{eq:alpha0}
\alpha_0(0)&=9.34 \frac{\gamma_3 }{\gamma_1+(5/2)\gamma_2} \ e^{-4i\varphi}\approx 2.631  \ e^{-4i\varphi} \ , \\
\label{eq:alphapi/2}
\alpha_0\!\left(\frac{\pi}{2}\right) &=\frac{a_1-a_2 \cos (4 \varphi )+a_3 \ (\gamma_3-\gamma_2)\cos(8\varphi) }{1-b_1 \  (\gamma_3-\gamma_2)\cos (4 \varphi )+b_2 \ (\gamma_3-\gamma_2)^2\cos (8 \varphi )} \ ,
\end{align}
\end{subequations}
where $a_i$ and $b_i$ are real functions of the Luttinger parameters, whose values for Silicon are shown in Table~\ref{tab:parameters_pi/2}. The numerical value $2.631$ in Eq.~\eqref{eq:alpha0} is obtained by using the Luttinger parameters of Si and describes the CA and the DRA.
From Eq.~\eqref{eq:alphapi/2}, we estimate $\alpha_0\!\left(\frac{\pi}{2}\right)\approx 2.354$ at $\varphi=-3\pi/4$, corresponding to the SA.

\begin{figure}[t]
\begin{center}
\includegraphics[width=0.45\textwidth]{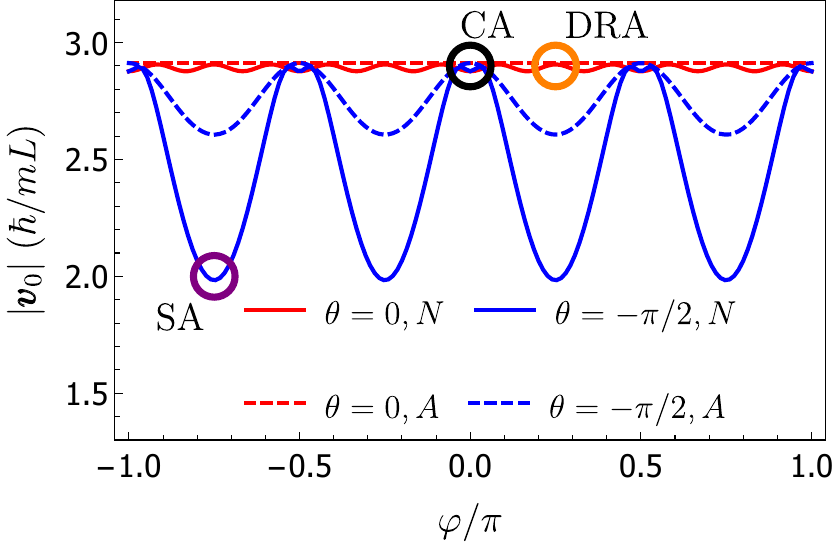}
\end{center}
\caption{ Intrinsic spin-orbit velocity $|\pmb{v}_0|$ without external electric fields in an equilateral FinFET. We compute $|\pmb{v}_0|$ as a function of the  angle $\varphi$ at $\theta=0$ (red lines) and  $\theta= -\pi/2$ (blue lines). The dashed lines are obtained by the approximate Eqs.~\eqref{eq:SO-Vector-Equilateral} and \eqref{eq:alpha_pt}, while the solid lines are obtained numerically by using Eq.~\eqref{eq:SO-Exact} and including  200 orbital states in Eq.~\eqref{eq:sol-LaplEq}. We mark with black, orange, and purple circles the results obtained for the relevant orientations of axes CA, DRA, and SA, respectively, see Fig.~\ref{fig:Geometry-finfet}b). }
	\label{fig:GD-dependence}
\end{figure}

\begin{table}[]
\centering
\begin{tabular}{|c|c|c|c|c|}
\hline
 $a_1$  & $a_2$ &  $a_3$ & $b_1$  & $b_2$\\ \hline
 2.468   & 0.683  &  0.013  &  0.298  & 0.011 \\ \hline
\end{tabular}
\caption{Parameters of the intrinsic spin-orbit vector $\pmb{v}_0$ in Eq.~\eqref{eq:alphapi/2} when $\theta=-\pi/2$. These parameters  describe the SA (CA) when $\varphi=-3\pi/4$ ($\varphi=0$).  \label{tab:parameters_pi/2} 
}
\end{table}

A comparison between the perturbative results in Eq.~\eqref{eq:SO-Vector-Equilateral}  and the exact spin-orbit velocity computed numerically by using Eq.~\eqref{eq:SO-Exact} and a larger number $N=200$ of the orbital basis states in Eq.~\eqref{eq:sol-LaplEq} is shown in Fig.~\ref{fig:GD-dependence}.
When the elevation angle is $\theta=0$, the wire extends along the $[001]$ direction, and the spin-orbit vector has a roughly constant amplitude, but  as a consequence of the four-fold rotational symmetry of the Luttinger-Kohn Hamiltonian, its direction oscillates as a function of the azimuthal angle $\varphi$ with period $\pi/2$. In contrast, when $\theta=\pm \pi/2$, the intrinsic spin-orbit vector has a constant direction ($\pmb{v}\parallel \textbf{e}_x$) and an oscillating amplitude that reaches its maximum when the wire is aligned to the main crystallographic axes $[100]$ or $[010]$. The minimal spin-orbit coupling occurs when the wire extends along the $[110]$ direction, which corresponds to the standard experimental growth direction (SA), see Fig.~\ref{fig:Geometry-finfet}b). 
We note that while Eq.~\eqref{eq:alpha0} agrees well with the numerically computed spin-orbit coupling,  Eq.~\eqref{eq:alphapi/2} captures the periodic oscillations but overestimates absolute value of the intrinsic coupling at the SA.

The dependence of the effective mass and of the spin-orbit length on the growth direction can also be  straightforwardly obtained, and a detailed analysis of these quantities is given in App.~\ref{app:Effective Parameters-GD}.
Importantly, the spin-orbit length defined in Eq.~\eqref{eq:SO-length} is of the order of the side  $L$ of the triangle, which is typically much smaller than the lateral size of the dot, and it reaches the minimal value $l_{so}\approx 0.83 L$ when $z\parallel[001]$. Consequently, we expect hole Si FinFETs to show effects due to  large spin-orbit interaction such as renormalization of the $g$-factor \cite{froning2020strong,PhysRevB.93.121408,voisin2016electrical,marx2020spin,PhysRevLett.120.137702,PhysRevB.87.161305} and ultrafast Rabi oscillations \cite{froning2020ultrafast}.

\subsection{Homogeneous electric field}
\label{Sec:Homogeneous_E}

\begin{figure}[t]
\centering
	\includegraphics[width=0.45\textwidth]{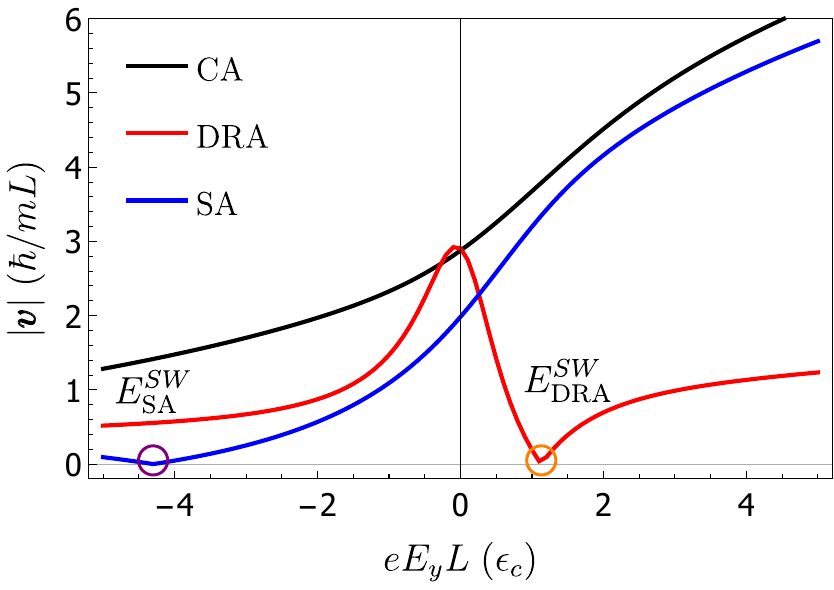}
	\caption{Spin-orbit velocity $|\pmb{v}|$ as a function of the homogeneous electric field $E_y$ in an equilateral FinFET. We neglect the inhomogeneous contribution to the electric field profile and we compute $|\pmb{v}|$ for the growth directions in Fig.~\ref{fig:Geometry-finfet}b). 
	At negative (positive) electric fields, i.e. when the hole wavefunction is pushed to the bottom (apex) of the triangle,  $|\pmb{v}|$ can be zero when the wire is grown along the SA (DRA). We show the sweet spot $E_\text{SA}^{SW}$  ($E_\text{DRA}^{SW}$) see Eq.~\eqref{eq:SW-Homogeneneous}, with a purple (orange) circle.
 }
	\label{fig:alpha_hom_E}
\end{figure}

We now analyze the effect of the electrostatic potential $V_E$ in Eq.~\eqref{eq:Multipole-E} on the spin-orbit velocity.
For Si wires, there are two distinct mechanisms that suppress the spin-orbit interactions and eventually remove them entirely. 
These mechanisms are the application of a homogeneous electric field $E_y$ pointing along the $y$ direction and the application of inhomogeneous electric fields $\delta\!E_{xx}$ and $\delta\!E_{yy}$ that harmonically confine the wavefunction inside the fin. 
Both these fields are controllable by the external potential, and their strengths strongly depend on the gate design, see App.~\ref{app:E-field} for a more detailed analysis.

Let us first consider the effect of homogeneous electric fields, which models  setups where the non-linearities of the fields are suppressed.
In realistic devices the top gate covers the wire rather symmetrically with respect to the $y$-direction. For this reason, the homogeneous electric field $E_x$ in the $x$-direction is zero and we will neglect it in the present analysis \footnote{The results presented in Sec.~\ref{Sec:Homogeneous_E} are qualitatively valid even in the presence of asymmetries that lead to a finite $E_x$. In particular, we find that the value of the spin-orbit velocity at $E_y=0$ is increased by $E_x$ and consequently the spin-orbit switch is pushed to higher values of $E_y$.}.

In Fig.~\ref{fig:alpha_hom_E}, we show the results of a numerical analysis showing the dependence of the spin-orbit velocity on the dipole energy $eE_y L$ for the growth directions given in Fig.~\ref{fig:Geometry-finfet}b). These results are obtained by  projecting the Hamiltonian~\eqref{eq:total-H} onto the first 200 orbital states in Eq.~\eqref{eq:sol-LaplEq} and using Eq.~\eqref{eq:SO-Exact}.
We observe that the electric field strongly influences the spin-orbit field and it can increase it or decrease it depending on the orientation of the wire. In particular, for the DRA and the SA, the spin-orbit field can be exactly switched off when the dipole energy $eE_yL$ becomes comparable to the confinement energy $\epsilon_c$.  More precisely, the spin-orbit switch occurs at
\begin{subequations}
\label{eq:SW-Homogeneneous}
\begin{align}
\label{eq:SW-Homogeneneous-DRA}
E_\text{DRA}^{SW} & \approx 1.13 \frac{\epsilon_c}{eL} \approx 19.42 \times  \frac{10^3\text{nm}^3}{L^3}  \text{V}/ \mu\text{m}  \ , \\
E_\text{SA}^{SW} & \approx -4.3 \frac{\epsilon_c}{eL} \approx - 73.9 \times  \frac{10^3\text{nm}^3}{L^3}  \text{V}/ \mu\text{m}  \ .
\end{align}
\end{subequations} 
for the DRA and SA case, respectively. 
For realistic cross-section with sides of a few tens of nanometers, these electric fields are of the order $\text{V}/ \mu\text{m}$, easily reachable in state-of-the-art devices.
For the DRA, the spin-orbit coupling is removed when the electric field is positive and the holes are pushed to the apex of the fin, while for the SA growth direction, a negative field is required and the holes are  pushed to the bottom of the triangle. Consequently, the SA is convenient in SOI FinFETs, where the wire is separated from the substrate by an oxide and the hard-wall boundary condition is a good approximation also at the bottom of the fin. In contrast, the DRA can also be suitable in bulk FinFETs because the large positive electric field confines the hole wavefunction in the fin and suppresses the leakage of the wavefunction into the substrate.

The suppression of the spin-orbit interaction comes from an interplay between the anisotropy of Si and the reduced symmetry of the cross-section.
For simplicity, we focus on wires where $z\parallel [001]$, i.e. $\theta=0$, and examine the dependence of $\pmb{v}$ on $\varphi$. In this limit, a reasonable description of the system is provided by the reduced Hamiltonian $H_{12}$ in Eq.~\eqref{Eq:12times12E}, which includes the lowest three orbital states given in Eq.~\eqref{eq:sol-LaplEq}. 
By applying a fourth order Schrieffer-Wolff transformation to $H_{12}$,  we find that $v_z=0$, and we can write the complex off-diagonal matrix element of the spin-orbit velocity $v=(\pmb{v})_x+i(\pmb{v})_y$ as the sum of an anisotropic term $\alpha_A$ and an isotropic direct Rashba-like term $\alpha_I$, i.e. 
\begin{equation}
\label{eq:v_E_funct_Ey}
v= \frac{\hbar}{m L}\Big[(\gamma_3-\gamma_2) \alpha_A(\varphi) + (\gamma_3+\gamma_2) \alpha_{I}\Big] \ .
\end{equation}
Importantly, $\alpha_I$ is a real function of $eE_y L$ and it does not depend on the azimuthal growth angle $\varphi$, while $\alpha_A$ is a complex function of $eE_y L$ and oscillates as a function of $\varphi$. In particular, for small electric fields and in Silicon we find
\begin{subequations}
\label{eq:alpha_E}
\begin{align}
\label{eq:alpha_E-aniso}
\alpha_A(\varphi)&\approx \alpha_0- 0.367 e^{-4i\varphi} \left(e E_y L/\epsilon_c\right)^2 \ , \\
\label{eq:alpha_E-iso}
\alpha_I&\approx  0.35 \left(e E_y L/\epsilon_c\right)+ 0.246 \left(e E_y L/\epsilon_c\right)^2 \ ,
\end{align}
\end{subequations}    
where
$\alpha_0$ is the zero field result in Eq.~\eqref{eq:alpha0} and in $\alpha_A$ we also neglected quantitatively small corrections linear in $e E_y L/\epsilon_c$ and proportional to different powers of $e^{-4i\varphi}$. For the general dependence of these parameters on the Luttinger parameters, see Eq.~\eqref{eq:so-velocity-expansion}. 
To better convey the importance of the shape of the cross-section, in App.~\ref{app:square-cross-sect}, we highlight the main differences in the spin-orbit coupling of triangular and square Si wires.

The spin-orbit interaction is fully switched off when $|\pmb{v}|=0$. From Eq.~\eqref{eq:alpha_E}, it follows that this cancellation can only occur at the growth angles $\varphi=\pi(2n+1)/4 $ where $\alpha_A$ is a real-valued function, and where $\alpha_0$ has a sign opposite to  all the electric field-dependent terms.
From Eqs.~\eqref{eq:v_E_funct_Ey} and \eqref{eq:alpha_E}, we estimate that the switch in Silicon occurs at $E_\text{DRA}^{SW}\approx 1.52 \epsilon_c/(eL) $. 
Note that our perturbative analysis provides good qualitative insights into the switching mechanism, and in addition, the numerical prefactor $1.52$ is reasonably close to the prefactor $1.13$, derived from the detailed numerical analysis including  higher orbital states.

\subsection{Inhomogeneous electric field}
\label{Sec:Inhomogeneous_E}

The electric field profile in a triangular FinFET comprises a large inhomogeneous component that significantly alters the spin-orbit velocity.  
In this section, we restrict ourselves to the analysis of Si FinFETs where a  negative gate potential $V_g$ is applied, such that the hole wavefunction is pushed to the apex of the triangle.

When the back gate is far from the wire, the inhomogeneous component of the electric field in the cross-section can be well-approximated by a linearly varying electric field parametrized by the tensor $\delta\!E_{ij}$, see Eq.~\eqref{eq:Multipole-E}.
As discussed in App.~\ref{app:E-field}, in typical devices, the cross-terms $\delta\!E_{xy}$ are very small. In contrast, the diagonal components $\delta\!E_{xx}$ and the $\delta\!E_{yy}$ are large and they vary linearly with the top gate potential $V_g$, with slopes that are comparable in absolute value, but have opposite signs.
In particular, we estimate  $\delta\!E_{xx}/\delta\!E_{yy}\approx -1.01$, and thus the saddle potential energy 
\begin{equation}
\label{eq:V_E_inhomogeneous}
V_E(x,y)= e \delta \! E(y^2-x^2)/2 \ ,
\end{equation} 
accurately describes the inhomogeneity of the electric field. Negative values of $V_g$ correspond to positive values of the parameter $\delta\!E$ and so $V_E(x,y)$ harmonically confines the holes in the $y$-direction and pushes their wavefunction to the sides of the triangle in the $x$-direction.

We first study separately the effect of the inhomogeneous coupling by setting the homogeneous electric field to zero, i.e. $E_y=0$. We remark that for inversion symmetric cross-sections such as cylindrical or rectangular wires, the potential in Eq.~\eqref{eq:V_E_inhomogeneous} does not induce any  spin-orbit interactions because $V_E({x,y})=V_E({-x,-y})$, and the results obtained in this section are specific for triangular wires. 

\begin{figure}[t]
\centering
\includegraphics[width=0.45\textwidth]{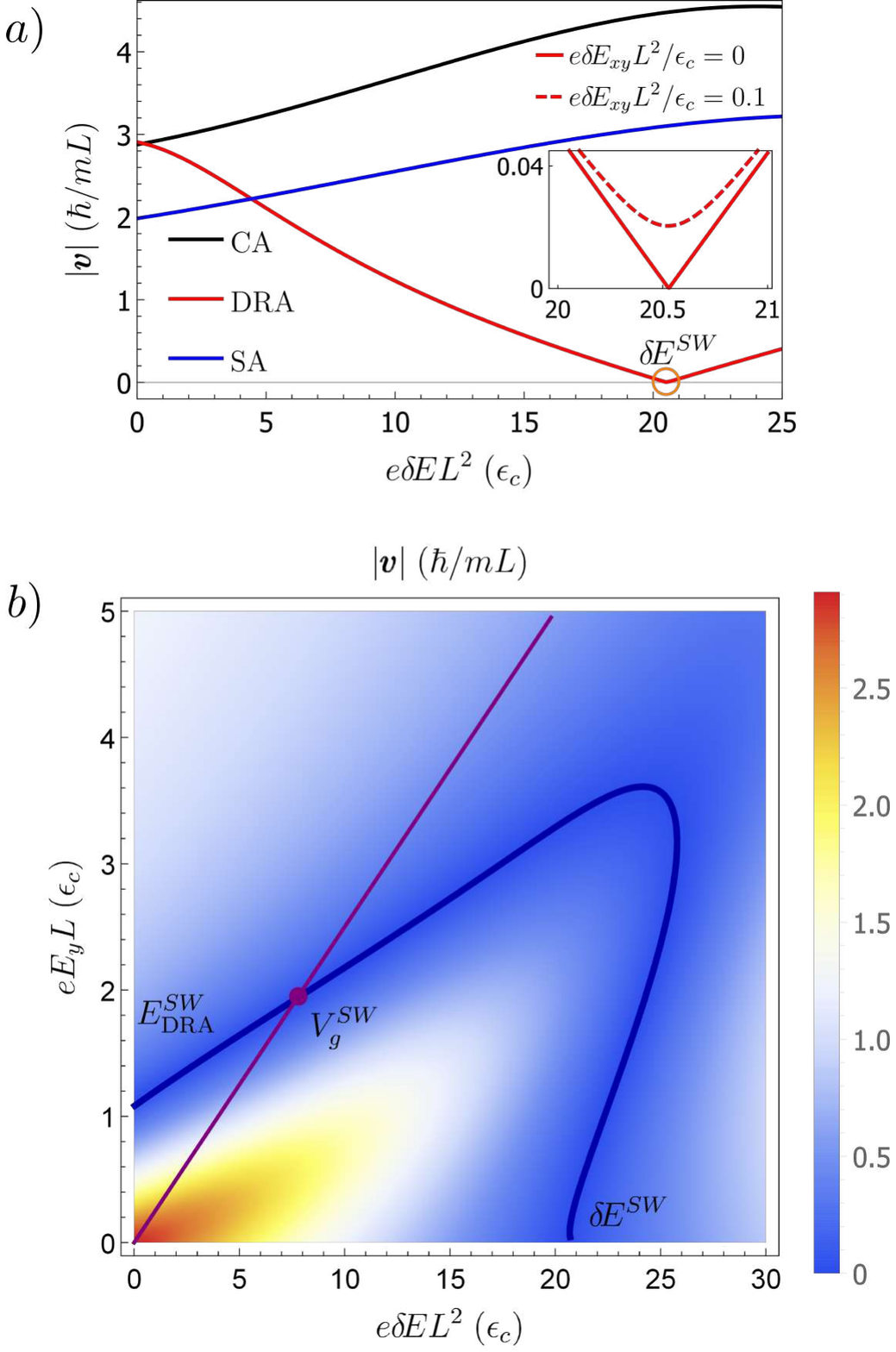}
\caption{Spin-orbit velocity $|\pmb{v}|$ as a function of the inhomogeneous electric field $\delta\!E$ in an equilateral FinFET. 
In a) we use the confinement potential in Eq.~\eqref{eq:V_E_inhomogeneous} without the homogeneous field $E_y$ and study how $|\pmb{v}|$ varies for the wire orientations in Fig.~\ref{fig:Geometry-finfet}b); for the DRA device, $|\pmb{v}|$ vanishes at the field $\delta\!E^{SW}$ given in Eq.~\eqref{eq:deltaEsw}. In the inset, we show the effect of $\delta\!E_{xy}$ for the DRA close to the switching field $\delta\!E^{SW}$.
In b) we show how $|\pmb{v}|$  varies for the DRA device when $\delta\!E$ and $E_y$ are tuned independently. Here, $|\pmb{v}|$ vanishes along the blue curve that connects $E_\text{DRA}^{SW}$ and $\delta\!E^{SW}$. In the device studied here,  $E_y$ and $\delta\!E$ are constrained on the purple line defined by Eq.~\eqref{eq:lineEy-deltaE}. From the intersection of the purple and blue lines, one finds the gate potential $V_g^{SW}$ of the spin-orbit switch.  }
	\label{fig:alpha_inhom_E}
\end{figure}

In Fig.~\ref{fig:alpha_inhom_E}a), we show the spin-orbit interactions as a function of $\delta\!E$ and for the orientations in Fig.~\ref{fig:Geometry-finfet}b).
We observe a qualitatively similar picture as discussed in Sec.~\ref{Sec:Homogeneous_E} for the homogeneous electric field: depending on the growth direction, the spin-orbit velocity varies with $\delta\!E$, and while it increases in the CA and SA, it decreases for the DRA, resulting in an operational sweet spot at
\begin{equation}
\label{eq:deltaEsw}
\delta\! E^{SW}\approx 20.5 \frac{\epsilon_c}{eL^2} \approx 35 \times 10^3\times \frac{10^4\text{nm}^4}{L^4}  \text{V}/ \mu\text{m}^2 \ ,
\end{equation} 
where $|\pmb{v}|$ vanishes.

This spin-orbit sweet spot remains present also when non-idealities of the electric fields are included. 
In App.~\ref{app:delta_Eii}, we show that the spin-orbit switch persists in rather general FinFET design, where $\delta\!E_{xx}\neq -\delta\!E_{yy}$.
Also, possible asymmetries of the gate design can lead to a small cross-coupling $\delta\!E_{xy}$.
The effect of  $\delta\!E_{xy}$ on the spin-orbit velocity is shown in the inset of Fig.~\ref{fig:alpha_inhom_E}a).
We find that the spin-orbit switch is removed by including $\delta\!E_{xy}$. However, we expect that in most setups, $\delta\!E_{xy}$ remains a few orders of magnitude smaller than $\delta\!E$ and thus, at the switch, the spin-orbit velocity is orders of magnitude smaller than the intrinsic velocity $|\pmb{v}_0|$, still providing a good working point where charge noise is strongly reduced. \\

In realistic devices, the effects of homogeneous and linearly varying electric fields cannot be easily decoupled and thus we now examine how their interplay affects the spin-orbit switch. 
 In App.~\ref{app:E-field}, we estimate that in the FinFET shown in Fig.~\ref{fig:Geometry-finfet}
\begin{equation}
\label{eq:lineEy-deltaE}
E_y  \approx -0.3 \frac{V_g}{d_B} \approx 0.25 L \delta\!E \ .
\end{equation}
While this constraint is strictly valid only for setups with top and back gates, we now explore a much larger parameter space where $E_y$ and $\delta\!E$ are varied independently. Consequently, our results are valid for a broad range of device designs, including for example Si FinFETs with gates placed  sequentially along the wire.

In Fig.~\ref{fig:alpha_inhom_E}b), we show the spin-orbit velocity in a FinFET grown along the DRA as a function of both $E_y$ and $\delta\!E$. 
We observe that  $|\pmb{v}|$ vanishes along a (blue) curve  that intersects the (purple)  line defining the constraint~\eqref{eq:lineEy-deltaE} at the point $(e\delta\!E L^2,eE_yL)=(7.8\epsilon_c,1.95\epsilon_c)$. 

The switching-off of the spin-orbit velocity is mostly driven by the homogeneous electric field $E_y$ and the inhomogeneous potential $\delta\!E$ only renormalizes the value of $E_y$ required to compensate for the intrinsic spin-orbit interaction.
Combining with Eq.~\eqref{eq:lineEy-deltaE}, this point corresponds to the potential $V_g^{SW}=-111.7 \times  \text{nm}^2(d_B/L^3)$~V and when $L=20$~nm and $d_B=100$~nm, one obtains the working point $V_g^{SW}=-1.4$~V.
 Perturbative expressions of $|\pmb{v}|$ as a function of  $E_y$ and $\delta\!E$ can be found by generalizing the treatment discussed in  Sec.~\ref{Sec:Homogeneous_E} and are given in App.~\ref{app:electric-field}, see Eq.~\eqref{eq:so-velocity-expansion}. \\

Another important feature of the spin-orbit switch is its robustness against moderate strain.
In semiconductor nanostructures, strain can play a relevant role by  renormalizing the response of the system to external fields \cite{lee2005strained}.
In Si wires, strain could be induced for example by nearby metallic gates \cite{liles2020electrical}, or by incoherent interfaces between $\text{SiO}_2$ and Si \cite{niquet2012effects,stan2010compressive}.
The precise strain profile is strongly device dependent and it can be engineered by a careful fabrication process, where details such as choice of the materials \cite{thorbeck2015formation} and dielectric thickness matter.
Instead of focusing on a specific device realization, here we  consider a simple strain model where the elements of the strain tensor $\epsilon_{ij}$ are homogeneous in the fin.
By using the Bir-Pikus Hamiltonian \cite{WinklerSpinOrbitCoupling2003,bir1974symmetry}, we estimate that the analysis provided above is qualitatively valid as long as 
\begin{subequations}
\label{eq:max-strain}
\begin{align}
{\epsilon_{xx}+\epsilon_{yy}}-2\epsilon_{zz} &\in[0.54\%,-1.6\%]\times{(L/10\ \text{nm})^{-2}} \ , \\
|\epsilon_{xx}-\epsilon_{yy}|&<0.12 \% \times{(L/10\ \text{nm})^{-2}} \ , \\
|\epsilon_{xz}|\ , \  |\epsilon_{yz}|&<0.08 \% \times{(L/10\ \text{nm})^{-2}} \ , \\
|\epsilon_{xy}|&<0.9 \% \times{(L/10\ \text{nm})^{-2}} \ .
\end{align}
\end{subequations}
Consequently, for typical cross-sections where $L$ is a few tens of nanometers, reasonable values of the strain parameters $\epsilon_{ij}\sim 0.1\%$ can still preserve the spin-orbit switch.
A detailed analysis of the effect of strain as well as a justification for the homogeneous model  is provided in App.~\ref{app:strain}.

\section{Effect of the SOHs}
\label{Sec:SOHS}

In Sec.~\ref{Sec:Equilateral}, the effective spin-orbit velocity  is computed by using the $4\times 4$ LK Hamiltonian in Eq.~\eqref{eq:LK_Hamiltonian_4}, which describes the mixing of heavy and light holes. In this case, nanowires with equilateral triangular cross-sections having different sides $L$ show the same qualitative behaviour, and $L$ only sets the scale of the spin-orbit velocity $v\propto \hbar/(mL)$ and of the confinement energy $ \epsilon_c\propto \hbar^2/(mL^2)$, see Eqs.~\eqref{eq:v_E_funct_Ey} and \eqref{eq:Confinement_energy}, respectively.
This model is valid for wires with a large cross-section, where  $\epsilon_c$ is the smallest energy scale and one can neglect the coupling to the split-off holes, gapped by a large energy $\Delta_0$. In contrast, for small Si wires, $\Delta_0$ is comparable with $\epsilon_c$ and because of the influence of the SOHs, the ground-state dynamics of the wire depends non-trivially on the side $L$ of the cross-section \cite{del2020vanishing}.
As anticipated in Sec.~\ref{Sec:Model}, to study this dependence we use the $6\times 6$ LK Hamiltonian \cite{WinklerSpinOrbitCoupling2003}, fully accounting for the SOHs. 

\begin{figure}[t]
	\includegraphics[width=0.45\textwidth]{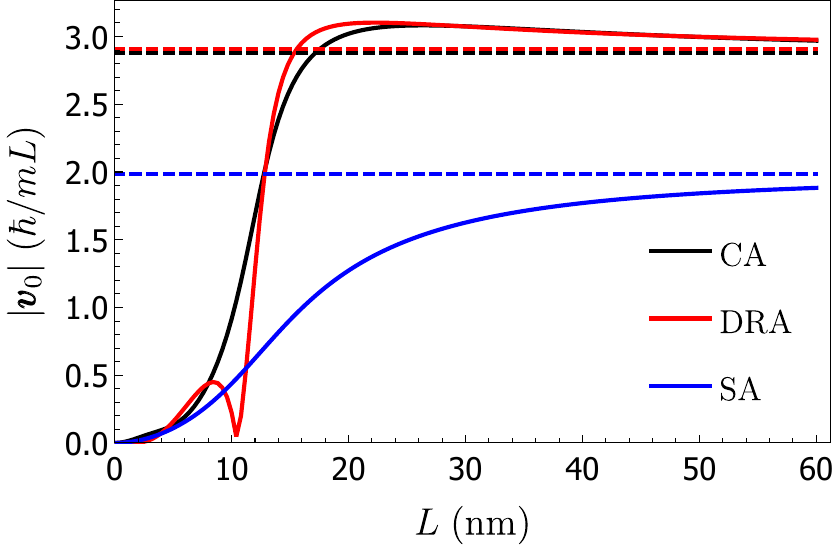}
	\caption{Intrinsic spin-orbit velocity $|\pmb{v}_0|$ as a function of the side length $L$ of an equilateral FinFET. We compare the effect of the SOHs in wires grown along the orientations in Fig~\ref{fig:Geometry-finfet}b) by showing with solid (dashed) lines the values of $|\pmb{v}_0|$ obtained by the $6\times 6$ ($4\times 4$) LK Hamiltonian that includes (neglects) the SOHs.   }
	\label{fig:Length_dependence}
\end{figure}

In Fig.~\ref{fig:Length_dependence}, we show how the intrinsic spin-orbit velocity  $|v_0|$ is modified by the SOHs in wires with different cross-section sides $L$. 
For large wires, with $L\gtrsim 35$~nm, the contribution of the SOHs is small and $|v_0|$ approaches the value in Eq.~\eqref{eq:SO-Vector-Equilateral} obtained with the $4\times 4$ LK Hamiltonian (dashed lines). 
In contrast, for very small wires, with $L\lesssim 10$~nm, the SOHs strongly suppress $|v_0|$.
For moderately small wires, because of the anisotropy of Si,  the effect of the SOHs strongly depends on the growth direction. In fact, while for the SA  $|v_0|$ decreases monotonically, when $\theta=0$, the spin-orbit velocity overshoots and reaches a maximum at $L\sim 20$~nm, where $|v_0|$ is larger than the value obtained for the $4\times 4$ LK Hamiltonian. In addition, at $L\sim 10$~nm we observe that for the DRA, the intrinsic spin-orbit interaction can be exactly cancelled, restoring the usual direct Rashba spin-orbit coupling typical of inversion symmetric cross-sections \cite{DRkloeffel1,DRkloeffel3}.

\begin{figure}
\centering
	\includegraphics[width=0.4\textwidth]{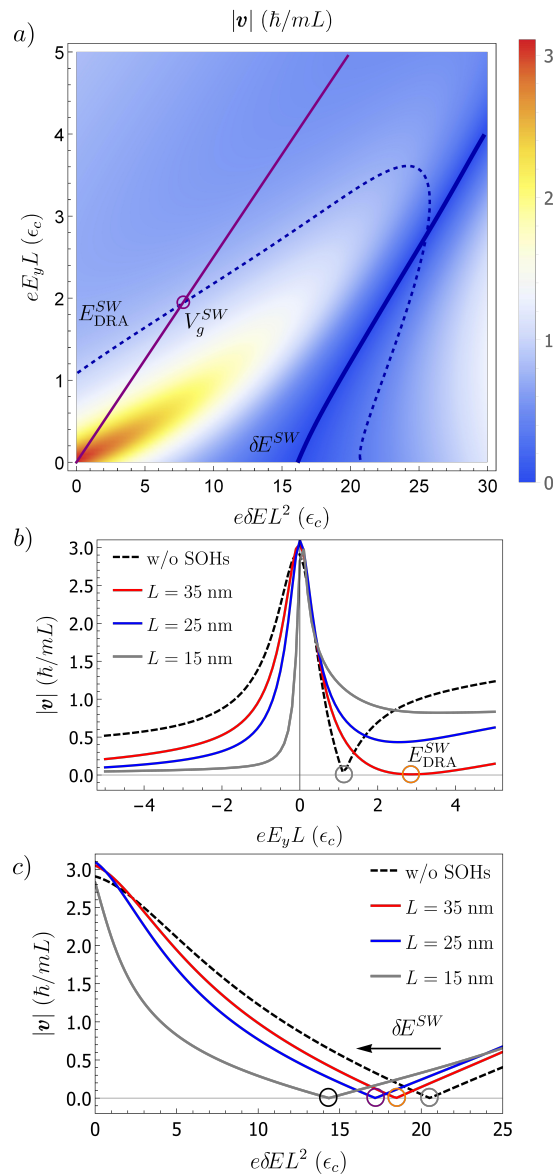}
	\caption{ Effect of the SOHs on the spin-orbit switch in small equilateral wires grown along the DRA. 	In a), we show how $|\pmb{v}|$ varies in a wire with cross-section $L=20$~nm as a function of homogeneous and inhomogeneous electric field, $E_y$ and $\delta\!E$, respectively. When the SOHs are included, the spin-orbit switch driven by $E_y$ is removed and $|\pmb{v}|$ vanishes only because of the inhomogeneous field $\delta\!E$. To facilitate the comparison with Fig.~\ref{fig:alpha_inhom_E}b), we show with a dashed blue line the curve along which $|\pmb{v}|$ vanishes in large wires. Importantly, when $\delta\!E$ and $E_y$ are constrained on the purple line [see Eq.~\eqref{eq:lineEy-deltaE}], the SOHs remove the spin-orbit switch at $V_g^{SW}$.
In b) and c) we study how the SOHs affect the dependence of  $|\pmb{v}|$ on $E_y$ and $\delta\!E$ when the cross-section side $L$ is varied. While the homogeneous spin-orbit switch $E_\text{DRA}^{SW}$ is removed for wires with $L\lesssim 35$~nm, the inhomogeneous switch $\delta\!E^{SW}$ remains and is pushed to lower values as the side length decreases. 
In the units used, the results obtained without the SOHs are independent of  $L$. }
	\label{fig:Length_dependence_E}
\end{figure}

The SOHs strongly affect the response of the system to external electric fields.
For example, we analyze here the spin-orbit velocity in wires with the DRA, where in Sec.~\ref{Sec:Equilateral}, we predict that the intrinsic spin-orbit interaction can be exactly cancelled by a positive homogeneous electric field $E_y$ and by an inhomogeneous field $\delta\!E$. 
In  Fig.~\ref{fig:Length_dependence_E}, we show how the spin-orbit switch-off mechanism discussed in Sec.~\ref{Sec:Homogeneous_E} and Sec.~\ref{Sec:Inhomogeneous_E} are modified by the SOHs.
In particular, in Fig.~\ref{fig:Length_dependence_E}a) we show the combined effect of $E_y$ and $\delta\!E$ on the spin-orbit velocity $v$ when the SOHs are accounted for.
We study here an equilateral triangle of side $L=20$~nm, which maximises the intrinsic spin-orbit coupling, see Fig.~\ref{fig:Length_dependence}, and is easily achievable in state-of-the-art devices \cite{geyer2020silicon,siliconFin,kuhlmann2018ambipolar}. 
Comparing to Fig.~\ref{fig:alpha_inhom_E}b), where $v$ is obtained by using the $4\times 4$ LK Hamiltonian, we observe that the SOHs drastically alter the response of the wire and they remove the charge noise sweet spot produced by the homogeneous electric field $E_y$, while maintaining the sweet spot resulting from the inhomogeneous field $\delta\!E$. This latter sweet spot persists also when a strong homogeneous electric field $E_y$ is present and $E_y$ only pushes the switch-off field $\delta\!E^{SW}$ to larger values. In the plot, we show with a dashed line the curve along which the spin-orbit vanishes when the SOHs are neglected. Importantly, for the simple gate design studied here, where $E_y$ and $\delta\!E$ are constrained along the purple line in the figure, the charge noise sweet spot is removed by the SOHs.

To have a better understanding of the system, we show in Figs.~\ref{fig:Length_dependence_E}b) and~\ref{fig:Length_dependence_E}c) the effect of $E_y$ and $\delta\!E$ in wires grown along the DRA and having different cross-section side $L$.
We observe that when $L\lesssim 35$~nm, the spin-orbit switch at $E^{SW}_\text{DRA}$ is removed by the SOHs. In contrast, the SOHs enhance the effect of $\delta\!E$ and the spin-orbit switch at  $\delta\!E^{SW}$ persists in small wires and is pushed to lower values. 
A more detailed analysis of the inhomogeneous electric field response, including  the  general dependence on $\delta\!E_{xx}$ and $\delta\!E_{yy}$ for different growth directions is given in App.~\ref{app:delta_Eii}. \\

A more careful device design can minimize the effect of the SOHs.
For example, in wires grown along the DRA, the spin-orbit switch is restored in wide isosceles FinFETs.
To understand this result, we analyze  the spin-orbit coupling in isosceles triangular wires with equal sides $L_y$ and base $L_x$, see Fig.~\ref{fig:Geometry-finfet}. 
The aspect ratio is
\begin{equation}
\label{eq:aspect-ratio}
r\equiv L_x/L_y=2\sin(\Theta/2)\in (0,2) \ .
\end{equation}  
The FinFET is equilateral when $r=1$, and we call wide and narrow FinFETs the devices with  $r>1$  and $r<1$, respectively [$\Theta$ is the apex angle of the fin]. For convenience, we also define an effective  length $\tilde{L}= L_y \sqrt{\sin(\Theta)/\sin(\pi/3)}$, that is the side of an ideal equilateral triangle with the same area of the isosceles triangle. We redefine the confinement energy $\epsilon_c$ in Eq.~\eqref{eq:Confinement_energy} by the substitution $L\rightarrow \tilde{L}$.

\begin{figure}[t]
\includegraphics[width=0.48\textwidth]{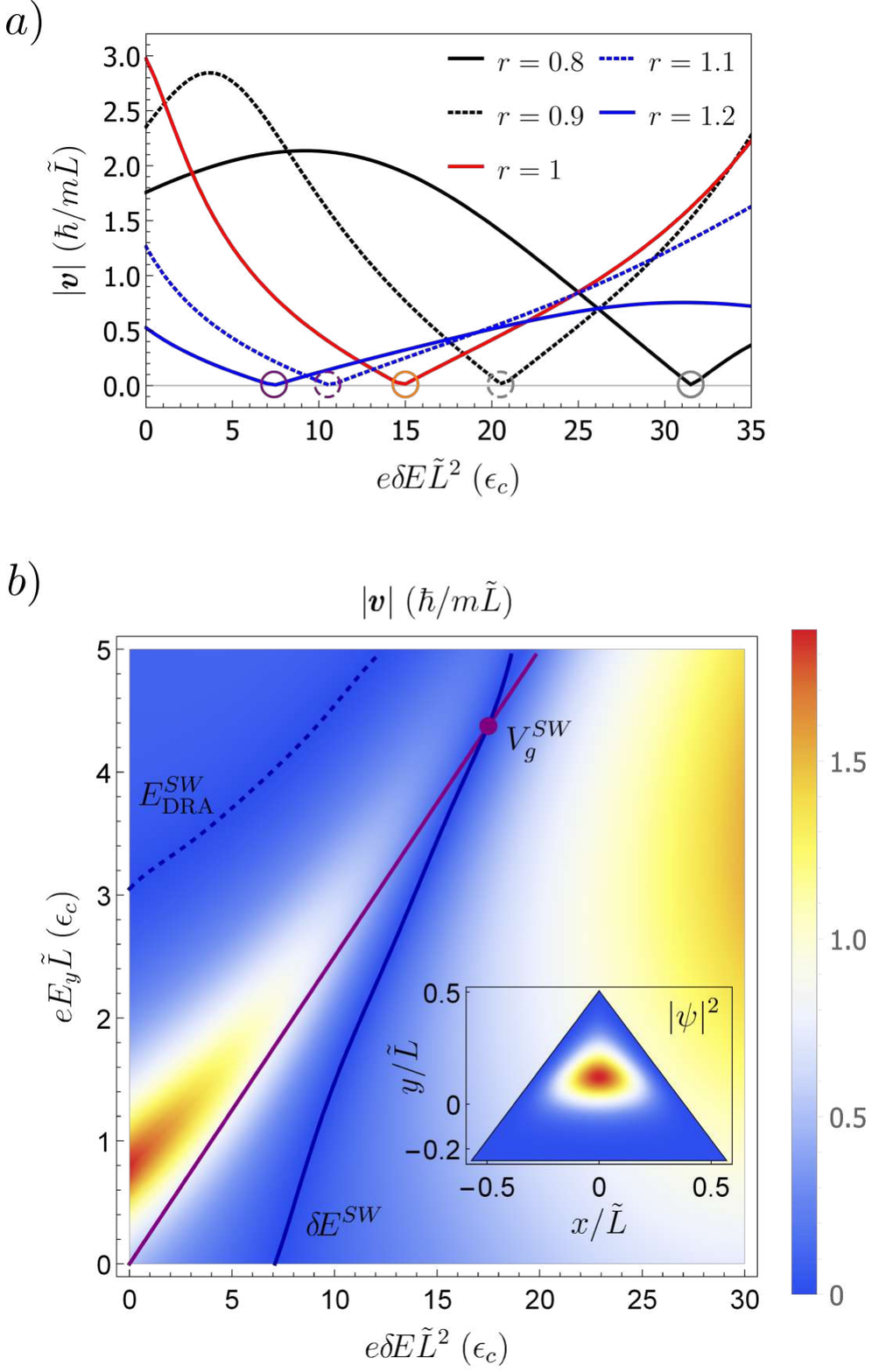}
	\caption{ Spin-orbit velocity $|\pmb{v}|$ in a wide FinFET grown along the DRA.
	In a) we show how the dependence of $|\pmb{v}|$ on the inhomogeneous electric field $\delta\!E$ varies as a function of the aspect ratio $r$ [see Eq.~\eqref{eq:aspect-ratio}] of the triangular cross-section. We compare triangles with the same area $\sqrt{3}\tilde{L}^2/4$, with effective length $\tilde{L}=20$~nm.  When $r>1$ ($r<1$) the FinFET is wide (narrow). 
	In b) we show $|\pmb{v}|$ as a function of $E_y$ and $\delta\!E$ for a device with $r=1.2$ and $\tilde{L}=20$~nm. The solid blue line shows the spin-orbit switch driven by the inhomogeneous field $\delta\!E$.  The spin-orbit switch  driven by the homogeneous field $E_y$ is also restored and $|\pmb{v}|$ vanishes along the dashed blue curve. In the  FinFET studied here, $\delta\!E$ and $E_y$ are constrained along the purple line defined by Eq.~\eqref{eq:lineEy-deltaE}.  
	In the inset, we show the density $|\psi|^2$ of the hole wavefunction in the cross-section at the switching point $V_g^{SW}$ where purple and blue lines intersect. The hole density vanishes in the blue region and attains maximal value in the red region.
}
	\label{fig:isosceles_dE_r}
\end{figure}

In Fig.~\ref{fig:isosceles_dE_r}a), we  focus on cross-sections with  $\tilde{L}=20$~nm and we examine  the dependence of $|\pmb{v}|$ on $\delta\!E$ when $E_y=0$ and when different values of $r$ are considered. 
These numerical results are obtained by using Eq.~\eqref{eq:SO-Exact} and by discretizing the $6\times 6$ LK Hamiltonian in isosceles triangular cross-sections. Here, we use the approximate potential in Eq.~\eqref{eq:V_E_inhomogeneous}; the limits of this approximation in isosceles fins are discussed in App.~\ref{app:delta_Eii}.
Importantly, we observe that $\delta\!E$ can remove the spin-orbit interactions for a broad range of $r$ and that $\delta\!E^{SW}$ is  significantly reduced when the fin is wide.  

This enhancement of the inhomogeneous field is crucial to restore the spin-orbit sweet spot. In fact, in Fig.~\ref{fig:isosceles_dE_r}b), we show the simultaneous effect of  $\delta\!E$ and  $E_y$ when $r=1.2$ and $\tilde{L}=20$~nm, i.e. $L_x=22.8$~nm and $L_y=19$~nm. In the FinFET design shown in Fig.~\ref{fig:Geometry-finfet}, where $\delta\!E$ and $E_y$ are constrained on the (purple) line defined by Eq.~\eqref{eq:lineEy-deltaE}, we find that the spin-orbit coupling can be switched off by the inhomogeneous electric field at the gate potential $V_g^{SW}\approx -249\times \text{nm}^2 d_B/\tilde{L}^3$~V. This value corresponds to $V_g^{SW}\approx -3.12$~V when $\tilde{L}=20$~nm and the back gate is $d_B=100$~nm apart from the center of mass of the wire. 
This gate potential  is rather large, but it can be reduced by placing a back gate closer to the fin. We believe that an optimized electrostatic design of the device can also reduce $V_g^{SW}$, but we do not investigate this aspect further.
Interestingly, in this setup, we recover also the spin-orbit sweet spot $E_\text{DRA}^{SW}$ driven by the homogeneous field $E_y$, see the dashed line in the figure. 
In addition, in the inset of Fig.~\ref{fig:isosceles_dE_r}b), we show the total hole density $|\psi|^2$ at $V_g^{SW}$. Because the  wavefunction is strongly confined in the fin and has no support close to the bottom boundary, we expect the results presented here to be valid for both SOI and bulk FinFETs. 

Finally, we estimate that in the range of parameters considered, the spin-orbit length in this setup can be pushed down to a minimal value of $l_{so}^\text{min}\approx 1.5 \tilde{L}$ by reducing the amplitude of the gate potential. While still rather short, this length is longer than in the equilateral triangle, resulting in a smaller maximal spin-orbit coupling. Other possible SOI and bulk FinFETs designs where the spin-orbit switch is restored are discussed in App.~\ref{Sec:Deviation}. Amongst the setups analyzed, we chose to focus on the wide DRA FinFET because it guarantees the largest spin-orbit coupling when the interaction is turned on.

\section{Suppressing charge noise in FinFET Qubits}
\label{Sec:Charge-noise}

\begin{figure}[t]
\includegraphics[width=0.45\textwidth]{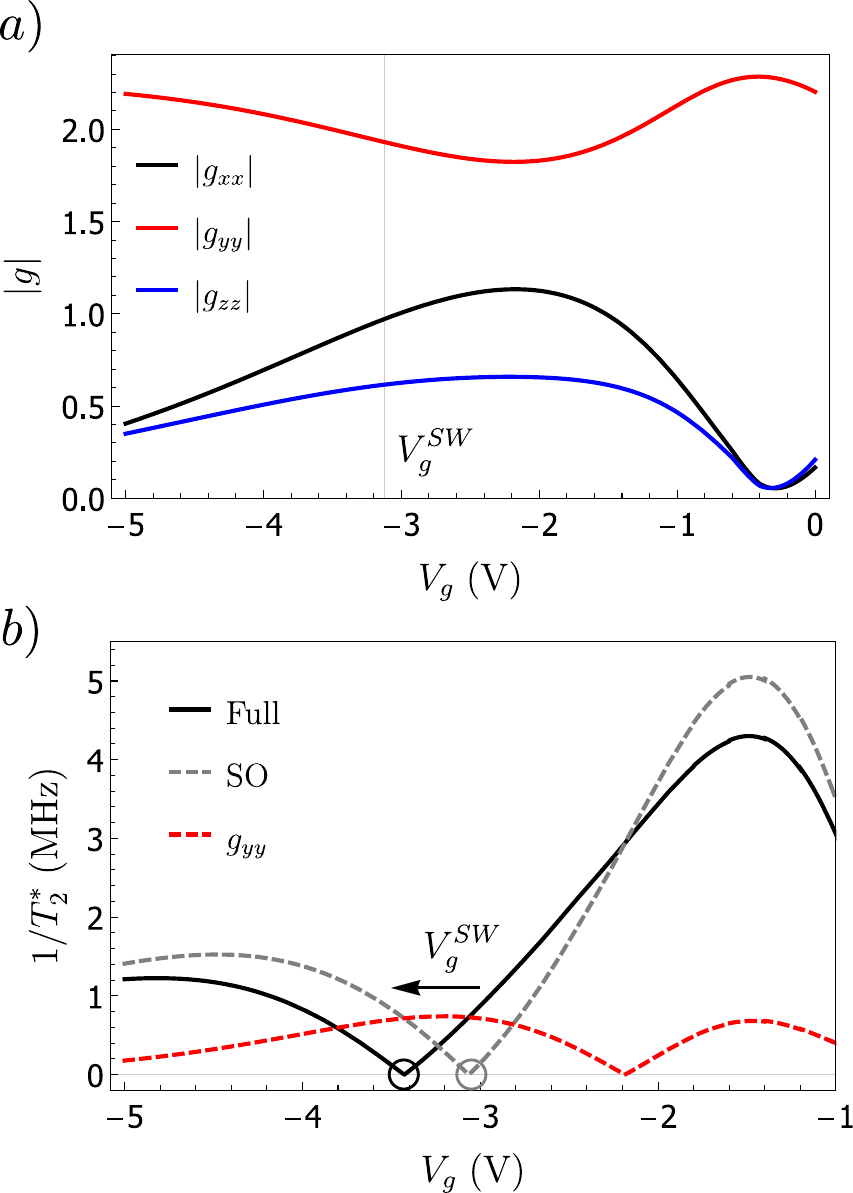}
	\caption{Anisotropic $g$-factor  and  dephasing rate $1/T_2^*$ of a Si FinFET qubit as a function of the gate potential $V_g$. We analyze a wide FinFET grown along the DRA with  $r=1.2$ and   $\tilde{L}=20$~nm.
	In a), we show the diagonal entries $g_{ii}$ of the wire $g$-factor matrix derived by considering a small magnetic field $B_i$ in the $i$-direction and diving $\pmb{\Delta}_i$ in Eq.~\eqref{eq:Delta_Zeeman} by $\mu_B B_i$. In this device, the off-diagonal components of the $g$-factor matrix vanish.
	 In b), the dephasing rate $1/T_2^*$ of the qubit caused by charge noise is obtained by combining Eqs.~\eqref{eq:charge-noise} and~\eqref{eq:T2-dep} when  $B_y=100$~mT. The black line represents the total dephasing of the qubit, while with dashed gray and red lines we show the contributions to $1/T_2^*$ of the spin-orbit coupling and of the $g$-factor fluctuations, respectively.  Because $\textbf{B}$ is applied along the $y$-direction, $\pmb{\Delta}_\parallel =0$ in Eq.~\eqref{eq:charge-noise} ($\pmb{v}$ points in the $x$-direction) and there is no relaxation ($\pmb{\Delta}_\perp'\parallel\pmb{\Delta}_\perp\parallel\textbf{Q}$). We fix $\delta\!E_{zz}/V_g$ such that at $V_g^{SW}=-3.12$~V  the lateral size of the dot in Eq.~\eqref{eq:l-harmonic} is $l=30$~nm.  The dependence of $l_{so}$ on $V_g$ is found by combining Eqs.~\eqref{eq:Multipole-E} and~\eqref{eq:lineEy-deltaE}. Here, we use $d_B=100$~nm, $\sqrt{\langle\delta V^2\rangle}=0.3$~$\text{mV}$ and $\omega_\text{ir}=1$~Hz.    \label{fig:t2}}	
\end{figure}

We now study the susceptibility to charge noise of an elongated quantum dot that defines a spin 1/2 qubit \cite{PhysRevA.57.120}.
All the results discussed in this section take fully into account the SOHs.
To define the dot, we include a confining potential in the direction of the wire $e\delta\!E_{zz}z^2/2$, which is assumed  to be much smoother than the cross-section side $L$; in this way, the effective wire Hamiltonian in Eq.~\eqref{eq:wire-Ham} accurately describes the system.
We emphasize that while here we discuss only elongated quantum dots, where $l\gg L$, our results apply also to FinFET-based spin qubits where $l\approx L$ \footnote{ 
A more precise condition for the validity of the nanowire Hamiltonian in Eq.~\eqref{eq:wire-Ham} is that the harmonic frequency $\hbar \omega_z=\hbar^2/(m^*l^2)$ is much smaller than the energy gap $\Delta E$ between the ground-state and the first excited state of the nanowire. For the devices considered and in the range of parameters shown the minimal energy gap is $\Delta E\sim 0.05 \epsilon_c$ at the switching field, and so we obtain the condition $1\gg \hbar\omega_z/\Delta E\sim {3L^2}/({0.05\times 16 \pi^2 \gamma_1 m^* l^2})\sim {L^2}/({\gamma_1 l^2})$, where we used the effective mass in Eq.~\eqref{eq:Effective_masses_zero_field}. Because of the prefactor $1/\gamma_1\sim 0.22$, the nanowire Hamiltonian works reasonably well when $L\approx l_{so}\approx l$.
}, such as the ones in \cite{geyer2020silicon,siliconFin}. 
A potential $e\delta\!E_{zz}z^2/2$ is typically generated by terminating the top gate above the FinFET such that it has a finite extension in the direction along the wire ($z$-direction). The precise value of $\delta\!E_{zz}$ depends on the size of the gate along $z$, on the distance $d_B$ of the back gate; also  $\delta\!E_{zz}$ varies linearly with the top gate potential $V_g$.

Without an external magnetic field  $\textbf{B}$, one can exactly gauge the spin-orbit coupling away by the unitary transformation $S=e^{-i \textbf{n}_{v} \cdot \sigma z/l_{so}}$ \cite{PhysRevB.67.115324}, where $\textbf{n}_v$ is the direction of the spin-orbit vector $\pmb{v}$, and the Hamiltonian reduces to a harmonic oscillator with frequency $\omega_z=\sqrt{e|\delta\!E_{zz}|/m^*}$, whose ground state wavefunction is a gaussian with standard deviation 
\begin{equation}
\label{eq:l-harmonic}
l=\sqrt{\frac{\hbar}{m^*\omega_z}}=\sqrt[4]{\frac{\hbar^2}{em^*|\delta\!E_{zz}|} } \ .
\end{equation}
Because $\delta\!E_{zz}\propto V_g$, the harmonic length $l$ depends on the gate potential as $l\propto |V_g|^{-1/4}$; the divergence of $l$ for $V_g\rightarrow 0$ is a consequence of the fact that in electrostatically defined quantum dots a finite value of the gate potential $V_g$ is required to confine the particles.

At finite values of $\textbf{B}$, the unitary $S$ leads to an effective magnetic field that oscillates as a function of the position along the wire. When projected onto the groundstate of the dot, these oscillations cause a spin-orbit coupling dependent renormalization of the $g$-tensor in the direction perpendicular to $\textbf{n}_v$ \cite{PhysRevB.87.161305,PhysRevB.93.121408,froning2020strong}.
In particular, decomposing the vector $\pmb{\Delta}$ defined in Eq.~\eqref{eq:Delta_Zeeman} into the sum of the two vectors $\pmb{\Delta}_\parallel$ and $\pmb{\Delta}_\perp$ that are parallel and perpendicular to the spin-orbit vector $\textbf{n}_v$, respectively, we obtain the qubit Hamiltonian
\begin{equation}
H_q=\frac{1}{2}\left(\pmb{\Delta}_\parallel+ e^{-\frac{l^2}{l_{so}^2}} \pmb{\Delta}_\perp  \right) \cdot\pmb{\sigma} \ .
\end{equation}

To study the effect of charge noise, we consider small fluctuations $\delta V$ of the gate potential around the fixed working point $V_g$. To linear order in   $\delta V$ the parameters of $H_q$ modify as $\pmb{\Delta}\rightarrow \pmb{\Delta}+ \pmb{\Delta}'\delta V$, $l\rightarrow l+ l'\delta V $ and $l_{so}\rightarrow l_{so}+l_{so}'\delta V$, leading to 
\begin{equation}
H_q\rightarrow H_q+\frac{\delta V}{2} \textbf{Q}\cdot \pmb{\sigma} \ ,
\end{equation}
where we define the vector with the units of charge 
\begin{equation}
\label{eq:charge-noise}
\textbf{Q}=\pmb{\Delta}_\parallel'+ e^{-\frac{l^2}{l_{so}^2}}\pmb{\Delta}_\perp' + 2\frac{l^2}{l_{so}^2}e^{-\frac{l^2}{l_{so}^2}} \left(\frac{l_{so}'}{l_{so}}-\frac{l'}{l}  \right)\pmb{\Delta}_\perp \ .
\end{equation}

Unless the device is operated at the sweet spot, in elongated hole quantum dots, $l$ is comparable with the spin-orbit length $l_{so}$, and the last term in Eq.~\eqref{eq:charge-noise} dominates.
In addition, we stress that in these systems the requirement of a vanishing first derivative $|\pmb{v}|'$ of the spin-orbit velocity $|\pmb{v}|$ at the sweet spot is not sufficient to remove spin-orbit-caused charge noise \cite{wang2019suppressing,PhysRevLett.116.246801}, because there is an additional large contribution coming from the variation of the dot size $l$. 
In contrast, at the spin-orbit switch point, where $l_{so}\rightarrow \infty$, the last term in Eq.~\eqref{eq:charge-noise} vanishes exactly and charge noise only affects the qubit by the fluctuations of the $g$-factor. \\

We now restrict ourselves to the analysis of the wide FinFET grown along the DRA that is discussed in Sec.~\ref{Sec:SOHS}.
Because of the symmetries of this device, the principal axes of the $g$-tensor are aligned to the coordinate system chosen in Fig.~\ref{fig:Geometry-finfet} \cite{PhysRevResearch.2.033036}, and the $i$th component of the Zeeman energy is $\pmb{\Delta}_i=g_{ii} \mu_B B_i$. 
Here, we consider a magnetic field pointing in the $y$-direction, i.e. $\textbf{B}=B_y \textbf{e}_y$.
This choice maximizes the Rabi frequency in electric spin dipole resonance experiments \cite{PhysRevB.74.165319} because the spin-orbit vector points in the $x$-direction, see Eq.~\eqref{eq:v_E_funct_Ey}.
In addition, this magnetic field direction minimizes the noise and provides the largest Zeeman energy gap. The dependence of the elements of the $g$-factor matrix on the gate potential $V_g$ for this device design is shown in Fig.~\ref{fig:t2}a). 
The value of the $g$-factors computed here and the strong anisotropy of the Zeeman energy depending on the direction of the magnetic field is in reasonable agreement with experiments \cite{maurand2016cmos,geyer2020silicon,siliconFin,voisin2016electrical}.

In general, the vector $\textbf{Q}$ can have a component pointing along the vector $\pmb{\Delta}$ and a component perpendicular to it; these components cause the dephasing and the relaxation of the qubit, respectively.
However, when the matrix of $g$-factors is diagonal and $\textbf{B}=B_y \textbf{e}_y$, the vector $\textbf{Q}$ is aligned to $\pmb{\Delta}$ and the qubit is only subjected to dephasing. 
From a Bloch-Redfield analysis and for $1/f$-type noise with spectrum $\mathcal{S}(\omega)=\langle\delta V^2\rangle/|\omega|$, the dephasing rate $1/T_2^*$ is given by \cite{PhysRevLett.93.016601,MAKHLIN2004315} 
\begin{equation}
\label{eq:T2-dep}
\frac{1}{T_2^*}= \frac{1}{\hbar}|\textbf{Q}|\sqrt{\langle\delta V^2\rangle}\sqrt{\frac{1}{2\pi}\log\!\left(\frac{|\textbf{Q}|\sqrt{\langle\delta V^2\rangle}}{\hbar\omega_\text{ir}} \right)} \ ,
\end{equation}
where $\omega_\text{ir}\sim 1$~Hz is a cut-off frequency depending on the experiment. Here, we only consider free induction decay, and do not account for echo pulses that can further improve the qubit lifetime.

The dependence of the dephasing rates due to charge noise on the gate potential for a wide FinFET qubit with $\tilde{L}=20$~nm and $r=1.2$ is shown in Fig.~\ref{fig:t2}b). 
For the plot, we consider a magnetic field  $B_y=100$~mT, which leads to a Zeeman energy of tens of $\mu$eV, comparable to the values measured in \cite{froning2020ultrafast}. Also, we consider dots with a fixed value of $\delta\!E_{zz}/V_g$, chosen such that the lateral size of the dot is $l=30$~nm at the spin-orbit switching point $V_g=V_g^{SW}=-3.12$~V. 
In addition, at $V_g=V_g^{SW}$, the longitudinal confinement energy is $\hbar\omega_z=0.14$~meV, an order of magnitude larger than the Zeeman gap $\Delta_y=11$~$\mu$eV and an order of magnitude smaller than the transverse subband gap $\Delta\!E=1.7$~meV; we then conclude that our approach is valid in this regime.
To estimate the fluctuations of the gate potential, we consider that the typical fluctuations of the energy levels are $\sqrt{\langle\hbar^2\omega_z^2\rangle}\sim 5$~$\mu\text{eV}$ \cite{PhysRevLett.105.246804}
and are connected to the fluctuations of the gate potential by the dimensionless lever arm $\alpha\equiv|\hbar\partial\omega_z/(e\partial V_g)|$, i.e. $\sqrt{\langle\delta V^2\rangle} =\alpha \sqrt{\langle\hbar^2\omega_z^2\rangle}/e $. At   $V_g=V_g^{SW}$, choosing $l=30$~nm, we obtain $\alpha\approx 65$, and we estimate $\sqrt{\langle\delta V^2\rangle}\approx 0.3 $~mV. 
We note that if $l=15$~nm at $V_g=V_g^{SW}$, the lever arm is $\alpha\approx 16$, in reasonable agreement with recent experiments where the lever arm is about 20~\cite{DZ_private}.

The black solid line in Fig.~\ref{fig:t2}b) represents the total dephasing rate of the FinFET qubit. 
We observe that charge noise leads to dephasing times ${T_2^*}$ of hundreds of nanoseconds,  in agreement with recent experimental data~\cite{DZ_private}, that can be pushed to infinity when the devices are tuned to work at the sweet spots.
Importantly, the sweet spot is close to the spin-orbit switching point $V_g^{SW}=-3.12$~V, but it does not exactly coincide with it.
To have a better understanding of this shift, we show with dashed red and gray lines the dephasing rates ${1}/{T_2^*}$ obtained by considering only the terms of the vector $\textbf{Q}$ in Eq.~\eqref{eq:charge-noise} that are related respectively to the fluctuations of the $g$-factor, i.e. $\propto \pmb{\Delta}'$, and to the spin-orbit coupling, i.e. $\propto l^2/l_{so}^2$. 
Because these different contributions in Eq.~\eqref{eq:charge-noise} can have a different sign depending on whether the $g$-factor and the spin-orbit coupling increases or decreases as a function of $V_g$, the small shift of the sweet spot is a result of the interference between the $g$-factor fluctuations and the spin-orbit coupling contribution to dephasing.
 In the wide  DRA FinFET, where the spin-orbit length is very short, the spin-orbit coupling contribution to dephasing is dominating and the $g$-factor fluctuations are relevant only very close to the spin-orbit switch. Working at the sweet spot leads to a clear practical advantage, completely removing the charge noise when the qubit is idle.

\section{Conclusion}

In this work, we present ways of suppressing charge noise in hole Si FinFET qubits.
The advantage of these structures compared to other nanowires is their triangular cross-section, which by symmetry permits large intrinsic spin-orbit interactions without external electric fields. When the device has a simple equilateral triangular cross-section, we find this effect to be a result of the interplay between the low symmetry of the cross-section and the anisotropy of Si. When an external gate potential is applied an extra tunable contribution to the spin-orbit coupling arises and depending on the growth direction of the wire, it can enhance or suppress the total spin-orbit interactions. This suppression leads to points where the spin-orbit velocity can be tuned exactly to zero, dramatically boosting up the coherence times of spin-1/2 qubits.

We study in detail the dependence of spin-orbit coupling on the gate potential by considering an inhomogeneous electric field profile, which matches numerical simulations of the electrostatics of realistic FinFET devices. We distinguish between different mechanisms that drive the switching-off of the spin-orbit coupling and that have a different behaviour as the cross-section area becomes smaller as a result of the spin-orbit split-off hole band.
These states generally degrade the performance of the FinFET and can even remove the spin-orbit switch.
We present more involved designs, e.g. wide FinFETs, that reduce their effect  and restore the spin-orbit switch. 

When a small external magnetic field is applied, hole nanowire qubits become susceptible also to fluctuations of the Zeeman energy caused by an electrically tunable $g$-factor.
By analyzing the response of FinFET devices to small magnetic fields, we find sweet spots where the charge noise can be completely removed to linear order in the fluctuations of the gate potential, providing an ideal working point where quantum information can be reliably stored in charge noise resilient spin qubits.

\begin{acknowledgments}
We would like to thank Dominik Zumb{\"u}hl, Dmitry Miserev, Christoph Kloeffel, and Peter Stano for useful discussions and comments.
 This work was supported by the Swiss National Science Foundation and NCCR SPIN.

\end{acknowledgments}

\begin{appendix}

\section{Electric field simulation}
\label{app:E-field}

\begin{figure}
 \centering
\includegraphics[width=0.45\textwidth]{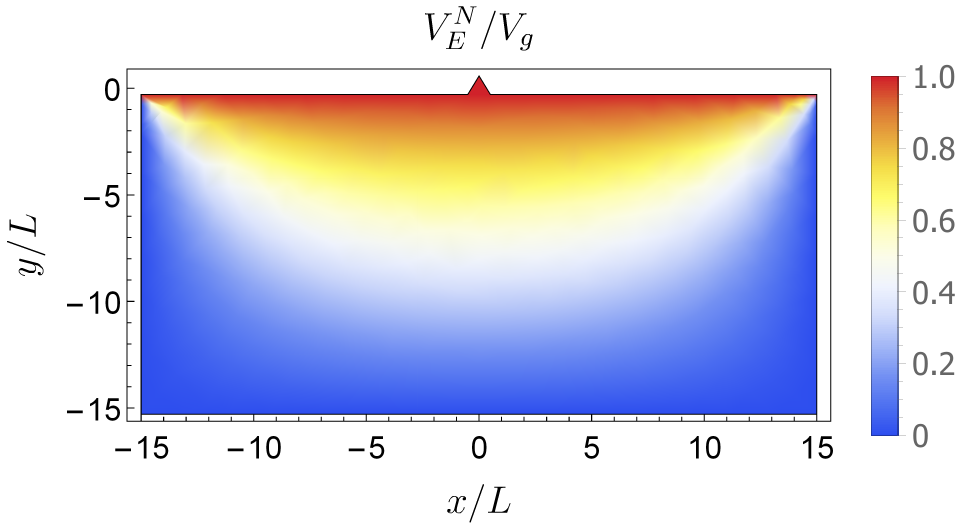}
	\caption{Electrostatic potential $V_E^N$ in the cross-section of a Si FinFET. For the simulation, we used $d_B=15L$ and a substrate $30L$ wide with an equilateral triangular fin of side $L$ placed on top of it. The top gate covers the whole upper part of the device and is fixed at the potential $V_g$; the rest of the boundary is grounded.}
	\label{fig:V-potential-full}
\end{figure}

\begin{figure*}
 \centering
\includegraphics[width=0.95\textwidth]{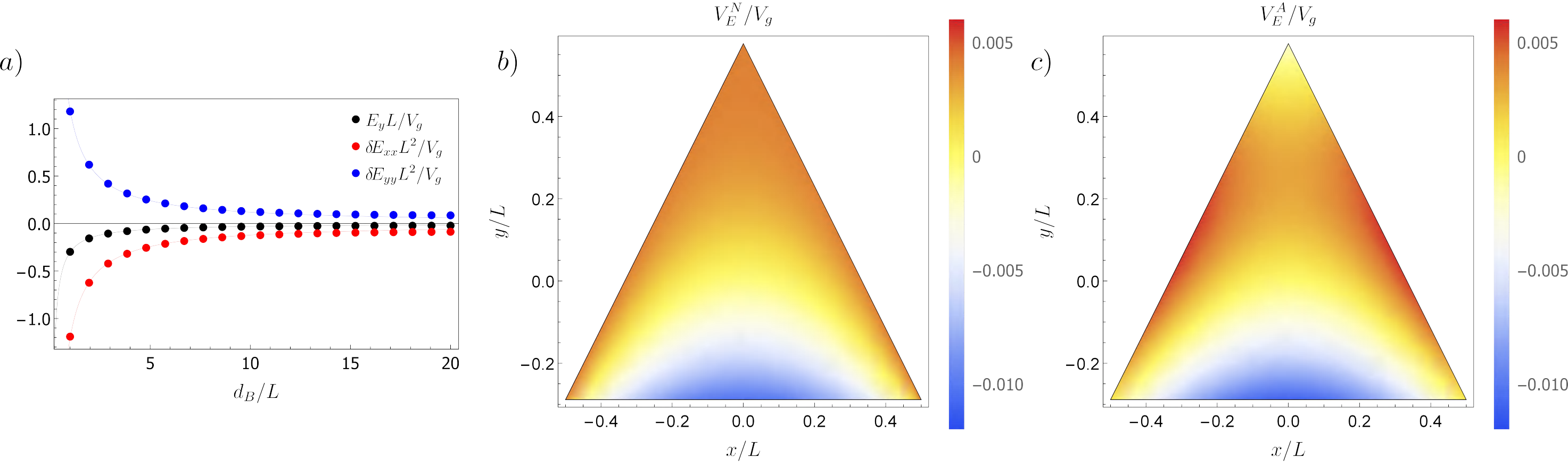}
	\caption{Comparison between the electrostatic potential $V_E^N$ simulated from the Laplace equation and the approximate potential $V_E^A$. In a) we show the dependence of the parameters $\delta\!E_{ii}$ and $E_y$ defining $V_E^A$ on the substrate thickness $d_B/L$. The dots are the results of the simulation, while the solid lines are obtained by combining Eqs.~\eqref{eq:multipole-parameters} and~\eqref{eq:multipole-parameters_1}. In~b) and c), we show a comparison between the potential in the fin computed numerically (b) and its approximation (c) when $d_B=15L$. To facilitate the comparison, in~b), we subtract the constant potential $V_0=0.996$ obtained by averaging the potential in the fin.
	  }
	\label{fig:V-potential}
\end{figure*}

We present the model of the electric field profile in Si FinFETs and compare the approximate potential in Eq.~\eqref{eq:Multipole-E} to the electrostatic potential obtained by solving the Laplace equation in the  FinFET sketched in Fig.~\ref{fig:Geometry-finfet}.
We study a fin with an equilateral triangular cross-section with side $L$.
For simplicity, we neglect the effect of the dielectric on the electric field lines. This approximation describes well devices that use thin high-k dielectric materials, with a dielectric constant similar to Si, but we expect our results to be at least qualitatively correct for a wider range of devices, including  devices made with $\text{SiO}_2$. In addition, we assume that the top and back gates extend to infinity along the wire (in the $z$-direction), such that we can restrict our analysis to a cross-section  of the FinFET in the $(x,y)$ plane.
We solve the Laplace equation considering that the top gate fixes the electrostatic potential of Si to $V_g$ and that the back gate is grounded to zero potential.  To model the lateral sides of the substrate, we consider a wide substrate, that extends symmetrically up to $x=\pm 15L$ from the position of the fin at $x=0$, and we set the potential at the sides below the top gate to ground. This approximation describes well the potential in the fin as long as the distance $d_B$ of the back gate from the fin is $\lesssim 30 L$.
In Fig.~\ref{fig:V-potential-full}, we show the potential energy $V_E^N$ simulated in this setup when the back gate is at $d_B=15L$ from the bottom of the fin. Note that the potential in this simple design varies linearly with $V_g$ and that the lengths are normalized against the side $L$ of the fin.

We  compare the numerical solution $V_E^N$ of the Laplace equation to the approximate potential in Eq.~\eqref{eq:Multipole-E};  in this section,  we call the approximate potential $V_E^A$ to distinguish it from the numerical solution $V_E^N$. To find the parameters $E_i$ and $\delta\!E_{ij}$, we compute the appropriate derivatives of the potential and find their average in the fin. For example, $\delta\!E_{xx}=-(1/A_\text{fin})\int_\text{fin} d\textbf{r}\partial_{xx}V_E^N(x,y)$, with $A_\text{fin}$ being the area of the triangle, where the averaging is performed.
Because the device is symmetric around $x=0$, the homogeneous electric field in the $x$-direction vanishes and $E_x=0$. 
Simulating  devices with different substrate thickness $d_B$, we find that  the homogeneous and inhomogeneous components of the electric field can be written to good approximation as
\begin{equation}
\label{eq:multipole-parameters}
E_y= \frac{c_{y}V_g}{d_B} \ \ \text{and} \ \ \delta\!E_{ij}= \frac{c_{ij}V_g}{ d_BL} \ ,
\end{equation}
where the dimensionless coefficients $c$ are of order one. By fitting these formulas against the results of the simulation, we find  a good fit when 
\begin{equation}
\label{eq:multipole-parameters_1}
c_y=-0.3  \ , \ \ \ c_{xx}=-1.21 \ , \ \ \ c_ {yy}=1.2 \ ,
\end{equation}
see Fig.~\ref{fig:V-potential}a). Note that $\delta\!E_{xx}\approx-\delta\!E_{yy}$, justifying the approximation in Eq.~\eqref{eq:Multipole-E}.
We also find that $c_{xy}=0$ in this geometry. When accounting for the finite size of the top gate, $c_{xy}$ acquires a finite value, however, it is reasonable to assume that $|c_{xy}|\ll|c_{ii}|$ as long as the top electrode fully covers the fin.
A comparison between the approximate potential $V_E^A$ and $V_E^N$ simulated from the Laplace equation for $d_B=15L$ is shown in Fig.~\ref{fig:V-potential}b) and c).
We observe that the electrostatic potential in the fin $V_E^N$ is  reasonably well approximated by $V_E^A$. 
We remark that the precise values of the coefficients $c$ can change for different devices, e.g. by including dielectric materials, changing the aspect ratio of the fin or including additional gates. While we do not expect drastic changes of our model, we do not investigate these effects in detail here.

\begin{figure*}
 \centering
  \includegraphics[width=0.95\textwidth]{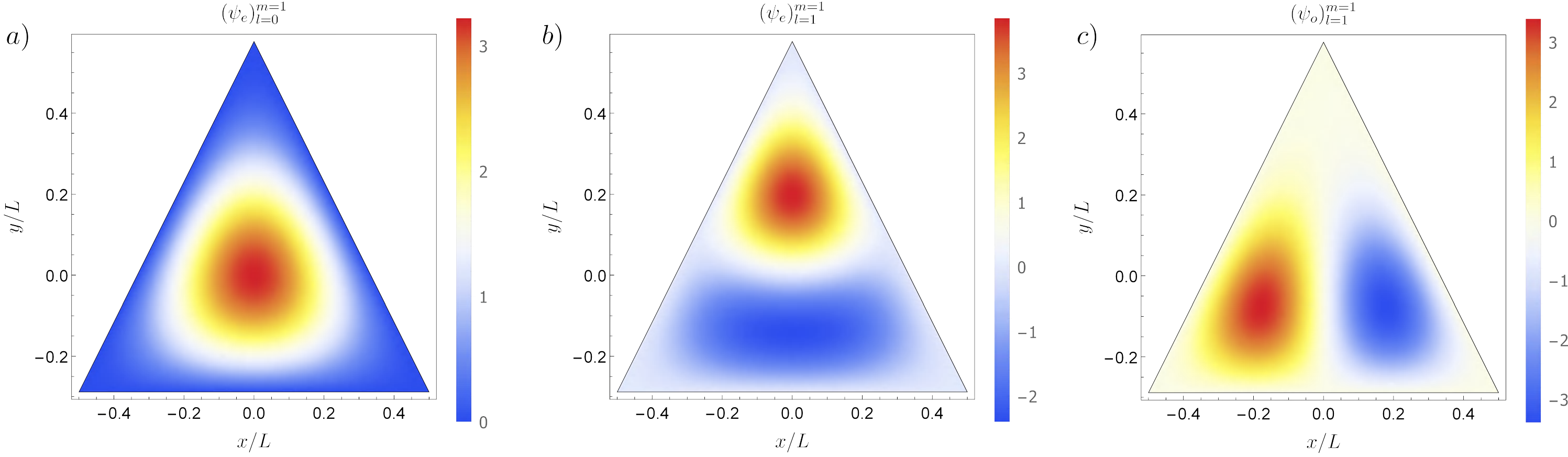}
  	\caption{First three orbital states in Eq.~\eqref{eq:sol-LaplEq}  of the Laplace equation in an equilateral triangle.  We consider the even solution with $(l,m)=(0,1)$ (a) and the solutions with $(l,m)=(1,1)$ and even (b) and odd (c) symmetry. The wavefunctions in this plot are normalized. These states are the ones used in App.~\ref{app:SO-pt}. }
	\label{fig:Laplace-equation-solution}
\end{figure*}

\section{Orbital eigenstates}
\label{app:Laplace}

A convenient basis to analyze wires with triangular cross-sections comprises the eigenstates of the two-dimensional Laplace operator $p_x^2+p_y^2$ vanishing at the boundary of an equilateral triangle of side $L$ \cite{mccartin2011laplacian}. 
These solutions can be chosen to be even ($e$) or odd ($o$) with respect to the height of the triangle at $x=0$ and they are  written compactly as
\begin{equation}
\label{eq:sol-LaplEq}
\psi_\lambda(x,y)= c_\lambda \textbf{f}_\lambda(x/L)\cdot \textbf{g}(y/L) \ 
\end{equation}
where where $\lambda=e,o$ indicates the parity and $c_\lambda$ is a normalization constant. We define the vectors
\begin{widetext}
\begin{subequations}
\label{eq:sol-Lap-vector}
\begin{align}
\textbf{f}_e(x)&= \left(\cos\left[\frac{2\pi(3m+l)}{3} x\right] \ , \  \cos\left[\frac{2\pi(3m+2l)}{3}x\right] \ , \ -\cos\left[\frac{2\pi l x}{3}\right] \right)\ , \\
\textbf{f}_o(x)&= \left(-\sin\left[ \frac{2\pi(3m+l)}{3} x\right] \ , \  \sin\left[\frac{2\pi(3m+2l)}{3}x\right] \ , \ \sin\left[\frac{2\pi l x}{3}\right] \right) \ , \\
\textbf{g}(y)&= \left(\sin\left[ \frac{2\pi(m+l)}{3}   \left(\sqrt{3} y-1\right) \right] \ , \  \sin\left[\frac{2\pi m}{3}   \left(\sqrt{3} y-1\right) \right] \ , \ \sin\left[\frac{2\pi(2m+l)}{3}   \left(\sqrt{3} y-1\right) \right] \right) \ .
\end{align}
\end{subequations} 
\end{widetext}
The quantum numbers $m,l$ are integers satisfying the conditions $m\geq 1 \ , \ l\geq 0$ and $m\geq 1 \ , \ l\geq 1$ for the even and odd solutions, respectively. These quantum numbers label the eigenvalues of the Laplace operator according to
\begin{equation}
\label{eq:Laplace-eigenvalues}
\left(p_x^2+p_y^2\right) \psi_{\lambda}=  \frac{\hbar^2 }{L^2}\frac{16\pi^2}{3}  \left(m^2+l m+\frac{l^2}{3}\right) \psi_{\lambda} \ .
\end{equation}
The states characterized by $l=0$ are three-fold rotationally symmetric and even with respect to the height of the triangle; no odd solutions with $l=0$ are allowed.
The remaining even and odd solutions labelled by the same quantum numbers $m$ and $l\neq 0$ are degenerate.
The lowest three normalized eigenfunctions are shown in Fig.~\ref{fig:Laplace-equation-solution}. These three states are  used in App.~\ref{app:SO-pt} to find an effective analytical model for the triangular FinFET.

\section{Spin-orbit interaction in perturbation theory}
\label{app:SO-pt}
\subsection{Intrinsic spin-orbit velocity and length}
\label{app:Effective Parameters-GD}

 \begin{figure*}
 \centering
\includegraphics[width=0.95\textwidth]{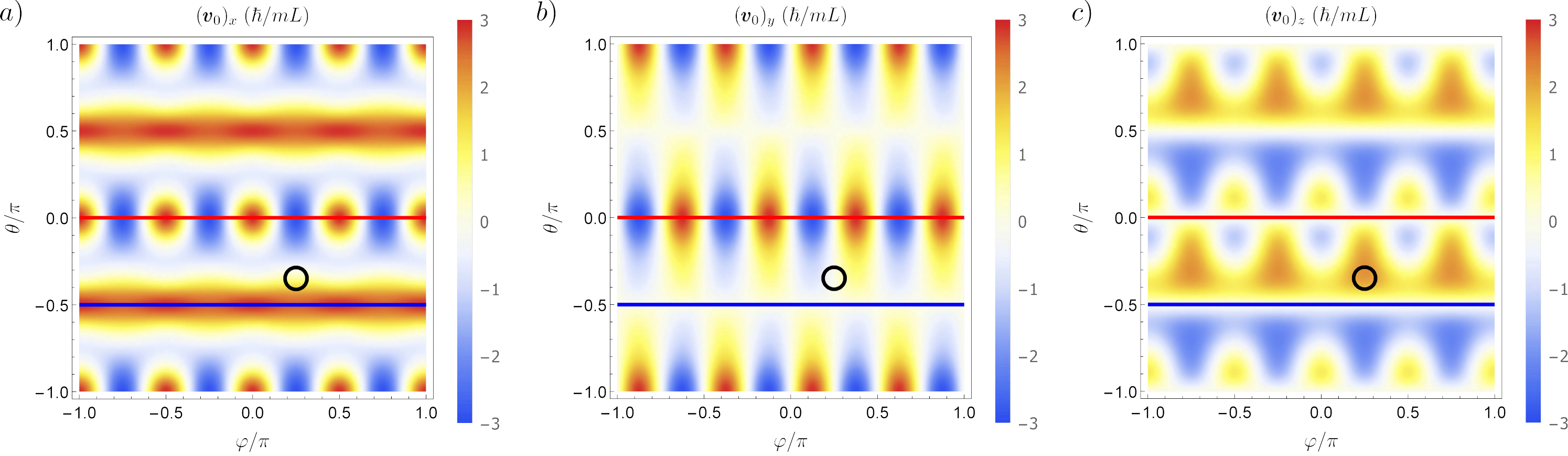}
	\caption{Components of the intrinsic spin-orbit vector $\pmb{v}_0$ in an equilateral FinFET as a function of the angles $\theta$ and $\varphi$ that parametrize the orientation of the wire with respect to the crystallographic axes, see Fig.~\ref{fig:Geometry-finfet}. The results shown here are obtained by using a second order Schrieffer-Wolff transformation and including only the lowest $3$ orbital states in Eq.~\eqref{eq:sol-LaplEq}.  The wavefunction of these states is shown in Fig.~\ref{fig:Laplace-equation-solution}. We indicate with red and blue lines the cuts $\theta=0$ and $\theta=-\pi/2$, respectively. These two cases are studied extensively in the main text. The black circle marks the point  $\varphi=\pi/4$ and $\theta=-\arctan\sqrt{2+\sqrt{3}}$, where the wire is grown along the $z\parallel [111]$ direction, and where $(\pmb{v}_0)_{x,y}=0$ and $(\pmb{v}_0)_{z}$ is maximal. }
	\label{fig:vel_GD}
\end{figure*}

In this section, we show the general dependence of the intrinsic spin-orbit vector $\pmb{v}_0$ in Eq.~\eqref{eq:SO-Vector-Equilateral} and of the spin-orbit length $l_{so}$ in Eq.~\eqref{eq:SO-length} on the orientation of the wire.
To account for the different growth directions, we rotate the $4\times 4$ LK Hamiltonian $H_\text{LK}'$ in Eq.~\eqref{eq:LK_Hamiltonian_4} by the unitary operator $U= e^{i\theta F_{y'}}e^{i\varphi F_{z'}}$, i.e. $H_\text{LK}'\rightarrow H_\text{LK}= U^\dagger H_\text{LK}' U$, where
$\textbf{F}=\textbf{J}'+ \textbf{x}'\times \textbf{p}'$ [with $\textbf{x}'=({x',y',z'})$ and $\textbf{J}'=(J_{x}',J_{y}',J_{z}')$] is the total angular momentum, and $\theta$ and $\varphi$ are the   angles between the crystallographic axes and the final coordinate system, see Fig.~\ref{fig:Geometry-finfet}. This unitary rotation aligns the coordinate system and the direction of the spin-matrices to the axes $x$, $y$ and $z$. 
More explicitly, the Hamiltonian in the rotated coordinate system is given by
\begin{equation}
\label{eq:Eq:LK_rotated}
H_\text{LK}=\left(\!\gamma_1+\frac{5}{2}\gamma_2\!\right)\!\frac{p^2}{2m}-\frac{\gamma_2}{m} \textbf{p}^2\cdot \textbf{J}^2-\frac{2\gamma_3}{m}  p_ip_j \left\lbrace J_i,J_j \right\rbrace+\text{cp}\ ,
\end{equation}
where the rotated momenta are
\begin{equation}
\label{eq:coordinate_system_rotated}
\textbf{p}=\left(\begin{array}{c}
\cos (\theta ) \cos (\varphi ) p_{x}'+\cos (\theta ) \sin (\varphi ) p_{y}'-\sin (\theta ) p_{z}'\\
\cos (\varphi ) p_{y}'-\sin (\varphi ) p_{x}'\\ 
\sin (\theta ) \left(\cos (\varphi ) p_{x}'+\sin (\varphi ) p_{y}'\right)+\cos (\theta ) p_{z}'
\end{array}\right) \ ,
\end{equation}
and $\textbf{p}^2=(p_x^2,p_y^2,p_z^2)$, $\textbf{J}^2=(J_x^2,J_y^2,J_z^2)$,  $p^2=p_x^2+p_y^2+p_z^2$. The spin 3/2 matrices $\textbf{J}$ are rotated in the same way.

We do not include electric or magnetic fields at the moment and we focus on the intrinsic spin-orbit coupling.
To obtain simple equations, we restrict the orbital space to the space spanned by the lowest three eigenstates of the Laplace equation in an equilateral triangle, see Fig.~\ref{fig:Laplace-equation-solution}. By projecting $H_\text{LK}$ onto this subspace, we obtain a $12\times 12$ reduced Hamiltonian  $H_{12}$ that parametrically depends on the angles $\varphi$ and $\theta$ and on the momentum  $p_z$.

Specifically, $H_{12}$ is found from $H_\text{LK}$ in Eq.~\eqref{eq:Eq:LK_rotated} by separating the different powers of $p_i p_j$, i.e. 
\begin{equation}
\label{eq:power-momenta}
H_\text{LK}\equiv\sum_{i,j} p_i p_j H_{ij}  \ ,
\end{equation} 
and using the matrix representation of the momenta operators in the basis $\Big(|1,0,e\rangle, |1,1,e\rangle, |1,1,o\rangle \Big) $, where the eigenstates  $ |m,l,\lambda\rangle$ are defined by Eqs.~\eqref{eq:sol-LaplEq} and~\eqref{eq:sol-Lap-vector}. 
Denoting the matrix representation of the momentum operators in this basis by a under bar $\underline{p_i p_j}$,  $H_{12}$ is given by
\begin{equation}
\label{eq:H12-general}
H_{12}=\sum_{i,j} \underline{p_i p_j} \otimes H_{ij}  \ ,
\end{equation}
 The expressions of the matrix elements of $H_{12}$ as a function of $\theta$ and $\varphi$ can be straightforwardly derived from Eq.~\eqref{eq:Eq:LK_rotated} and we do not give them here. 
 We report instead the matrices of the momenta: 
\begin{subequations}
\label{eq:momenta-3x3}
\begin{align}
\underline{p_x^2}&=\frac{\hbar^2}{L^2} \left(
\begin{array}{ccc}
 \frac{8 \pi ^2}{3} & \frac{k}{\sqrt{2}} & 0 \\
 \frac{k}{\sqrt{2}} & \frac{56 \pi ^2}{9}-\frac{21 }{25} k & 0 \\
 0 & 0 & \frac{56 \pi ^2}{9}+\frac{21 }{25} k \\
\end{array}
\right)  \ , \\
\underline{p_x p_y} &=\frac{\hbar^2}{L^2} \left(
\begin{array}{ccc}
 0 & 0 & -\frac{k}{\sqrt{2}} \\
 0 & 0 & -\frac{21}{25}  k \\
 -\frac{k}{\sqrt{2}} & -\frac{21}{25}  k & 0 \\
\end{array}
\right) \ , \\
\underline{p_x p_z} &= \frac{9 \hbar k p_z}{10 \sqrt{2} \pi L}\left(
\begin{array}{ccc}
 0 & 0 & i \\
 0 & 0 & 0 \\
 -i & 0 & 0 \\
\end{array}
\right) \ , \\
\underline{p_y p_z} &=\frac{9 \hbar k p_z}{10 \sqrt{2} \pi L} \left(
\begin{array}{ccc}
 0 & -i & 0 \\
 i & 0 & 0 \\
 0 & 0 & 0 \\
\end{array}
\right) \ , 
\end{align}
\end{subequations}
and $\underline{p_y^2}=\left.\underline{p_x^2}\right|_{k\rightarrow-k} $, $
\underline{p_z^2}= p_z^2 \mathcal{I}_3 $; also  $k=2187/112\approx 19.53$ and $\mathcal{I}_3$ is the 3-dimensional identity matrix. In the simple case $\theta=0$, the $12\times 12$ Hamiltonian, also including electric fields, is explicitly given in Eq.~\eqref{Eq:12times12E}.

We  resort to perturbation theory on $H_{12}$ to derive a low-energy description of the system.
With a second order Schrieffer-Wolff transformation, see e.g. Appendix B of \cite{WinklerSpinOrbitCoupling2003} or \cite{bravyi2011schrieffer}, we obtain a $2\times 2 $ effective Hamiltonian that acts  on the ground state subspace of the wire. Expanding the matrix elements of this Hamiltonian  up to second order in  $p_z$, we find an effective wire Hamiltonian as in Eq.~\eqref{eq:wire-Ham}.
The terms linear in  $p_z$ are related to the spin-orbit velocity $\pmb{v}$, while the effective mass $m^*$ is twice of the inverse of the diagonal term quadratic in $p_z$. \\

For a Si wire, the dependence of the components of the intrinsic spin-orbit velocity vector $\pmb{v}_0$ on the growth angles obtained in this way is shown in Fig.~\ref{fig:vel_GD}.
Importantly, we find that the $\pmb{v}_0$ is in general proportional to the anisotropy of the material $\gamma_3-\gamma_2$ and to the velocity $\hbar/mL$, and thus it can be written as in Eq.~\eqref{eq:SO-Vector-Equilateral}. 
In addition, there is no spin-orbit coupling in the direction along the wire when $\theta=0$ and $\theta=\pm \pi/2$ (red and blue lines, respectively), and in these cases $(\pmb{v}_0)_z=0$. While these results were found by perturbation theory, we find numerically that they hold generally. 
Interestingly, the off-diagonal components of the spin-orbit velocity $(\pmb{v}_0)_{x,y}$ vanish e.g. when $\varphi=\pi/4$ and $\theta=-\arctan\sqrt{2+\sqrt{3}}$, where the diagonal spin-orbit $(\pmb{v}_0)_z$ is maximal. This orientation corresponds to $z\parallel [111]$, along which the Si lattice has a three-fold rotational symmetry, and is marked with  black circles in the figures. 

By taking the limits $\theta=0$ and $\theta=\pm \pi/2$, the expressions of the spin-orbit vector simplify notably and are given in Eqs.~\eqref{eq:alpha_pt}. 
When $\theta=\pm \pi/2$,  the spin-orbit vector points along the $x$-direction and it has an oscillating amplitude, while when $\theta=0$, the direction is oscillating as a function of $\varphi$. A plot of the absolute values of $\pmb{v}_0$ in these cases and a comparison between the perturbation theory and a more detailed numerical solution comprising 200  orbital states in Eq.~\eqref{eq:sol-LaplEq} is shown in Fig.~\ref{fig:GD-dependence}.\\

\begin{figure}
 \centering
  \includegraphics[width=0.45\textwidth]{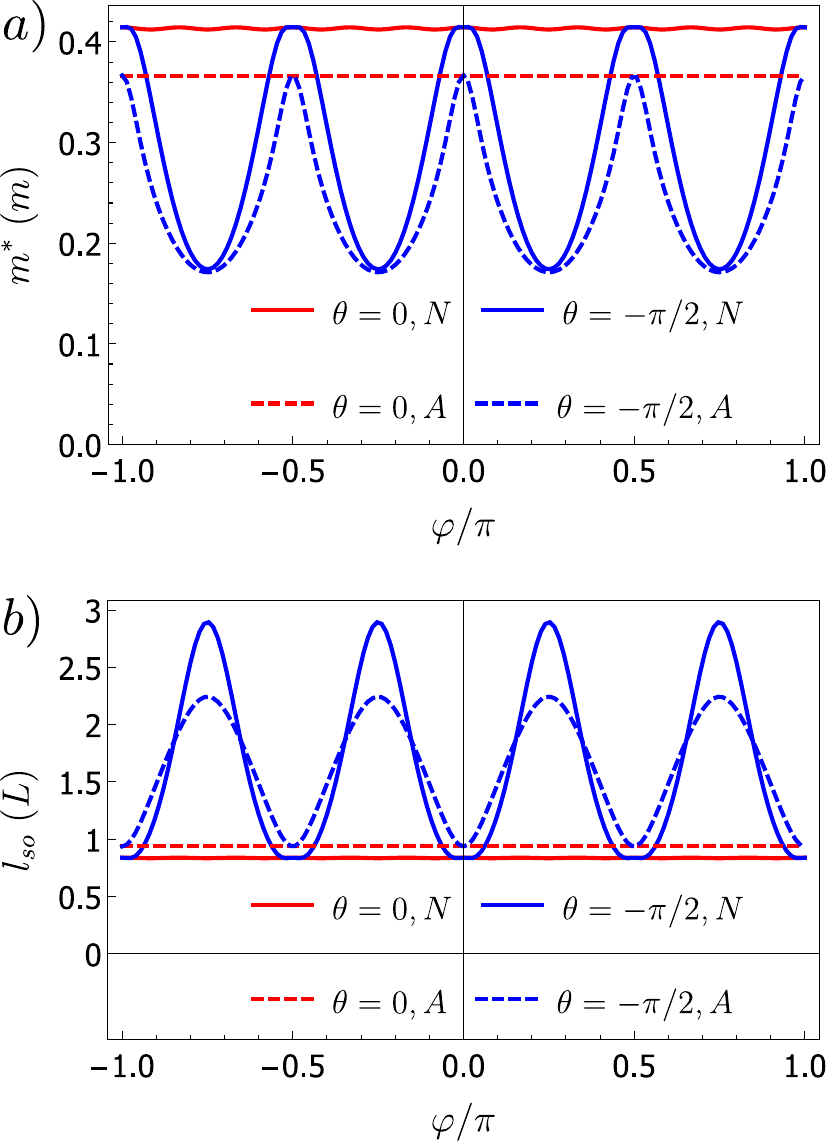}
  	\caption{ a) Effective mass $m^*$ and b) intrinsic spin-orbit length $l_{so}$ at zero electric field as a function of  $\varphi$ in an equilateral FinFET. We show with solid lines the results obtained numerically by using Eq.~\eqref{eq:mass-Exact} for $m^*$ and by combining Eqs.~\eqref{eq:SO-Exact}, \eqref{eq:SO-length} and~\eqref{eq:mass-Exact}  for $l_{so}$. The dashed lines show the approximate values obtained by a second order Schrieffer-Wolff transformation and including only the lowest three orbital states in Eq.~\eqref{eq:sol-LaplEq}. For the $m^*$, we use Eq.~\eqref{eq:Effective_masses_zero_field} and for $l_{so}$ we combine Eqs.~\eqref{eq:SO-length}, \eqref{eq:Effective_masses_zero_field}, \eqref{eq:SO-Vector-Equilateral} and~\eqref{eq:alpha_pt}. Red and blue lines show results obtained at  $\theta=0$ and $\theta=-\pi/2$, respectively. 
}
	\label{fig:intrinsic_par}
\end{figure}

With this approach, we also find the effective mass $m^*$ from the diagonal elements of the effective Hamiltonian. Along the $\theta=0$ and $\theta=\pm \pi/2$ directions, $m^*$ can be compactly written as 
\begin{widetext}
\begin{subequations}
\label{eq:Effective_masses_zero_field}
\begin{align}
\frac{m}{m^*(\theta=0)}&=\gamma_1+2\gamma_2-\frac{5.35 \gamma_1 \gamma_3^2+1.86 \gamma_2^3+15.23 \gamma_2 \gamma_3^2}{(\gamma_1+5 \gamma_2/2)^2}\approx 2.73 \ , \\
\label{eq:Effective_masses_zero_field-SA}
\frac{m}{m^*(\theta=\pm\pi/2)}&\approx\frac{5.12-5.57 \cos (4 \varphi )+1.14 \cos (8 \varphi )-0.09 \cos (12 \varphi )}{1-0.91 \cos (4 \varphi )+0.13 \cos (8 \varphi )} \ .
\end{align}
\end{subequations}
\end{widetext}
where in the second equation we discarded higher harmonic components oscillating with a small amplitude and a fast period in $\varphi$; $m$ is the bare electron mass.
In Fig.~\ref{fig:intrinsic_par}a), we show the effective mass as a function of  $\varphi$, comparing the approximate Eq.~\eqref{eq:Effective_masses_zero_field} with a more precise numerical result.  
This numerical result is obtained by extending the general perturbation theory developed in Sec.~\ref{Sec:Model} [see in particular Eq.~\eqref{eq:SO-Exact}]. By considering $H_\text{LK}=H_0+H_1 p_z+H_2 p_z^2$ and defining the matrix  $M_E$ of column eigenvectors of $H_0$, the effective mass is given by 
\begin{equation}
\label{eq:mass-Exact}
\frac{1}{2m^*}= \Big(M_E^\dagger H_2 M_E\Big)_{ii}+\sum_{k\neq i,j} \frac{\Big|\left(M_E^\dagger H_1 M_E\right)_{ik}\Big|^2}{\epsilon_E^i-\epsilon_E^k} \ ;
\end{equation}
the indexes $i,j$ label the ground state Kramers partners and, on the left-hand side of the equation, we omitted them because  $m^*_i=m^*_j\equiv m^*$. In analogy to above, $M_E$ is computed numerically by  accommodating $200$ orbital states given in Eq.~\eqref{eq:sol-LaplEq}.

We observe that the simple analytical results capture well the oscillating behaviour of the mass as a function of $\varphi$, but they  underestimate the amplitude of the oscillations, leading to a smaller mass, especially when the wire extends along a crystallographic axis, e.g  when $\theta=0$ or when $\theta=-\pi/2$  and $\varphi=0$.
For the SA, where $\theta=-\pi/2$ and $\varphi=-3\pi/4$, Eq.~\eqref{eq:Effective_masses_zero_field-SA} works well and gives an effective mass $m^*_\text{SA}\approx 0.17 m$, while for the CA and DRA, the numerical analysis is more precise and it gives  an effective mass $m^*_\text{CA}\approx m^*_\text{DRA} \approx 0.41 m$.\\

Analytical expressions for the intrinsic spin-orbit length $l_{so}$ defined in Eq.~\eqref{eq:SO-length} can be obtained  when $\theta=0$ and $\theta=\pi/2$  by combining Eqs.~\eqref{eq:SO-Vector-Equilateral}, \eqref{eq:alpha_pt} and~\eqref{eq:Effective_masses_zero_field}.  
A comparison between these expressions and the numerically computed values of $l_{so}$ is shown in Fig.~\ref{fig:intrinsic_par}b). Because of the underestimation of the effective mass, the spin-orbit length predicted by perturbation theory is larger than the numerical values when the wire extends along a crystallographic axis. In contrast, for the SA, the perturbative result is smaller than the numerical result because of the overestimation of the spin-orbit velocity, see Fig.~\ref{fig:GD-dependence}. From the numerical analysis, we find that
\begin{subequations}
\begin{align}
l_{so}(\theta=0)&\approx 0.83 L \ , \\
l_{so}(\theta=\pm\pi/2)&\in \big[0.83 L \ , 2.89 L \big] \ .
\end{align}
\end{subequations}
Importantly, the intrinsic spin-orbit length is always of the order of the side of the triangle, typically much shorter than the confinement length of the quantum dot along the wire, leading to large spin-orbit interactions. \\

Finally, we point out that while our quantitative analysis here is limited to Si, our results can apply also to other semiconductors. For example, in~\cite{doi:10.1002/adma.201906523}, the spin-orbit interaction in hole Ge hut-wires is studied and an intrinsic spin-orbit field of $\hbar |{v}_0|\sim 10$~$\text{meV}\cdot\text{nm}$ was measured for a triangular device with width $L_x=80$~nm and height $H=4$~nm.
An estimate of the amplitude of the intrinsic spin-orbit interaction related to the triangular cross-section can be found by using the equilateral triangle equation for the crystallographic growth direction,
\begin{equation}
|v_0|=9.34 \frac{\hbar}{m\tilde{L}}\frac{ \gamma_3 (\gamma_3-\gamma_2)}{\gamma_1+(5/2)\gamma_2}\ ,
\end{equation}
 obtained by combining Eqs.~\eqref{eq:SO-Vector-Equilateral} and~\eqref{eq:alpha0}. To better compare with the experiment, we consider a fictitious equilateral triangle of side $\tilde{L}$ having the same area of the cross-section of the hut-wire; from the condition $\sqrt{3}\tilde{L}^2/4= L_x H/2$, we find that the effective side of the fictitious equilateral triangle is $\tilde{L}\approx 19$~nm.
Using the Luttinger  parameters of Ge \cite{WinklerSpinOrbitCoupling2003}, we estimate an intrinsic spin-orbit field $\hbar|{v}_0|\approx 12.65 $~$\text{meV}\cdot\text{nm}$, in very good agreement with the experiment~\cite{doi:10.1002/adma.201906523}.

\subsection{Electric field dependence}
\label{app:electric-field}

Here, we focus on wires with $\theta=0$, i.e. wires grown along the [001] crystallographic direction, and we study the dependence of the spin-orbit velocity on the electric field. In this case, the spin-orbit velocity vector has no component along the $z$-direction and is off-diagonal. We define then the complex off-diagonal component of the spin-orbit velocity $v=(\pmb{v })_x+i(\pmb{v })_y$.
This quantity can be decomposed into the sum of an isotropic direct Rashba-like component that vanishes when the external electrostatic potential in Eq.~\eqref{eq:Multipole-E} is turned off, and an anisotropic component that varies as a function of the  angle $\varphi$, see Eq.~\eqref{eq:v_E_funct_Ey}.
At finite values of the electrostatic potential, $v$ depends on the homogeneous electric field $E_y$ and on the inhomogeneous fields $\delta\!E_{xx}$ and $\delta\!E_{yy}$. Here, we do not account for the effect of the terms $E_x$ and $\delta\!E_{xy}$ and we introduce the sum and difference of the inhomogeneous fields $\delta\!E=(\delta\!E_{xx}-\delta\!E_{yy})/2$ and $\Sigma\!E=(\delta\!E_{xx}+\delta\!E_{yy})/2$. For simplicity of notation, we  rescale the electric field by the confinement energy to obtain dimensionless quantities, i.e. $eE_y L /\epsilon_c\rightarrow E_y$, $e\delta\!E L^2 /\epsilon_c\rightarrow \delta\!E$ and $e\Sigma\!E L^2 /\epsilon_c\rightarrow \Sigma\!E$, and we define the  vector $\pmb{\epsilon}=\Big(E_y\ , \ \delta\!E \ , \ \Sigma\!E  \Big)$. 

We neglect the SOHs and in analogy to the treatment in Sec.~\ref{app:Effective Parameters-GD}, we study the $12\times 12$ Hamiltonian obtained from the $4\times 4$ total Hamiltonian in Eq.~\eqref{eq:total-H} rotated by the angle $\varphi$ and projected onto the subspace spanned by the lowest three orbital states in Eq.~\eqref{eq:sol-LaplEq}.
Explicitly,
\begin{equation}
\label{Eq:12times12E}
H_{12}=\frac{\hbar^2}{m L^2}\left(
\begin{array}{ccc}
H_{00} & H_{0e}& H_{0o} \\
H_{0e}^\dagger & H_{ee}& H_{eo} \\
H_{0o}^\dagger & H_{eo}^\dagger & H_{oo}
\end{array}
\right) \ .
\end{equation}
By introducing the quantities $\gamma_\perp^\pm =\gamma_1\pm\gamma_2$, $\gamma_\parallel^\mp =\gamma_1\mp 2\gamma_2$, 
$C_\pm= \frac{\sqrt{3}}{2} k \left[e^{4i\varphi}\left(\gamma_3-\gamma_2\right)\pm \left(\gamma_3+\gamma_2\right)\right] $, $
a_E= \frac{59049}{4480 \pi }\gamma_1\approx 4.186 \gamma_1 $, $a_\delta=\left(\frac{3969}{2000 \pi }-\frac{7 \sqrt{3}}{25}\right) a_E \approx 0.616 \gamma_1$, $b_\delta = \left(\frac{1107}{280 \sqrt{2} \pi}-\sqrt{\frac{2}{3}} \right) a_E\approx 0.308 \gamma_1$, $K= \frac{9}{10\pi}\sqrt{\frac{3}{2}} k \gamma_3\approx 6.851 \gamma_3$,
%\begin{subequations}
%\begin{align}
%C_\pm&= \frac{\sqrt{3}}{2} k \left[e^{4i\varphi}\left(\gamma_3-\gamma_2\right)\pm \left(\gamma_3+\gamma_2\right)\right] \ , \\
%a_E&= \frac{59049}{4480 \pi }\gamma_1\approx 4.186 \gamma_1 \ , \\
%a_\delta&=\left(\frac{3969}{2000 \pi }-\frac{7 \sqrt{3}}{25}\right) a_E \approx 0.616 \gamma_1  \, \\
%b_\delta &= \left(\frac{1107}{280 \sqrt{2} \pi}-\sqrt{\frac{2}{3}} \right) a_E\approx 0.308 \gamma_1 \ , \\
%K&= \frac{9}{10\pi}\sqrt{\frac{3}{2}} k \gamma_3\approx 6.851 \gamma_3 \ ,
%\end{align}
%\end{subequations}
[$k\approx 19.53$, see Eq.~\eqref{eq:momenta-3x3}], we can write the dimensionless blocks as 
\begin{widetext}
\begin{subequations}
\begin{align}
H_{00}&=\text{diag}\left(\frac{8\pi^2}{3}\gamma_\perp^++\frac{p_z^2}{2}\gamma_\parallel^-,\frac{8\pi^2}{3}\gamma_\perp^-+\frac{p_z^2}{2}\gamma_\parallel^+,\frac{8\pi^2}{3}\gamma_\perp^-+\frac{p_z^2}{2}\gamma_\parallel^+,\frac{8\pi^2}{3}\gamma_\perp^++\frac{p_z^2}{2}\gamma_\parallel^-\right) \ , \\
H_{ee}&=\left(
\begin{array}{cccc}
\frac{56 \pi^2}{9}  \gamma_\perp^+ +\frac{p_z^2}{2}\gamma_\parallel^- & 0 & -\frac{21}{25} C_- & 0 \\
 0 & \frac{56 \pi^2}{9}  \gamma_\perp^-+ \frac{p_z^2}{2}\gamma_\parallel^+ & 0 & -\frac{21}{25} C_- \\
 -\frac{21}{25} C_-^* & 0 & \frac{56 \pi^2}{9}  \gamma_\perp^- + \frac{p_z^2}{2}\gamma_\parallel^+ & 0 \\
 0 & -\frac{21}{25} C_-^* & 0 & \frac{56 \pi^2}{9}   \gamma_\perp^+ + \frac{p_z^2}{2}\gamma_\parallel^-\\
\end{array}
\right) -  \mathcal{I}_4 \left(  \frac{21}{25} a_E E_y-a_\delta \delta\!E+\frac{16}{27} \gamma_1 \Sigma\!E \right) \ , \\
H_{0e}&=\left(
\begin{array}{cccc}
 0 & K p_z & \frac{1}{\sqrt{2}} C_- & 0 \\
 - K p_z & 0 & 0 & \frac{1}{\sqrt{2}} C_- \\
 \frac{1}{\sqrt{2}} C_-^* & 0 & 0 & - K p_z\\
 0 & \frac{1}{\sqrt{2}} C_-^* &  K p_z & 0 \\
\end{array}
\right) - \mathcal{I}_4 \left( \sqrt{2}a_E E_y - b_\delta \delta\!E \right)\ , \\
H_{eo}&=\left(
\begin{array}{cccc}
 0 & 0 & -\frac{21 }{25}i C_+ & 0 \\
 0 & 0 & 0 & -\frac{21 }{25}i C_+ \\
 \frac{21}{25}i C_+^* & 0 & 0 & 0 \\
 0 & \frac{21}{25}iC_+^* & 0 & 0 \\
\end{array}
\right) \ .
\end{align}
\end{subequations}
\end{widetext}
Also, $H_{oo}=  \left. H_{ee}\right|_{C_-\rightarrow -C_-, \delta\!E\rightarrow -\delta\!E, E_y\rightarrow -E_y} $, $H_{0o}=\left. H_{0e}\right|_{C_-\rightarrow -iC_+ ,   K\rightarrow -iK , \delta\!E\rightarrow 0  ,  E_y\rightarrow 0}$, and  $\mathcal{I}_4$ is  a $4\times 4$ identity matrix. 
Here, $p_z$ is given in units of $\hbar/L$.

In contrast to Sec.~\ref{app:Effective Parameters-GD}, a second-order Schrieffer-Wolff transformation  does not capture accurately the dependence of the spin-orbit coupling on the electric field and  thus we increase the accuracy of our calculation by using a fourth-order  order transformation. To obtain compact equations, we also Taylor expand the spin-orbit velocity to second order in the vector $\pmb{\epsilon}$, leading to
\begin{widetext}
\begin{subequations}
\label{eq:so-velocity-expansion}
\begin{align}
\alpha_I&\approx\frac{\gamma_3}{\gamma_1+5\gamma_2/2}\left( \frac{\gamma_1}{\gamma_1+5\gamma_2/2}\pmb{\alpha}_I^{(1)} \cdot \pmb{\epsilon}+\frac{\gamma_1^2}{(\gamma_1+5\gamma_2/2)(\gamma_1-\gamma_2) } \pmb{\epsilon}\cdot \underline{\alpha}_I^{(2)} \cdot \pmb{\epsilon}\right) \ , \\
\alpha_A(\varphi)&\approx \frac{  \gamma_3}{\gamma_1+5\gamma_2/2}e^{-4i\varphi}\left(\alpha_A^{(0)}+\frac{\gamma_1}{\gamma_1+5\gamma_2/2} \pmb{\alpha}_A^{(1)} \cdot \pmb{\epsilon}+ \frac{\gamma_1^2}{(\gamma_1+5\gamma_2/2)(\gamma_1-\gamma_2)}\pmb{\epsilon}\cdot \underline{\alpha}_A^{(2)} \cdot \pmb{\epsilon}\right) \ , \\
\pmb{\alpha}_I^{(1)}&\approx\left(  \frac{0.928 (\gamma_1+6 \gamma_2)}{\gamma_1-\gamma_2}\ , \ \frac{0.066 \gamma_1-0.407\gamma_2}{\gamma_1-\gamma_2} \ , \ 0 \right) \ , \\ 
\underline{\alpha}_I^{(2)}&\approx \left(
\begin{array}{ccc}
 \frac{0.811 \gamma _1+0.938 \gamma _2}{\gamma_1-\gamma_2} & -\frac{0.107 \gamma _1+0.155 \gamma _2}{\gamma_1-\gamma_2} &  \frac{0.0317\gamma _1^2+0.2692 \gamma _2 \gamma _1+0.0871 \gamma _2^2}{(\gamma_1-\gamma_2)(\gamma_1+5\gamma_2/2)} \\
 0 & \frac{0.0034 \gamma _1+0.0055 \gamma _2}{\gamma_1-\gamma_2} & \frac{0.0022 \gamma _1^2-0.0217 \gamma _2 \gamma _1-0.0006 \gamma _2^2}{(\gamma_1-\gamma_2)(\gamma_1+5\gamma_2/2)} \\
 0 & 0 & 0 \\
\end{array}
\right)\ , \\ 
\alpha_A^{(0)}&=\frac{14}{45}\left(\frac{k}{2\pi}\right)^3\approx 9.338 \ , \\ 
\pmb{\alpha}_A^{(1)}&\approx\Big( 0. \ , \ 0. \ , \ 0.158 \Big) \ , \\ 
\underline{\alpha}_A^{(2)}&\approx \left(
\begin{array}{ccc}
 -\frac{1.251 \gamma _1^2+4.231 \gamma _2 \gamma _1+1.604 \gamma _2^2}{\left(\gamma _1-\gamma _2\right)(\gamma_1+5\gamma_2/2)} & \frac{0.13 \gamma _1^2+0.52 \gamma _2 \gamma _1+0.0855 \gamma _2^2}{\left(\gamma _1-\gamma _2\right)(\gamma_1+5\gamma_2/2)} & 0. \\
 0 & -\frac{0.0019 \gamma _1^2+0.0185 \gamma _2 \gamma _1-0.0013 \gamma _2^2}{\left(\gamma _1-\gamma _2\right)(\gamma_1+5\gamma_2/2)} & 0. \\
 0 & 0 & 0.0027 \frac{\gamma_1-\gamma_2}{\gamma_1+5\gamma_2/2}\\
\end{array}
\right) \ .
\end{align}
\end{subequations}
\end{widetext}
To simplify further the expressions, we kept only the terms with the lowest possible power in $1/\gamma_1$. This approximation allows for a good qualitative understanding of the system while keeping the equations short. As a result of this approximation, the $\alpha$ parameters are independent of $\gamma_3$ and we discard small terms -at least one order of magnitude smaller than the dominant terms- in $\alpha_A$ that do not oscillate as $e^{-4i\varphi}$, i.e. terms proportional to $e^{+4i\varphi}$ and to $(\gamma_3-\gamma_2)e^{- 8i\varphi}$. To specify where these terms have been neglected in Eq.~\eqref{eq:so-velocity-expansion}, we use the notation $0.$ and $0$ to distinguish between terms that are negligibly small  but finite ($0.$) from terms that are exactly zero ($0$). 
For Si, Eq.~\eqref{eq:so-velocity-expansion} reduces to
\begin{widetext}
\begin{subequations}
\label{eq:v_iso_aniso_Si}
\begin{align}
\alpha_I&\approx 0.35E_y+0.246 E_y^2 +0.0086 \delta\!E + 0.0011 \delta\!E^2-0.0331 E_y  \delta\!E+0.0124 E_y  \Sigma\!E +0.00011 \delta\!E \Sigma\!E  \ , \\
\alpha_A(\varphi)&\approx e^{-4 i \varphi } \left(2.631-0.367 E_y^2-0.00078 \delta\!E^2+0.0372 \Sigma\!E+0.00053 \Sigma\!E^2+0.0398 \delta\!E E_y\right) \ .
\end{align} 
\end{subequations}
\end{widetext}

These expressions give valuable insights into the dependence of the spin-orbit coupling on the external fields and allow for a qualitative understanding of the numerical results presented in Sec.~\ref{Sec:Homogeneous_E} and~\ref{Sec:Inhomogeneous_E}. 
For example, let us take the limit $\Sigma\!E=0$ as in the main text. 
To obtain $v=0$, the  intrinsic and electric field dependent spin-orbit couplings need to have  opposite signs.  
By looking at Eq.~\eqref{eq:v_iso_aniso_Si}, it is clear that the DRA, with $\varphi=\pi/4$, can drive the switch  because in this case all the terms varying with the electric field have the opposite sign with respect to the intrinsic coupling.
In this case, when $\delta\!E=0$, we find  that $v=0$ at $E_\text{DRA}^{SW}\approx 1.52$ and when  $E_y=0$, $v=0$ at $\delta\!E^{SW}\approx 29.5$.
Quantitatively, the values of these critical fields are only accurate up to prefactors of order one, see Eqs.~\eqref{eq:SW-Homogeneneous-DRA} and~\eqref{eq:deltaEsw}. 

The spin-orbit coupling vanishes also when the fields $E_y$ and $\delta\!E$ are both present. The precise shape of the curve along which this occurs strongly depends on the numerical values of the $\alpha$ parameters. Using Eq.~\eqref{eq:v_iso_aniso_Si}, one predicts $v=0$ along two separate lines that do not intersect, while numerically we observe that the two lines merge together, see the blue curve in Fig.~\ref{fig:alpha_inhom_E}b). 
This discrepancy is a consequence of the numerical inaccuracy of the approximation used: slight variations of the cross-coupling terms $\propto E_y \delta\!E$ can drastically change  the behaviour of the switching curve. We note that the correct qualitative behaviour of $v$ is restored by including higher powers of $\pmb{\epsilon}$ in the expansion in Eq.~\eqref{eq:so-velocity-expansion}, but we do not give explicit expressions for these terms here.

\subsection{Comparison with square cross-section}
\label{app:square-cross-sect}
We summarize now the key qualitative differences between Si FinFETs with equilateral triangular cross-section and Si wires with an inversion symmetric cross-section. In particular, here we focus on wires with a square cross-section; a detailed analysis of the direct Rasbha spin-orbit coupling in the these wires can be found in Ref.~\cite{DRkloeffel3}.
The first key difference is that without external fields, the inversion symmetry of a square cross-section prohibits the presence of an intrinsic spin-orbit coupling, and $\pmb{v}_0^\text{sq}=0$. 
Another important difference between the two systems is that the amplitude of the spin-orbit velocity in an square wire is a symmetric function of the homogeneous electric field, and $|\pmb{v}^\text{sq}(E_y)|=|\pmb{v}^\text{sq}(-E_y)|$. In contrast, in a triangular wire, the spin-orbit coupling does not need to be symmetric and the spin-orbit velocity $v$ is modified in different ways when the hole wavefunction is pushed to the bottom or to the apex of the triangle, see Fig.~\ref{fig:alpha_hom_E}.
We notice, however, that the amplitude of the spin-orbit velocity in triangular wires is still a symmetric function of the homogeneous field $E_x$.

To make a more quantitative comparison, we consider a Si wire with side $L$ grown along the $z\parallel [001]$ direction. To linear order in the electric field, the direct Rashba spin-orbit coupling can be written as~\cite{DRkloeffel3}
\begin{equation}
\label{eq:squared-v}
v^\text{sq}\approx  \frac{\hbar}{mL} \left[-0.41\left(\gamma_3-\gamma_2\right)e^{-4i\varphi}+0.38\left(\gamma_3+\gamma_2\right)\right]   \frac{eE_y L}{\epsilon_c^\text{sq}},
\end{equation}
where $\epsilon_c^\text{sq}=\hbar^2\pi^2\gamma_1/mL^2$ is the characteristic confinement energy for a particle in a square cross-section. 
To obtain this expression, we combined Eqs.~(77), (78), (79), and (80) in Ref.~\citep{DRkloeffel3} and used the Luttinger parameter of Si. To facilitate the comparison with our Eqs.~\eqref{eq:v_E_funct_Ey} and~\eqref{eq:alpha_E} obtained for an equilateral triangle, we also introduced the imaginary spin-orbit velocity $v^\text{sq}=\pmb{v}_x^\text{sq}+i\pmb{v}_y^\text{sq}$, we expanded the function $\chi(\varphi)={0.36}/\Big({1+0.16(\gamma_3-\gamma_2)\cos(4\varphi)}\Big)$ [directly related to the function in Eq.~(79) of~\cite{DRkloeffel3}] to linear order in $(\gamma_3-\gamma_2)\cos(4\varphi)$, and we discarded the small terms oscillating as $e^{4i\varphi}$ and $e^{-8i\varphi}$.
The overall minus sign of the spin-orbit velocity here compared to Eq.~(80) in Ref.~\cite{DRkloeffel3} is a result of the field being applied in the $y$-direction instead of  the $x$-direction.
  
Comparing Eqs.~\eqref{eq:v_E_funct_Ey} and~\eqref{eq:squared-v}, we observe that in square and triangular wires the spin-orbit coupling is a sum of an isotropic term and an anisotropic term, proportional to $\gamma_3+\gamma_2$ and to $\gamma_3-\gamma_2$, respectively.
To linear order in $E_y$, the isotropic contributions in  both cross-sections are in good quantitative agreement, but the anisotropic terms are qualitatively different, see  Eq.~\eqref{eq:alpha_E}.
In fact, while in square wires the anisotropic term varies linearly with $E_y$   and its contribution to the overall direct Rasbha spin-orbit velocity is  roughly equal to the isotropic contribution, in a triangular wire $\alpha_A(\varphi)$ comprises a constant intrinsic term and has a negligible linear dependence on $E_y$.

Including higher powers in the electric field, we find an additional qualitative difference between the spin-orbit coupling in the two different cross-sections. In fact, in a square wire, the spin-orbit velocity has no corrections quadratic in $E_y$ and the next order corrections are proportional to $E_y^3$. In contrast, in a triangular FinFET, both $\alpha_{I}$ and  $\alpha_{A}$ present quadratic terms proportional to $E_y^2$, that make $v$ asymmetric in $E_y$, see Eq.~\eqref{eq:alpha_E}.

These qualitative differences are crucial here because, as discussed in Sec.~\ref{Sec:Homogeneous_E}, in triangular wires the presence of a spin-orbit switch at a finite value of the electric field is a result of the competition between the intrinsic spin-orbit coupling and the direct Rashba-like spin-orbit interaction dependent on the electric field. In a square Si wire there is no intrinsic spin-orbit velocity and  the spin-orbit coupling only vanishes when $E_{x,y}=0$ (or when $E_{x,y}\rightarrow \infty$, see Eq.~(86) in Ref.~\cite{DRkloeffel3}), a much inconvenient working point for electrostatically defined quantum dots.

Finally, another difference between triangular and square cross-sections comes from the sensitivity of the spin-orbit coupling to the quadratic potential $\delta\!E_{ij} r_i r_j/2$, see Eq.~\eqref{eq:Multipole-E}. Without an homogeneous electric field, such a potential is inversion symmetric and does not produce spin-orbit coupling in square wires. In contrast, in Sec.~\ref{Sec:Inhomogeneous_E}, we show that this potential can produce another spin-orbit switch in triangular FinFETs, where the inversion symmetry is broken by the cross-section.

\section{Spin-orbit coupling against $\delta\!E_{ii}$}
\label{app:delta_Eii}

\begin{figure*}
 \centering
\includegraphics[width=0.8\textwidth]{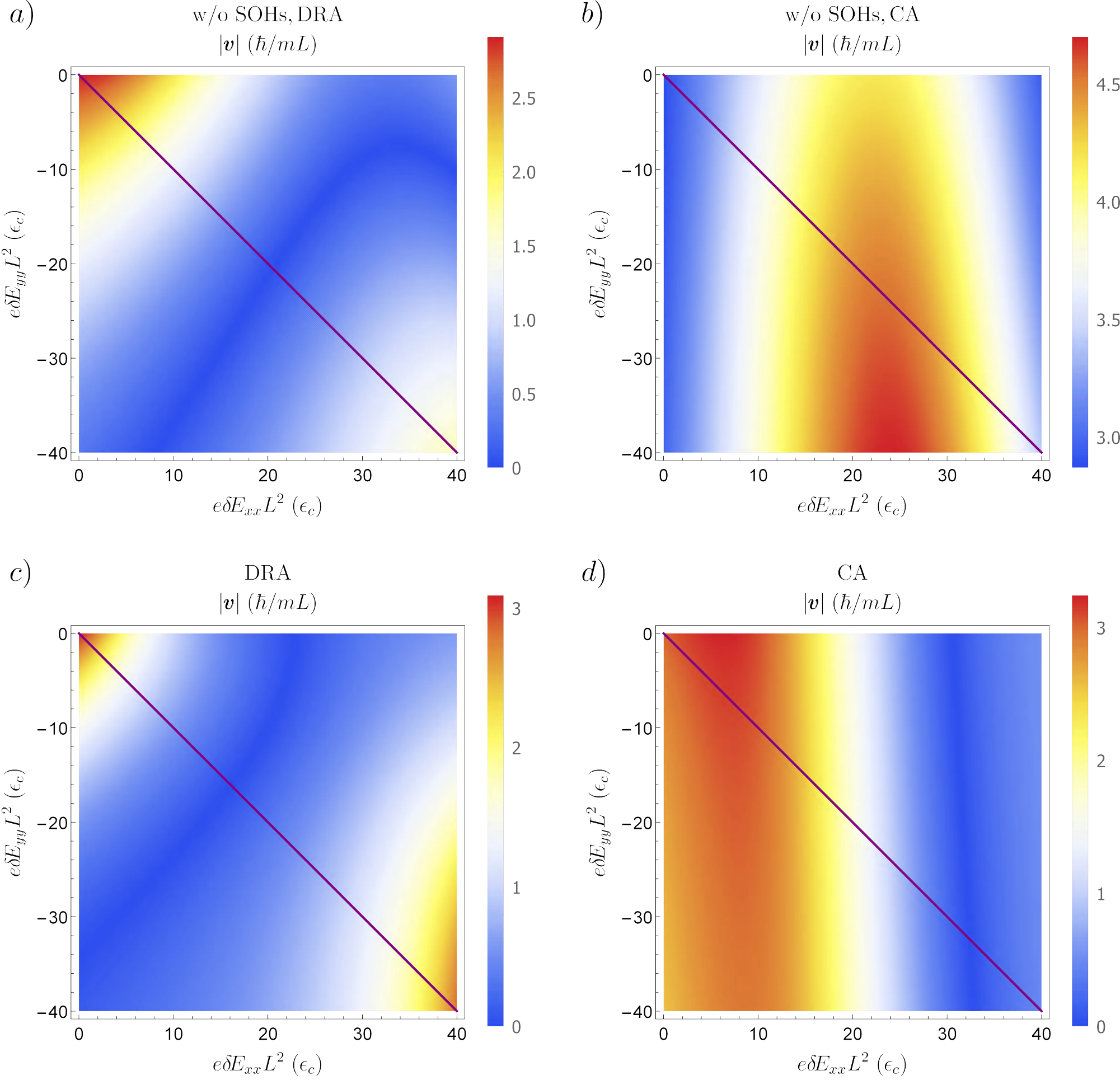} 
	\caption{ Spin-orbit velocity $|\pmb{v}|$ as a function of $\delta\!E_{xx}$ and $\delta\!E_{yy}$ when $\theta=0$. Here, we consider an equilateral triangular cross-section. In a) and b) we neglect the SOHs and show results obtained for the DRA and the CA orientations, respectively.
	In c) and d) we include the SOHs and simulate a wire with cross-section of side $L=20$~nm. We show results obtained for the DRA and CA orientations, respectively. 
	 The purple lines  show the constraint $\delta\!E_{xx}=-\delta\!E_{yy}$.  }
	\label{fig:alpha_inhom_6}
\end{figure*}

\begin{figure*}
 \centering
\includegraphics[width=0.8\textwidth]{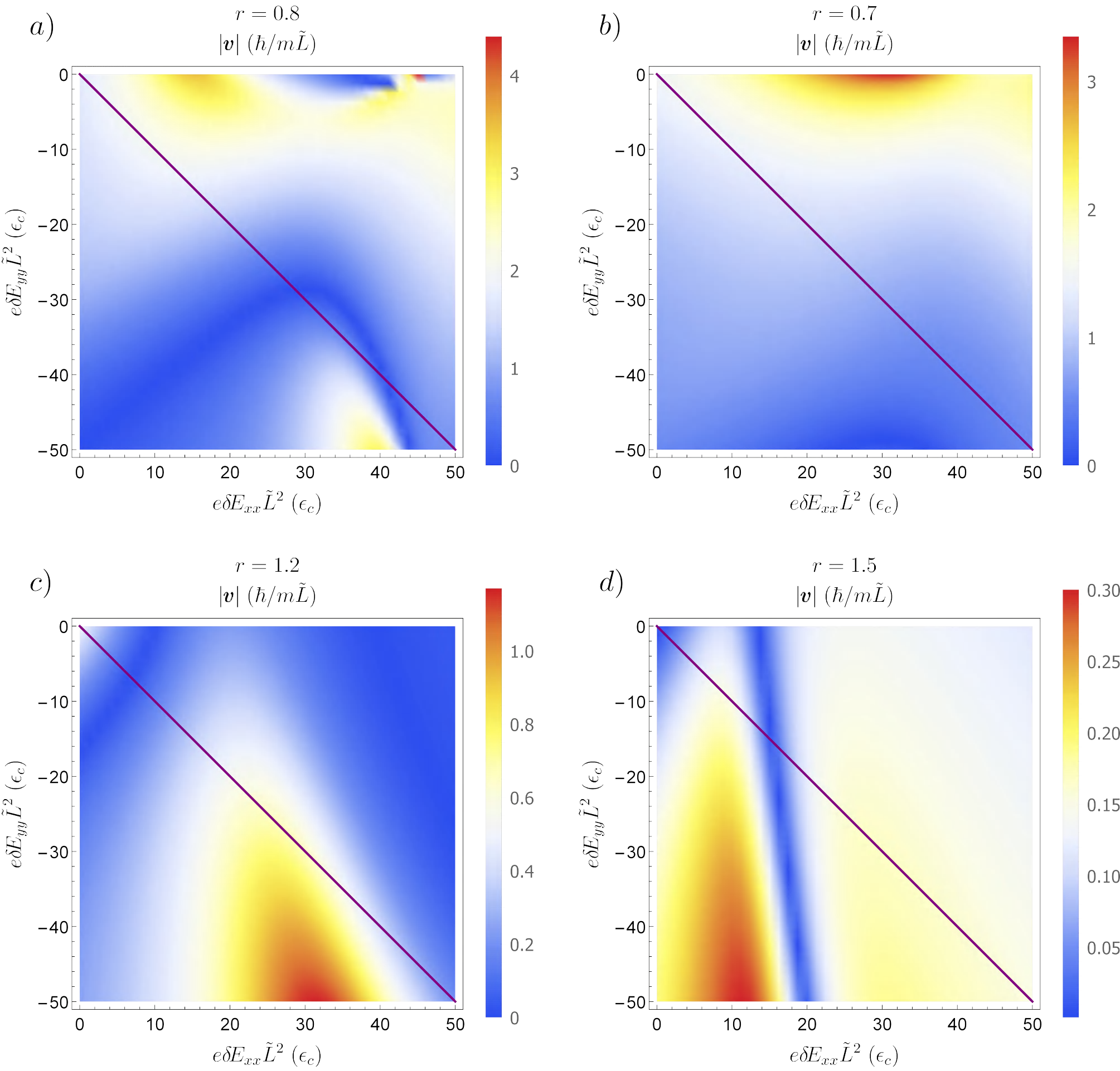}
	\caption{  Spin-orbit velocity $|\pmb{v}|$ as a function of $\delta\!E_{xx}$ and $\delta\!E_{yy}$ for isosceles triangles. Here, we consider wires grown along the DRA orientation whose cross-sections have an effective length $\tilde{L}=20$~nm and different aspect ratios $r=L_y/L_x$. The SOHs are fully included in these results. In a) and b) we show the spin-orbit coupling obtained for narrow triangles ($r<1$), while in c) and d) we show the results obtained for wide triangles ($r>1$). The purple lines  show the constraint $\delta\!E_{xx}=-\delta\!E_{yy}$. 
		}
	\label{fig:alpha_inhom_6_ISO}
\end{figure*}

In the main text, we assume that the inhomogeneous electric field tensor is diagonal and $\delta\!E_{xx}\approx-\delta\!E_{yy}\equiv  \delta\!E$. Here, we show that the presence of the spin-orbit switch is not related to this approximation by examining separately the effect of the fields $\delta\!E_{xx}$ and $\delta\!E_{yy}$ on the spin-orbit velocity. We restrict ourselves to the analysis of FinFETs grown along the [001] direction, with $\theta=0$. In this section, we consider $E_y=0$.

In Fig.~\ref{fig:alpha_inhom_6}, we study the spin-orbit velocity in equilateral triangles.
In the top Figs.~\ref{fig:alpha_inhom_6}a) and \ref{fig:alpha_inhom_6}b), we show the results obtained for large wires neglecting the SOHs and using the $4\times 4$ LK Hamiltonian. The two figures correspond to the DRA and CA, respectively.
The purple line marks the approximation $\delta\!E_{xx}=-\delta\!E_{yy}$ used in the main text. We observe that in the DRA, $|\pmb{v}|=0$ also by considering a more general relation between the inhomogeneous fields $\delta\!E_{yy}=c_{yy} \delta\!E_{xx}/c_{xx}$, see Eq.~\eqref{eq:multipole-parameters}. The spin-orbit coupling is not suppressed only when $|c_{yy} /c_{xx}|\ll 1$. We remark that the parameters $c_{yy}$ and $c_{xx}$ can  vary depending on the device design.
In the same regime of parameters, we do not find a comparable sweet spots for the CA.
In the bottom Figs.~\ref{fig:alpha_inhom_6}c) and \ref{fig:alpha_inhom_6}d), we show the results obtained by including the SOHs for an equilateral triangular cross-section of side $L=20$~nm. For the DRA, the spin-orbit coupling vanishes for any value of the ratio $|c_{yy} /c_{xx}|$ and interestingly also when $|c_{yy} /c_{xx}|\ll 1$, in contrast to when the SOHs are neglected.
In addition, the SOHs  modify the response when the wire is grown along the crystallographic orientation. In fact, in Fig.~\ref{fig:alpha_inhom_6}d) we observe that the spin-orbit velocity vanishes along the vertical line $\delta\!E_{xx}\approx 30 \epsilon_c/(eL^2)$. While this result shows that also this orientation might present suitable working points where  charge noise is suppressed, we do not investigate this possibility further.\\

We now study how these results change when we consider wires with an isosceles triangular cross-section as the ones studied in Sec.~\ref{Sec:SOHS}.
We consider a FinFET in the DRA and in Fig.~\ref{fig:alpha_inhom_6_ISO}, we compare the spin-orbit velocity in isosceles triangles with the same area $\sqrt{3}\tilde{L}^2/4$ and different aspect ratios $r=L_x/L_y$. Here, we fully account for the SOHs and we use an effective side length of $\tilde{L}=20$~nm. In the top Figs.~\ref{fig:alpha_inhom_6_ISO}a) and~\ref{fig:alpha_inhom_6_ISO}b), we show the spin-orbit velocity when the triangle is narrow.  When $r=0.8$, the  spin-orbit coupling is still suppressed when $\delta\!E_{yy}=c_{yy} \delta\!E_{xx}/c_{xx}=-\delta\!E_{xx}$ (purple line), but when the ratio  $|c_{yy} /c_{xx}|\lesssim 1$, the spin-orbit coupling does not vanish and the charge noise sweet spot is removed. Even worse, when $r=0.7$, the spin-orbit coupling vanishes only when the ratio $|c_{yy} /c_{xx}|\gg 1$, away from the limit studied in the main text.
In the bottom Figs.~\ref{fig:alpha_inhom_6_ISO}c) and~\ref{fig:alpha_inhom_6_ISO}d), we  show the spin-orbit velocity when the triangle is wide. 
In contrast to the narrow triangle, here the spin-orbit coupling vanishes for any values of the ratio $|c_{yy} /c_{xx}|$. By  increasing $r$, we observe that the line where $|\pmb{v}|=0$ is pushed towards lower values of $\delta\!E$ and when $r=1.5$, an additional line where the spin-orbit coupling vanishes appears. However, we also note that the maximal spin-orbit velocity in these devices decreases compared to the equilateral FinFET, see Fig.~\ref{fig:alpha_inhom_6}c).

\section{Effect of strain}
\label{app:strain}

\begin{figure*}
 \centering
\includegraphics[width=0.8\textwidth]{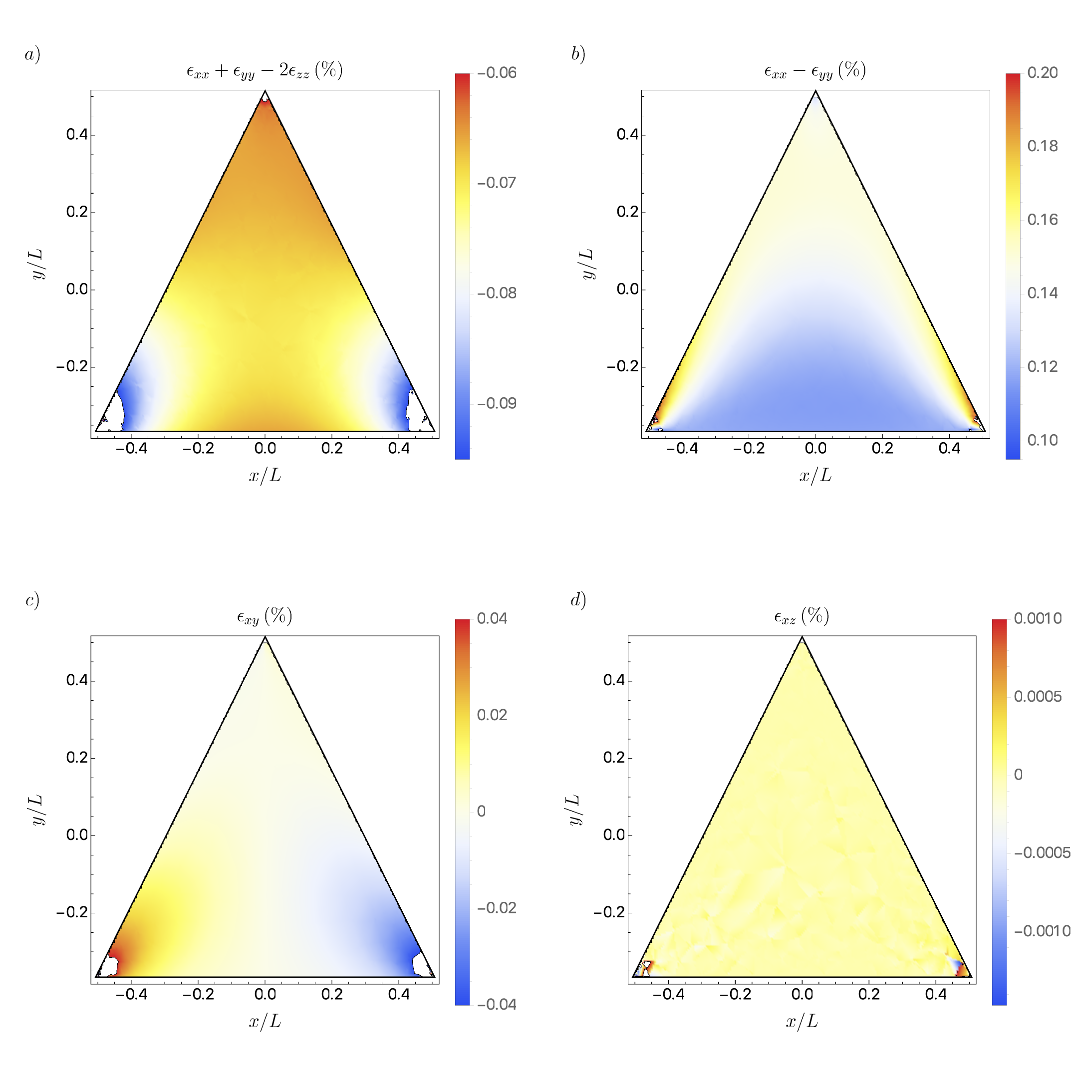} 
	\caption{ Simulation of the strain profile in a Si FinFET. We consider here the strain in a $L=20$~nm triangular fin caused by a pressure of 100 MPa applied on the top interface (the two surfaces with side length $L_y$ in Fig.~\ref{fig:Geometry-finfet}), and show the distribution of the relevant combinations of the $\epsilon_{ij}$ parameters, see Eq.~\eqref{eq:max-strain}.
	The values provided here are given in percentages.
	}
	\label{fig:strain-profile}
\end{figure*}

Here, we examine in detail how strain modifies the spin-orbit coupling. In particular, we extract the maximal strain that the system can support before the spin-orbit switch is removed. We restrict ourselves to the analysis of heavy and light holes of a fin grown along the DRA, where the effect of the strain tensor elements $\epsilon_{ij}$ is well-described by the Bir-Pikus Hamiltonian \cite{WinklerSpinOrbitCoupling2003,bir1974symmetry}
\begin{equation}
\label{eq:BPH}
H_\text{BP}^\text{DRA}= \epsilon_0^S J_z^2+ \left(
\begin{array}{cccc}
 0 & \epsilon _1^S & \epsilon _2^S & 0 \\
 \left(\epsilon _1^S\right){}^* & 0 & 0 & \epsilon _2^S \\
 \left(\epsilon _2^S\right){}^* & 0 & 0 & -\epsilon _1^S \\
 0 & \left(\epsilon _2^S\right){}^* & -\left(\epsilon _1^S\right){}^* & 0 \\
\end{array}
\right) \ ,
\end{equation}
with
\begin{subequations}
\label{eq:strain-energies}
\begin{align}
\epsilon_0^S&= -2b \left(\epsilon_{xx}+\epsilon_{yy}-2\epsilon_{zz} \right) \ , \\
\epsilon_1^S&= d \left(\epsilon_{xz}-i\epsilon_{yz} \right) \ , \\
\epsilon_2^S&= \frac{\sqrt{3}}{2}b \left(\epsilon_{xx}-\epsilon_{yy} \right)+i d \epsilon_{xy} \ .
\end{align}
\end{subequations}
For Si, the parameters $b=-2.2$~eV and $d=-5.1$~eV can be found e.g. in Ref.~\cite{WinklerSpinOrbitCoupling2003}.

\begin{figure*}
 \centering
\includegraphics[width=0.95\textwidth]{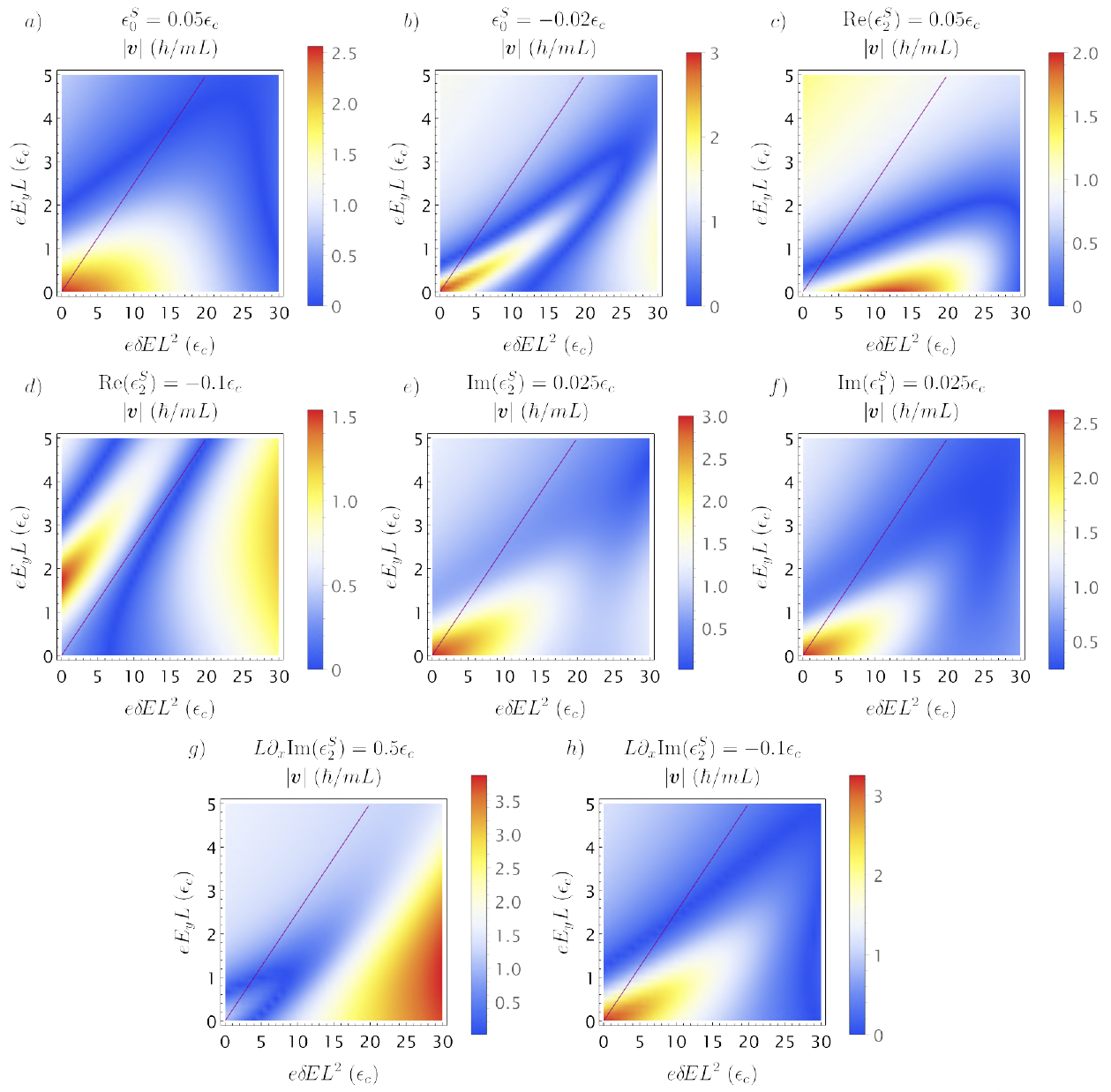} 
	\caption{ Effect of strain on the spin-orbit coupling. We use here the Bir-Pikus Hamiltonian in Eq.~\eqref{eq:BPH}, that is parameterized by the strain energies $\epsilon_i^S$ related to the strain tensor via Eq.~\eqref{eq:strain-energies}. The values used in the simulation are given in units of $\epsilon_c$, see Eq.~\eqref{eq:Confinement_energy}. In a) and b), we show how  the dependence of $|\pmb{v}|$ as a function of $E_y$ and $\delta\!E$ changes by  positive and negative values of the energy $\epsilon_0^S$, respectively.  In c), d) and e), f), we study the effect of  positive and negative values of $\text{Re}(\epsilon_2^S)$ and the effect of $\text{Im}(\epsilon_2^S)$ and $\text{Im}(\epsilon_1^S)$, respectively.  In g) and h), we show how the inhomogeneous strain field $x \partial_x \text{Im}(\epsilon_2^S) $ acts on the spin-orbit coupling for positive and negative values of the gradient.  }
	\label{fig:strain-homo}
\end{figure*}

In general, the strain elements $\epsilon_{ij}$ are functions of position, resulting in a complicated spin-dependent potential. We estimate the strain profile in the Si FinFET shown in Fig.~\ref{fig:Geometry-finfet},  when a pressure of $100$~MPa is applied pushing the top interface downwards in the $y$-direction. For the simulation, we used the Structural Mechanics module of COMSOL Multiphysics{\textregistered} \cite{COMSOL}, and considered an equilateral triangular cross-section of side $L=20$~nm, and a substrate thickness of $d_B= 20$~nm. We imposed free boundary condition on the substrate in the $x$-direction, while the bottom interface is kept fixed. The total width of the substrate in the $x$ direction is $50$~nm, with the fin being placed in the middle.

The relevant combination of the strain tensors are shown in Fig.~\ref{fig:strain-profile}. With this simple model, we observe that most  terms are rather homogeneous in the cross-section, and thus we study the effect of constant values of the $\epsilon_i^S$ energies. 
The homogeneous approximation is reasonable for the diagonal elements $\epsilon_{ii}$, but it is more debatable for the cross-terms $\epsilon_{ij\neq i}$. In particular, the term $\epsilon_{xy}$ has also a component that varies linearly in the $x$-direction, and so we extend our analysis by using $\text{Im}[\epsilon_2^S(x)]\approx \text{Im}(\epsilon_2^S)+x \partial_x \text{Im}(\epsilon_2^S)$. Note however that in our simulation we are applying the pressure directly on the top of the Si structure. Applying the pressure on the electrode could potentially reduce the strain close to the boundaries of the Si triangle, due to the effect of the gate oxide.
We also point out that the terms $\epsilon_{zz}$ and $\epsilon_{iz}$ are likely to have a $z$-dependence when qubits are defined and the top gate is terminated along the $z$-direction. 
However, we expect the strain field to be strongly peaked in a narrow region close to the edges of the electrodes, where the hole density is small, and, in the following, we neglect these inhomogeneities.

By studying the effect of each term independently, we find a reasonable estimation of the strain that the system can support. 
The results of this analysis are given in Fig.~\ref{fig:strain-homo}, where we show how the spin-orbit velocity dependence on $E_y$ and $\delta\!E$ is affected by the strain energies $\epsilon_i^S$ given in Eq.~\eqref{eq:strain-energies}.
Here, we express the energies $\epsilon_i^S$ in terms of the confinement energy $\epsilon_c\approx 17.2/L^2 \ \text{eV}\times\text{nm}^2$.
A positive and negative diagonal strain $\epsilon_0^S$ is examined in Figs.~\ref{fig:strain-homo}a) and~\ref{fig:strain-homo}b), respectively. The spin-orbit switch is robust against $\epsilon_0^S$, but the shape of the curve along which $|\pmb{v}|=0$ changes and, in particular, negative (positive) values of the strain narrow (widen) the curve.
We extract bounds on the maximal strain allowed by considering that when $\epsilon_0^S/\epsilon_c\gtrsim 0.1$, the $|\pmb{v}|=0$ curve is too wide and the inhomogeneous switch $\delta\!E^{SW}$ disappears from the range of parameters considered. In contrast, when $\epsilon_0^S/\epsilon_c\lesssim -0.035$, the curve becomes too narrow and results in a finite spin-orbit coupling in the whole parameter space.

Similar physics appears when the contribution of $\text{Re}(\epsilon_2^S)\propto \epsilon_{xx}-\epsilon_{yy}$ is examined, see Figs.~\ref{fig:strain-homo}c) and~\ref{fig:strain-homo}d) for positive and negative values of $\text{Re}(\epsilon_2^S)$, respectively. In fact, the spin-orbit switch persists, but the $|\pmb{v}|=0$ curve is modified by the strain. 
In particular, when $\text{Re}(\epsilon_2^S)<0$  the inhomogeneous field $\delta\!E$ is strongly enhanced and the switching field $\delta\!E^{SW}$ is pushed towards lower values.
For this reason, a moderate negative strain might be helpful to compensate for the SOHs, in analogy to the wide FinFET discussed in Sec.~\ref{Sec:SOHS}. We do not explore this intriguing possibility in more detail. 
In analogy to before, we extract the bounds $\text{Re}(\epsilon_2^S)/\epsilon_c\in [-0.1,0.1]$ by verifying when the switch is pushed outside the range of parameters studied.

In contrast, the homogeneous components of the cross-couplings $\epsilon_{ij}$ remove the spin-orbit switch, but the shape of the curve where $|\pmb{v}|$ is minimal does not change.
In Figs.~\ref{fig:strain-homo}e) and~\ref{fig:strain-homo}f), we show how the terms $\text{Im}(\epsilon_2^S)\propto \epsilon_{xy}$ and $\text{Im}(\epsilon_1^S)\propto \epsilon_{yz}$ influence $|\pmb{v}|$. The effect of $\text{Re}(\epsilon_1^S)\propto \epsilon_{xz}$ is analogous to the effect of $ \epsilon_{yz}$ and is not reported here.
To find bounds on the maximal strain allowed, we estimate that when $|\text{Im}(\epsilon_1^S)|/\epsilon_c<0.025$ and $|\text{Im}(\epsilon_2^S)|/\epsilon_c<0.025$, the minimal value of the spin-orbit coupling is lower than $10\%$ of the maximal coupling. 
We report the limiting values of the tensor elements $\epsilon_{ij}$ in Eq.~\eqref{eq:max-strain}.

Finally, in Figs.~\ref{fig:strain-homo}g) and~\ref{fig:strain-homo}h), we show how the largest inhomogeneous strain component $x \partial_x \text{Im}(\epsilon_2^S)$ alters $|\pmb{v}|$.
In analogy to $\epsilon_0^{S}$ and $\text{Re}(\epsilon_2^S)$, the spin-orbit coupling vanishes along a curve that is rescaled by strain.
We estimate that the physics described in the main text remains qualitatively valid when $L \partial_x \text{Im}(\epsilon_2^S)/\epsilon_c\in [-0.15,0.5]$, from which it follows that $L\partial_x\epsilon_{xy}\in [-1.7\%,0.5\%]\times (L/10\ \text{nm})^{-2}$. 

From Fig.~\ref{fig:strain-profile},   we see that when $L=20$~nm, the cross-term $\epsilon_{xy}\in[-0.06\%,0.06\%]$, resulting in  $|L\partial_x\epsilon_{xy}|\sim 0.1\%$. This value is smaller but still comparable to the estimated bound. However, we observe that most of the inhomogeneity of the strain comes from hotspots at the boundary of the triangle, where the wavefunction has no support, and thus the effective value of $|L\partial_x\epsilon_{xy}|$ is even smaller in more realistic scenarios.  In addition, while the homogeneous part of the cross-terms $\epsilon_{ij\neq i}$ remove the spin-orbit switch, from the simple simulations shown in Fig.~\ref{fig:strain-profile}, we expect those terms to be rather small. Thus, we believe that the conditions on the homogeneous part of the diagonal elements $\epsilon_{ii}$ are the most stringent ones.

\section{Compensating for the SOHs}
\label{Sec:Deviation}

\begin{figure}[t]
\includegraphics[width=0.48\textwidth]{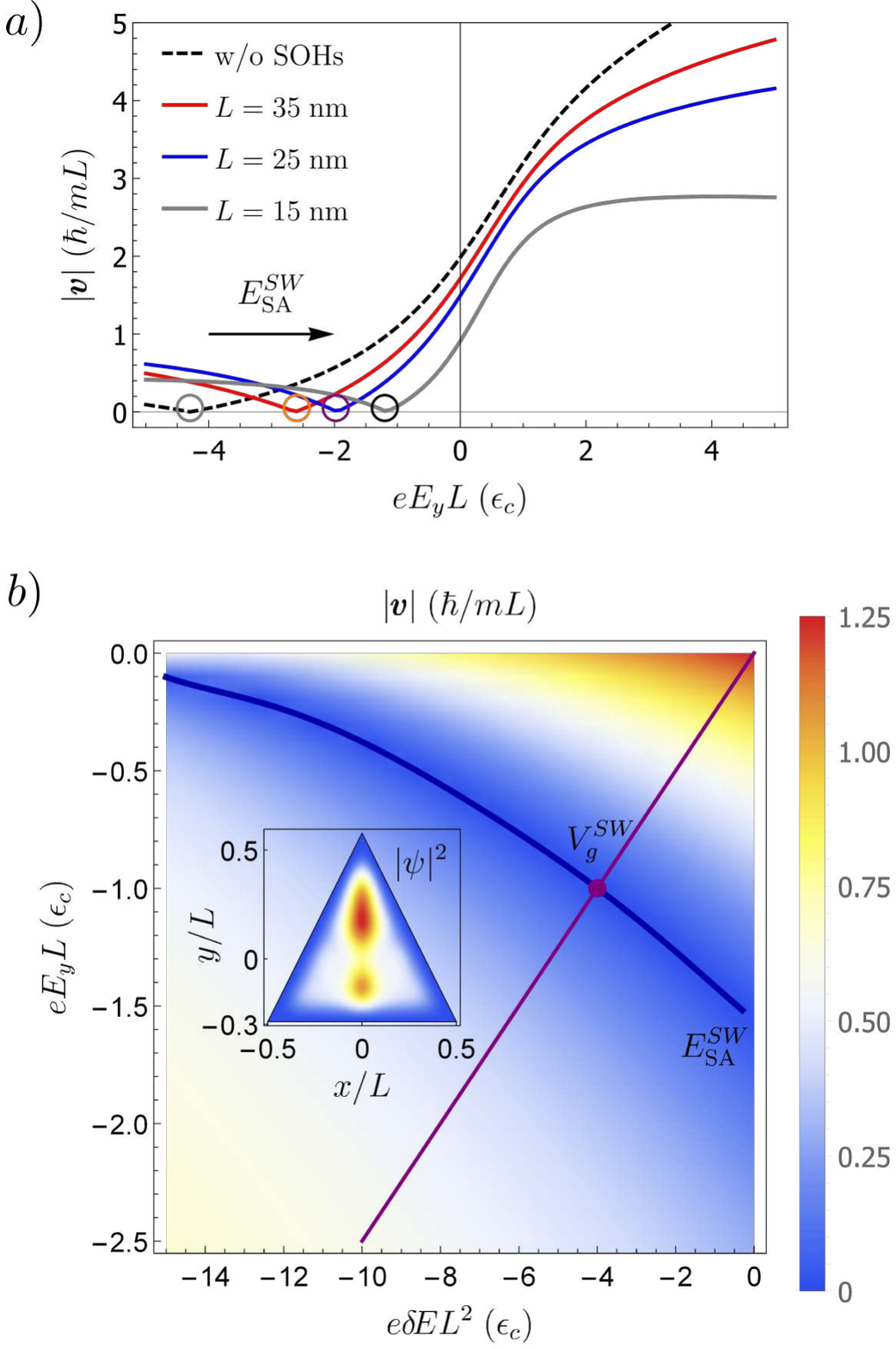}
	\caption{Spin-orbit velocity $|\pmb{v}|$ in an equilateral SOI FinFET grown along the SA.
	In a), we show how the dependence of $|\pmb{v}|$ on  $E_y$ varies as a function of $L$. The switching field $E_\text{SA}^{SW}$ is pushed towards lower values as the triangle becomes smaller. In this device, the spin-orbit coupling is removed by a negative electric field generated by a positive gate potential $V_g$ that pushes the hole wavefunction to the bottom of the triangle. In the units used, the results obtained without including the SOHs are independent of  $L$.
	In b) we show $|\pmb{v}|$ as a function of $E_y$ and $\delta\!E$ for a device with $L=20$~nm. The spin-orbit coupling vanishes along the blue line. The purple line indicates the constraint in Eq.~\eqref{eq:lineEy-deltaE}.
	In the inset, we show the total density $|\psi|^2$ of the hole wavefunction in the cross-section at  $V_g^{SW}$, where purple and blue lines intersect. The density vanishes (is maximal) in the blue (red) region.}
	\label{fig:SOI_Exp_SOHs}
\end{figure}
 
\begin{figure}[t]
\includegraphics[width=0.45\textwidth]{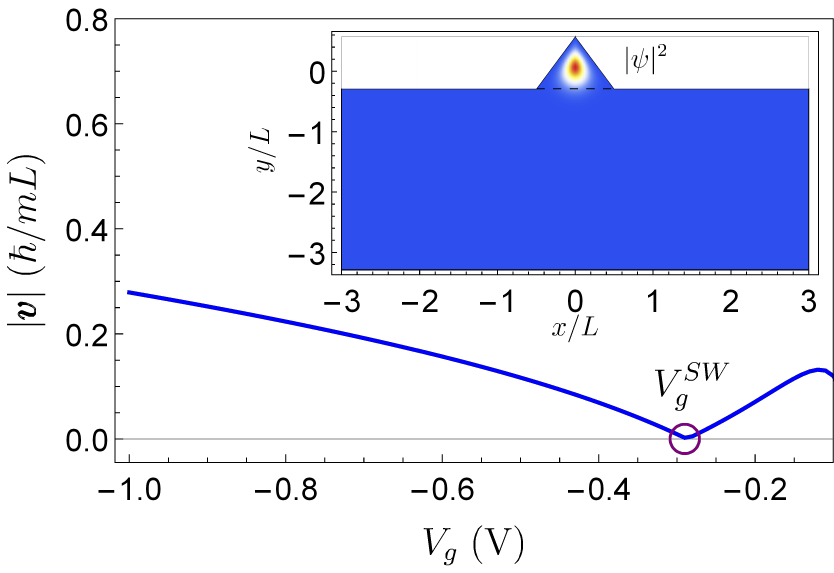}
	\caption{Spin-orbit velocity $|\pmb{v}|$ in a bulk equilateral FinFET grown along the DRA.
	We show $|\pmb{v}|$ as a function of the gate potential $V_g$ in a device where an equilateral triangle with side $L=20$~nm is placed on top of a substrate $6L$ wide and $3L$ high. The electrostatic potential in this calculation is simulated by solving the Laplace equation as discussed in App.~\ref{app:E-field}.  In the inset, we show the total density $|\psi|^2$ of the hole wavefunction at $V_g=-0.28$~V, where the spin-orbit coupling vanishes. The hole density vanishes in the blue region and attains maximal value in the red region. At this potential, the holes are localised in the fin.  }
	\label{fig:substrate-v}
\end{figure}

In Sec.~\ref{Sec:SOHS}, we show that the SOHs can remove the spin-orbit switch in small wires and wide DRA FinFETs are proposed to restore the sweet spot. 
Here, we discuss other possible design concepts valid for both SOI and bulk Si FinFETs that can compensate for the SOHs. 
In particular, in equilateral SOI FinFETs  the spin-orbit can be switched-off in wires grown along the $[110]$ direction (SA)  by pushing the hole wavefunction at the bottom of the triangles, while in bulk FinFETs, the spin orbit switch is naturally recovered by considering the leakage of the hole wavefunction into the Si substrate; a summary of the different designs considered is given in  Table~\ref{tab:FinFETS}.

\begin{table}[]
\begin{tabular}{|c|c|c|c|c|}
\hline
 FinFET   & Orientation  &  Aspect   ratio   & $l_{so}^\text{min}$ & $V_g^{SW}$  \\ \hline
 SOI/Bulk  & DRA & $r=1.2$   & $1.5 \tilde{L}$ & $-3.12$~V   \\ \hline
 SOI  &  SA  & $r=1 \ \ $ & $4L$               & $+0.72$~V   \\ \hline
 Bulk  &  DRA  &  $r=1 \ \ $ &   $10L$                & $-0.28$~V   \\ \hline
\end{tabular}
\caption{Examples of FinFETs designs where the spin-orbit sweet spot is restored in the small  cross-section limit. To estimate the values of minimal spin-orbit length we consider devices with $L=\tilde{L}=20$~nm. For the first two designs, we consider a back gate at $d_B=100$~nm, while for the last FinFET, we consider a cross-section $120$~nm wide and $60$~nm high. $V_g^{SW}$ increases by increasing $d_B$ or by decreasing $L$.  \label{tab:FinFETS}}
\end{table}

In small equilateral SOI FinFETs, the spin-orbit velocity can also be suppressed by pushing the hole wavefunction towards the bottom of the fin by a positive gate potential. As discussed in Sec.~\ref{Sec:Homogeneous_E}, in a wire grown along the $[110]$ direction (SA) with an equilateral triangular cross-section, the spin-orbit velocity vanishes by the effect of a negative homogeneous electric field $E_\text{SA}^{SW}<0$, see Eq.~\eqref{eq:SW-Homogeneneous}.
In Fig.~\ref{fig:SOI_Exp_SOHs}a), we show the dependence of this switching field on the size of the equilateral cross-section. In this case, the  SOHs do not remove the sweet spot, but they push it to lower values of the homogeneous electric field, that can be reached by a smaller gate potential.
In Fig.~\ref{fig:SOI_Exp_SOHs}b), we study the spin-orbit coupling in an equilateral wire with side $L=20$~nm  when also $\delta\!E$ is included.
In this case, we observe that, along the purple line defined by Eq.~\eqref{eq:lineEy-deltaE}, $|\pmb{v}|$ vanishes at the gate potential $V_g^{SW}\approx + 57.3\times \text{nm}^2 d_B/\tilde{L}^3$~V, corresponding to $V_g^{SW}\approx +0.72$~V when $\tilde{L}=20$~nm and $d_B=100$~nm, easily achievable in state-of-the-art devices. The total hole density $|\psi|^2$ at the switching potential $V_g^{SW}$ is shown in the inset of Fig.~\ref{fig:SOI_Exp_SOHs}b).
 Comparing to the wide DRA FinFET, we find that in the regime of parameters examined, the SA device has a larger minimal spin-orbit length $l_{so}^\text{min}\approx 4 L$, leading to  smaller spin-orbit interactions when the qubit is operational.  \\

So far, we focused on FinFETs, where the triangular fin is well-separated from the bulk and we modelled these systems by using hard-wall boundary conditions  at the edges of the triangle. This approximation is valid as long as the hole wavefunction is well-confined inside the fin, such that the substrate can be neglected. 
In bulk Si FinFETs, there is a thick substrate that is strongly coupled to the holes in the wire and the confinement potential that localizes the holes in the fin is  provided by the negative potential $V_g$ applied to the top gate.
By fully simulating this cross-section, in Fig.~\ref{fig:substrate-v}, we show that a convenient working point where the spin-orbit velocity can be completely removed at smaller values of the gate potential [see Table~\ref{tab:FinFETS}] emerges naturally also in these systems when the wire is grown in the DRA.
To obtain this result, we simulate a cross-section composed of an equilateral triangular fin with side $L=20$~nm symmetrically placed on top of a rectangular substrate $120$~nm wide and $60$~nm high. For this simulation, we used the $6\times 6$ LK Hamiltonian and the electrostatic potential generated by the top gate is calculated by solving the Laplace equation with the boundary conditions described in App.~\ref{app:E-field}.
When $|V_g|\gtrsim 0.1$~V, the holes are confined in the fin and  the spin-orbit velocity shows a behaviour that is in qualitative agreement with our treatment, see e.g. Fig.~\ref{fig:alpha_hom_E}.  The results obtained for lower values of the gate potential $|V_g|<0.1$~V, where the hole wavefunction is largely spread in the substrate are inaccurate and have been removed from the figure.
In this system, $|\pmb{v}|=0$ at $V_g^{SW}=-0.28$~V, where the wavefunction is strongly localized into the fin, see the inset of the figure; in larger devices, we expect again that the potential will scale roughly as $V_g^{SW}\propto d_B/L^3$. While the spin-orbit coupling can be conveniently switched off in this setup, we also estimate that the minimal spin-orbit length $l_{so}^\text{min}\sim 10L$ is larger than in the other designs. \\

Finally, in App.~\ref{app:strain}, it is shown that strain can enhance the effect of the inhomogeneous  electric field $\delta\!E$ on the spin-orbit velocity. Consequently, we expect that  one could recover the spin-orbit switch also by appropriately engineering the strain field in the device. However, here we do not analyze this possibility more quantitatively.

\end{appendix}
\bibliography{references}

\begin{thebibliography}{93}%
\makeatletter
\providecommand \@ifxundefined [1]{%
 \@ifx{#1\undefined}
}%
\providecommand \@ifnum [1]{%
 \ifnum #1\expandafter \@firstoftwo
 \else \expandafter \@secondoftwo
 \fi
}%
\providecommand \@ifx [1]{%
 \ifx #1\expandafter \@firstoftwo
 \else \expandafter \@secondoftwo
 \fi
}%
\providecommand \natexlab [1]{#1}%
\providecommand \enquote  [1]{``#1''}%
\providecommand \bibnamefont  [1]{#1}%
\providecommand \bibfnamefont [1]{#1}%
\providecommand \citenamefont [1]{#1}%
\providecommand \href@noop [0]{\@secondoftwo}%
\providecommand \href [0]{\begingroup \@sanitize@url \@href}%
\providecommand \@href[1]{\@@startlink{#1}\@@href}%
\providecommand \@@href[1]{\endgroup#1\@@endlink}%
\providecommand \@sanitize@url [0]{\catcode `\\12\catcode `\$12\catcode
  `\&12\catcode `\#12\catcode `\^12\catcode `\_12\catcode `\%12\relax}%
\providecommand \@@startlink[1]{}%
\providecommand \@@endlink[0]{}%
\providecommand \url  [0]{\begingroup\@sanitize@url \@url }%
\providecommand \@url [1]{\endgroup\@href {#1}{\urlprefix }}%
\providecommand \urlprefix  [0]{URL }%
\providecommand \Eprint [0]{\href }%
\providecommand \doibase [0]{http://dx.doi.org/}%
\providecommand \selectlanguage [0]{\@gobble}%
\providecommand \bibinfo  [0]{\@secondoftwo}%
\providecommand \bibfield  [0]{\@secondoftwo}%
\providecommand \translation [1]{[#1]}%
\providecommand \BibitemOpen [0]{}%
\providecommand \bibitemStop [0]{}%
\providecommand \bibitemNoStop [0]{.\EOS\space}%
\providecommand \EOS [0]{\spacefactor3000\relax}%
\providecommand \BibitemShut  [1]{\csname bibitem#1\endcsname}%
\let\auto@bib@innerbib\@empty
%</preamble>
\bibitem [{\citenamefont {Winkler}(2003)}]{WinklerSpinOrbitCoupling2003}%
  \BibitemOpen
  \bibfield  {author} {\bibinfo {author} {\bibfnamefont {R.}~\bibnamefont
  {Winkler}},\ }\href {\doibase 10.1007/b13586} {\emph {\bibinfo {title}
  {Spin--{{Orbit Coupling Effects}} in {{Two}}-{{Dimensional Electron}} and
  {{Hole Systems}}}}},\ edited by\ \bibinfo {editor} {\bibfnamefont
  {G.}~\bibnamefont {H{\"o}hler}}, \bibinfo {editor} {\bibfnamefont {J.~H.}\
  \bibnamefont {K{\"u}hn}}, \bibinfo {editor} {\bibfnamefont {T.}~\bibnamefont
  {M{\"u}ller}}, \bibinfo {editor} {\bibfnamefont {J.}~\bibnamefont
  {Tr{\"u}mper}}, \bibinfo {editor} {\bibfnamefont {A.}~\bibnamefont
  {Ruckenstein}}, \bibinfo {editor} {\bibfnamefont {P.}~\bibnamefont
  {W{\"o}lfle}}, \ and\ \bibinfo {editor} {\bibfnamefont {F.}~\bibnamefont
  {Steiner}},\ \bibinfo {series} {Springer {{Tracts}} in {{Modern Physics}}},
  Vol.\ \bibinfo {volume} {191}\ (\bibinfo  {publisher} {Springer Berlin
  Heidelberg},\ \bibinfo {address} {Berlin, Heidelberg},\ \bibinfo {year}
  {2003})\BibitemShut {NoStop}%
\bibitem [{\citenamefont {Loss}\ and\ \citenamefont
  {DiVincenzo}(1998)}]{PhysRevA.57.120}%
  \BibitemOpen
  \bibfield  {author} {\bibinfo {author} {\bibfnamefont {D.}~\bibnamefont
  {Loss}}\ and\ \bibinfo {author} {\bibfnamefont {D.~P.}\ \bibnamefont
  {DiVincenzo}},\ }\href {\doibase 10.1103/PhysRevA.57.120} {\bibfield
  {journal} {\bibinfo  {journal} {Phys. Rev. A}\ }\textbf {\bibinfo {volume}
  {57}},\ \bibinfo {pages} {120} (\bibinfo {year} {1998})}\BibitemShut
  {NoStop}%
\bibitem [{\citenamefont {Hanson}\ \emph {et~al.}(2007)\citenamefont {Hanson},
  \citenamefont {Kouwenhoven}, \citenamefont {Petta}, \citenamefont {Tarucha},\
  and\ \citenamefont {Vandersypen}}]{RevModPhys.79.1217}%
  \BibitemOpen
  \bibfield  {author} {\bibinfo {author} {\bibfnamefont {R.}~\bibnamefont
  {Hanson}}, \bibinfo {author} {\bibfnamefont {L.~P.}\ \bibnamefont
  {Kouwenhoven}}, \bibinfo {author} {\bibfnamefont {J.~R.}\ \bibnamefont
  {Petta}}, \bibinfo {author} {\bibfnamefont {S.}~\bibnamefont {Tarucha}}, \
  and\ \bibinfo {author} {\bibfnamefont {L.~M.~K.}\ \bibnamefont
  {Vandersypen}},\ }\href {\doibase 10.1103/RevModPhys.79.1217} {\bibfield
  {journal} {\bibinfo  {journal} {Rev. Mod. Phys.}\ }\textbf {\bibinfo {volume}
  {79}},\ \bibinfo {pages} {1217} (\bibinfo {year} {2007})}\BibitemShut
  {NoStop}%
\bibitem [{\citenamefont {Nadj-Perge}\ \emph {et~al.}(2010)\citenamefont
  {Nadj-Perge}, \citenamefont {Frolov}, \citenamefont {Bakkers},\ and\
  \citenamefont {Kouwenhoven}}]{nadj2010spin}%
  \BibitemOpen
  \bibfield  {author} {\bibinfo {author} {\bibfnamefont {S.}~\bibnamefont
  {Nadj-Perge}}, \bibinfo {author} {\bibfnamefont {S.}~\bibnamefont {Frolov}},
  \bibinfo {author} {\bibfnamefont {E.}~\bibnamefont {Bakkers}}, \ and\
  \bibinfo {author} {\bibfnamefont {L.~P.}\ \bibnamefont {Kouwenhoven}},\
  }\href {\doibase 10.1038/nature09682} {\bibfield  {journal} {\bibinfo
  {journal} {Nature}\ }\textbf {\bibinfo {volume} {468}},\ \bibinfo {pages}
  {1084} (\bibinfo {year} {2010})}\BibitemShut {NoStop}%
\bibitem [{\citenamefont {Nowack}\ \emph {et~al.}(2007)\citenamefont {Nowack},
  \citenamefont {Koppens}, \citenamefont {Nazarov},\ and\ \citenamefont
  {Vandersypen}}]{Nowack1430}%
  \BibitemOpen
  \bibfield  {author} {\bibinfo {author} {\bibfnamefont {K.~C.}\ \bibnamefont
  {Nowack}}, \bibinfo {author} {\bibfnamefont {F.~H.~L.}\ \bibnamefont
  {Koppens}}, \bibinfo {author} {\bibfnamefont {Y.~V.}\ \bibnamefont
  {Nazarov}}, \ and\ \bibinfo {author} {\bibfnamefont {L.~M.~K.}\ \bibnamefont
  {Vandersypen}},\ }\href {\doibase 10.1126/science.1148092} {\bibfield
  {journal} {\bibinfo  {journal} {Science}\ }\textbf {\bibinfo {volume}
  {318}},\ \bibinfo {pages} {1430} (\bibinfo {year} {2007})}\BibitemShut
  {NoStop}%
\bibitem [{\citenamefont {Hendrickx}\ \emph
  {et~al.}(2020{\natexlab{a}})\citenamefont {Hendrickx}, \citenamefont
  {Franke}, \citenamefont {Sammak}, \citenamefont {Scappucci},\ and\
  \citenamefont {Veldhorst}}]{hendrickx2020fast}%
  \BibitemOpen
  \bibfield  {author} {\bibinfo {author} {\bibfnamefont {N.}~\bibnamefont
  {Hendrickx}}, \bibinfo {author} {\bibfnamefont {D.}~\bibnamefont {Franke}},
  \bibinfo {author} {\bibfnamefont {A.}~\bibnamefont {Sammak}}, \bibinfo
  {author} {\bibfnamefont {G.}~\bibnamefont {Scappucci}}, \ and\ \bibinfo
  {author} {\bibfnamefont {M.}~\bibnamefont {Veldhorst}},\ }\href {\doibase
  10.1038/s41586-019-1919-3} {\bibfield  {journal} {\bibinfo  {journal}
  {Nature}\ }\textbf {\bibinfo {volume} {577}},\ \bibinfo {pages} {487}
  (\bibinfo {year} {2020}{\natexlab{a}})}\BibitemShut {NoStop}%
\bibitem [{\citenamefont {Bulaev}\ and\ \citenamefont
  {Loss}(2005)}]{PhysRevLett.95.076805}%
  \BibitemOpen
  \bibfield  {author} {\bibinfo {author} {\bibfnamefont {D.~V.}\ \bibnamefont
  {Bulaev}}\ and\ \bibinfo {author} {\bibfnamefont {D.}~\bibnamefont {Loss}},\
  }\href {\doibase 10.1103/PhysRevLett.95.076805} {\bibfield  {journal}
  {\bibinfo  {journal} {Phys. Rev. Lett.}\ }\textbf {\bibinfo {volume} {95}},\
  \bibinfo {pages} {076805} (\bibinfo {year} {2005})}\BibitemShut {NoStop}%
\bibitem [{\citenamefont {Bulaev}\ and\ \citenamefont
  {Loss}(2007)}]{PhysRevLett.98.097202}%
  \BibitemOpen
  \bibfield  {author} {\bibinfo {author} {\bibfnamefont {D.~V.}\ \bibnamefont
  {Bulaev}}\ and\ \bibinfo {author} {\bibfnamefont {D.}~\bibnamefont {Loss}},\
  }\href {\doibase 10.1103/PhysRevLett.98.097202} {\bibfield  {journal}
  {\bibinfo  {journal} {Phys. Rev. Lett.}\ }\textbf {\bibinfo {volume} {98}},\
  \bibinfo {pages} {097202} (\bibinfo {year} {2007})}\BibitemShut {NoStop}%
\bibitem [{\citenamefont {Kloeffel}\ and\ \citenamefont
  {Loss}(2013)}]{doi:10.1146/annurev-conmatphys-030212-184248}%
  \BibitemOpen
  \bibfield  {author} {\bibinfo {author} {\bibfnamefont {C.}~\bibnamefont
  {Kloeffel}}\ and\ \bibinfo {author} {\bibfnamefont {D.}~\bibnamefont
  {Loss}},\ }\href {\doibase 10.1146/annurev-conmatphys-030212-184248}
  {\bibfield  {journal} {\bibinfo  {journal} {Annual Review of Condensed Matter
  Physics}\ }\textbf {\bibinfo {volume} {4}},\ \bibinfo {pages} {51} (\bibinfo
  {year} {2013})}\BibitemShut {NoStop}%
\bibitem [{\citenamefont {Marx}\ \emph {et~al.}(2020)\citenamefont {Marx},
  \citenamefont {Yoneda}, \citenamefont {Rubio}, \citenamefont {Stano},
  \citenamefont {Otsuka}, \citenamefont {Takeda}, \citenamefont {Li},
  \citenamefont {Yamaoka}, \citenamefont {Nakajima}, \citenamefont {Noiri},
  \citenamefont {Loss}, \citenamefont {Kodera},\ and\ \citenamefont
  {Tarucha}}]{marx2020spin}%
  \BibitemOpen
  \bibfield  {author} {\bibinfo {author} {\bibfnamefont {M.}~\bibnamefont
  {Marx}}, \bibinfo {author} {\bibfnamefont {J.}~\bibnamefont {Yoneda}},
  \bibinfo {author} {\bibfnamefont {{\'A}.~G.}\ \bibnamefont {Rubio}}, \bibinfo
  {author} {\bibfnamefont {P.}~\bibnamefont {Stano}}, \bibinfo {author}
  {\bibfnamefont {T.}~\bibnamefont {Otsuka}}, \bibinfo {author} {\bibfnamefont
  {K.}~\bibnamefont {Takeda}}, \bibinfo {author} {\bibfnamefont
  {S.}~\bibnamefont {Li}}, \bibinfo {author} {\bibfnamefont {Y.}~\bibnamefont
  {Yamaoka}}, \bibinfo {author} {\bibfnamefont {T.}~\bibnamefont {Nakajima}},
  \bibinfo {author} {\bibfnamefont {A.}~\bibnamefont {Noiri}}, \bibinfo
  {author} {\bibfnamefont {D.}~\bibnamefont {Loss}}, \bibinfo {author}
  {\bibfnamefont {T.}~\bibnamefont {Kodera}}, \ and\ \bibinfo {author}
  {\bibfnamefont {S.}~\bibnamefont {Tarucha}},\ }\href@noop {} {\bibfield
  {journal} {\bibinfo  {journal} {arXiv preprint arXiv:2003.07079}\ } (\bibinfo
  {year} {2020})}\BibitemShut {NoStop}%
\bibitem [{\citenamefont {Scappucci}\ \emph {et~al.}(2020)\citenamefont
  {Scappucci}, \citenamefont {Kloeffel}, \citenamefont {Zwanenburg},
  \citenamefont {Loss}, \citenamefont {Myronov}, \citenamefont {Zhang},
  \citenamefont {De~Franceschi}, \citenamefont {Katsaros},\ and\ \citenamefont
  {Veldhorst}}]{scappucci2020germanium}%
  \BibitemOpen
  \bibfield  {author} {\bibinfo {author} {\bibfnamefont {G.}~\bibnamefont
  {Scappucci}}, \bibinfo {author} {\bibfnamefont {C.}~\bibnamefont {Kloeffel}},
  \bibinfo {author} {\bibfnamefont {F.~A.}\ \bibnamefont {Zwanenburg}},
  \bibinfo {author} {\bibfnamefont {D.}~\bibnamefont {Loss}}, \bibinfo {author}
  {\bibfnamefont {M.}~\bibnamefont {Myronov}}, \bibinfo {author} {\bibfnamefont
  {J.-J.}\ \bibnamefont {Zhang}}, \bibinfo {author} {\bibfnamefont
  {S.}~\bibnamefont {De~Franceschi}}, \bibinfo {author} {\bibfnamefont
  {G.}~\bibnamefont {Katsaros}}, \ and\ \bibinfo {author} {\bibfnamefont
  {M.}~\bibnamefont {Veldhorst}},\ }\href {\doibase 10.1038/s41578-020-00262-z}
  {\bibfield  {journal} {\bibinfo  {journal} {Nature Reviews Materials}\ ,\
  \bibinfo {pages} {1}} (\bibinfo {year} {2020})}\BibitemShut {NoStop}%
\bibitem [{\citenamefont {Maurand}\ \emph {et~al.}(2016)\citenamefont
  {Maurand}, \citenamefont {Jehl}, \citenamefont {Kotekar-Patil}, \citenamefont
  {Corna}, \citenamefont {Bohuslavskyi}, \citenamefont {Lavi{\'e}ville},
  \citenamefont {Hutin}, \citenamefont {Barraud}, \citenamefont {Vinet},
  \citenamefont {Sanquer},\ and\ \citenamefont
  {De~Franceschi}}]{maurand2016cmos}%
  \BibitemOpen
  \bibfield  {author} {\bibinfo {author} {\bibfnamefont {R.}~\bibnamefont
  {Maurand}}, \bibinfo {author} {\bibfnamefont {X.}~\bibnamefont {Jehl}},
  \bibinfo {author} {\bibfnamefont {D.}~\bibnamefont {Kotekar-Patil}}, \bibinfo
  {author} {\bibfnamefont {A.}~\bibnamefont {Corna}}, \bibinfo {author}
  {\bibfnamefont {H.}~\bibnamefont {Bohuslavskyi}}, \bibinfo {author}
  {\bibfnamefont {R.}~\bibnamefont {Lavi{\'e}ville}}, \bibinfo {author}
  {\bibfnamefont {L.}~\bibnamefont {Hutin}}, \bibinfo {author} {\bibfnamefont
  {S.}~\bibnamefont {Barraud}}, \bibinfo {author} {\bibfnamefont
  {M.}~\bibnamefont {Vinet}}, \bibinfo {author} {\bibfnamefont
  {M.}~\bibnamefont {Sanquer}}, \ and\ \bibinfo {author} {\bibfnamefont
  {S.}~\bibnamefont {De~Franceschi}},\ }\href {\doibase 10.1038/ncomms13575}
  {\bibfield  {journal} {\bibinfo  {journal} {Nature communications}\ }\textbf
  {\bibinfo {volume} {7}},\ \bibinfo {pages} {1} (\bibinfo {year}
  {2016})}\BibitemShut {NoStop}%
\bibitem [{\citenamefont {Li}\ \emph {et~al.}(2015)\citenamefont {Li},
  \citenamefont {Hudson}, \citenamefont {Dzurak},\ and\ \citenamefont
  {Hamilton}}]{LiPauliSpinBlockade2015}%
  \BibitemOpen
  \bibfield  {author} {\bibinfo {author} {\bibfnamefont {R.}~\bibnamefont
  {Li}}, \bibinfo {author} {\bibfnamefont {F.~E.}\ \bibnamefont {Hudson}},
  \bibinfo {author} {\bibfnamefont {A.~S.}\ \bibnamefont {Dzurak}}, \ and\
  \bibinfo {author} {\bibfnamefont {A.~R.}\ \bibnamefont {Hamilton}},\ }\href
  {\doibase 10.1021/acs.nanolett.5b02561} {\bibfield  {journal} {\bibinfo
  {journal} {Nano Lett.}\ }\textbf {\bibinfo {volume} {15}},\ \bibinfo {pages}
  {7314} (\bibinfo {year} {2015})}\BibitemShut {NoStop}%
\bibitem [{\citenamefont {Hendrickx}\ \emph
  {et~al.}(2020{\natexlab{b}})\citenamefont {Hendrickx}, \citenamefont
  {Lawrie}, \citenamefont {Petit}, \citenamefont {Sammak}, \citenamefont
  {Scappucci},\ and\ \citenamefont
  {Veldhorst}}]{Hendrickxsingleholespinqubit2019}%
  \BibitemOpen
  \bibfield  {author} {\bibinfo {author} {\bibfnamefont {N.}~\bibnamefont
  {Hendrickx}}, \bibinfo {author} {\bibfnamefont {W.}~\bibnamefont {Lawrie}},
  \bibinfo {author} {\bibfnamefont {L.}~\bibnamefont {Petit}}, \bibinfo
  {author} {\bibfnamefont {A.}~\bibnamefont {Sammak}}, \bibinfo {author}
  {\bibfnamefont {G.}~\bibnamefont {Scappucci}}, \ and\ \bibinfo {author}
  {\bibfnamefont {M.}~\bibnamefont {Veldhorst}},\ }\href {\doibase
  10.1038/s41467-020-17211-7} {\bibfield  {journal} {\bibinfo  {journal}
  {Nature communications}\ }\textbf {\bibinfo {volume} {11}},\ \bibinfo {pages}
  {1} (\bibinfo {year} {2020}{\natexlab{b}})}\BibitemShut {NoStop}%
\bibitem [{\citenamefont {Kloeffel}\ \emph {et~al.}(2011)\citenamefont
  {Kloeffel}, \citenamefont {Trif},\ and\ \citenamefont {Loss}}]{DRkloeffel1}%
  \BibitemOpen
  \bibfield  {author} {\bibinfo {author} {\bibfnamefont {C.}~\bibnamefont
  {Kloeffel}}, \bibinfo {author} {\bibfnamefont {M.}~\bibnamefont {Trif}}, \
  and\ \bibinfo {author} {\bibfnamefont {D.}~\bibnamefont {Loss}},\ }\href
  {\doibase 10.1103/PhysRevB.84.195314} {\bibfield  {journal} {\bibinfo
  {journal} {Phys. Rev. B}\ }\textbf {\bibinfo {volume} {84}},\ \bibinfo
  {pages} {195314} (\bibinfo {year} {2011})}\BibitemShut {NoStop}%
\bibitem [{\citenamefont {Kloeffel}\ \emph {et~al.}(2018)\citenamefont
  {Kloeffel}, \citenamefont {Ran\ifmmode \check{c}\else
  \v{c}\fi{}i\ifmmode~\acute{c}\else \'{c}\fi{}},\ and\ \citenamefont
  {Loss}}]{DRkloeffel3}%
  \BibitemOpen
  \bibfield  {author} {\bibinfo {author} {\bibfnamefont {C.}~\bibnamefont
  {Kloeffel}}, \bibinfo {author} {\bibfnamefont {M.~J.}\ \bibnamefont
  {Ran\ifmmode \check{c}\else \v{c}\fi{}i\ifmmode~\acute{c}\else \'{c}\fi{}}},
  \ and\ \bibinfo {author} {\bibfnamefont {D.}~\bibnamefont {Loss}},\ }\href
  {\doibase 10.1103/PhysRevB.97.235422} {\bibfield  {journal} {\bibinfo
  {journal} {Phys. Rev. B}\ }\textbf {\bibinfo {volume} {97}},\ \bibinfo
  {pages} {235422} (\bibinfo {year} {2018})}\BibitemShut {NoStop}%
\bibitem [{\citenamefont {Froning}\ \emph
  {et~al.}(2021{\natexlab{a}})\citenamefont {Froning}, \citenamefont
  {Camenzind}, \citenamefont {van~der Molen}, \citenamefont {Li}, \citenamefont
  {Bakkers}, \citenamefont {Zumb{\"u}hl},\ and\ \citenamefont
  {Braakman}}]{froning2020ultrafast}%
  \BibitemOpen
  \bibfield  {author} {\bibinfo {author} {\bibfnamefont {F.~N.~M.}\
  \bibnamefont {Froning}}, \bibinfo {author} {\bibfnamefont {L.~C.}\
  \bibnamefont {Camenzind}}, \bibinfo {author} {\bibfnamefont {O.~A.~H.}\
  \bibnamefont {van~der Molen}}, \bibinfo {author} {\bibfnamefont
  {A.}~\bibnamefont {Li}}, \bibinfo {author} {\bibfnamefont {E.~P. A.~M.}\
  \bibnamefont {Bakkers}}, \bibinfo {author} {\bibfnamefont {D.~M.}\
  \bibnamefont {Zumb{\"u}hl}}, \ and\ \bibinfo {author} {\bibfnamefont {F.~R.}\
  \bibnamefont {Braakman}},\ }\href {\doibase 10.1038/s41565-020-00828-6}
  {\bibfield  {journal} {\bibinfo  {journal} {Nature Nanotechnology}\ }\textbf
  {\bibinfo {volume} {16}},\ \bibinfo {pages} {1} (\bibinfo {year}
  {2021}{\natexlab{a}})}\BibitemShut {NoStop}%
\bibitem [{\citenamefont {Wang}\ \emph {et~al.}(2020)\citenamefont {Wang},
  \citenamefont {Xu}, \citenamefont {Gao}, \citenamefont {Liu}, \citenamefont
  {Ma}, \citenamefont {Zhang}, \citenamefont {Zhang}, \citenamefont {Cao},
  \citenamefont {Wang}, \citenamefont {Zhang}, \citenamefont {Hu},
  \citenamefont {Jiang}, \citenamefont {Li}, \citenamefont {Guo},\ and\
  \citenamefont {Guo}}]{wang2020ultrafast}%
  \BibitemOpen
  \bibfield  {author} {\bibinfo {author} {\bibfnamefont {K.}~\bibnamefont
  {Wang}}, \bibinfo {author} {\bibfnamefont {G.}~\bibnamefont {Xu}}, \bibinfo
  {author} {\bibfnamefont {F.}~\bibnamefont {Gao}}, \bibinfo {author}
  {\bibfnamefont {H.}~\bibnamefont {Liu}}, \bibinfo {author} {\bibfnamefont
  {R.-L.}\ \bibnamefont {Ma}}, \bibinfo {author} {\bibfnamefont
  {X.}~\bibnamefont {Zhang}}, \bibinfo {author} {\bibfnamefont
  {T.}~\bibnamefont {Zhang}}, \bibinfo {author} {\bibfnamefont
  {G.}~\bibnamefont {Cao}}, \bibinfo {author} {\bibfnamefont {T.}~\bibnamefont
  {Wang}}, \bibinfo {author} {\bibfnamefont {J.-J.}\ \bibnamefont {Zhang}},
  \bibinfo {author} {\bibfnamefont {X.}~\bibnamefont {Hu}}, \bibinfo {author}
  {\bibfnamefont {H.-W.}\ \bibnamefont {Jiang}}, \bibinfo {author}
  {\bibfnamefont {H.-O.}\ \bibnamefont {Li}}, \bibinfo {author} {\bibfnamefont
  {G.-C.}\ \bibnamefont {Guo}}, \ and\ \bibinfo {author} {\bibfnamefont
  {G.-P.}\ \bibnamefont {Guo}},\ }\href@noop {} {\bibfield  {journal} {\bibinfo
   {journal} {arXiv preprint arXiv:2006.12340}\ } (\bibinfo {year}
  {2020})}\BibitemShut {NoStop}%
\bibitem [{\citenamefont {de~Vries}\ \emph {et~al.}(2018)\citenamefont
  {de~Vries}, \citenamefont {Shen}, \citenamefont {Skolasinski}, \citenamefont
  {Nowak}, \citenamefont {Varjas}, \citenamefont {Wang}, \citenamefont
  {Wimmer}, \citenamefont {Ridderbos}, \citenamefont {Zwanenburg},
  \citenamefont {Li}, \citenamefont {Koelling}, \citenamefont {Verheijen},
  \citenamefont {Bakkers},\ and\ \citenamefont {Kouwenhoven}}]{de2018spin}%
  \BibitemOpen
  \bibfield  {author} {\bibinfo {author} {\bibfnamefont {F.~K.}\ \bibnamefont
  {de~Vries}}, \bibinfo {author} {\bibfnamefont {J.}~\bibnamefont {Shen}},
  \bibinfo {author} {\bibfnamefont {R.~J.}\ \bibnamefont {Skolasinski}},
  \bibinfo {author} {\bibfnamefont {M.~P.}\ \bibnamefont {Nowak}}, \bibinfo
  {author} {\bibfnamefont {D.}~\bibnamefont {Varjas}}, \bibinfo {author}
  {\bibfnamefont {L.}~\bibnamefont {Wang}}, \bibinfo {author} {\bibfnamefont
  {M.}~\bibnamefont {Wimmer}}, \bibinfo {author} {\bibfnamefont
  {J.}~\bibnamefont {Ridderbos}}, \bibinfo {author} {\bibfnamefont {F.~A.}\
  \bibnamefont {Zwanenburg}}, \bibinfo {author} {\bibfnamefont
  {A.}~\bibnamefont {Li}}, \bibinfo {author} {\bibfnamefont {S.}~\bibnamefont
  {Koelling}}, \bibinfo {author} {\bibfnamefont {M.~A.}\ \bibnamefont
  {Verheijen}}, \bibinfo {author} {\bibfnamefont {E.~P. A.~M.}\ \bibnamefont
  {Bakkers}}, \ and\ \bibinfo {author} {\bibfnamefont {L.~P.}\ \bibnamefont
  {Kouwenhoven}},\ }\href {\doibase 10.1021/acs.nanolett.8b02981} {\bibfield
  {journal} {\bibinfo  {journal} {Nano letters}\ }\textbf {\bibinfo {volume}
  {18}},\ \bibinfo {pages} {6483} (\bibinfo {year} {2018})}\BibitemShut
  {NoStop}%
\bibitem [{\citenamefont {Higginbotham}\ \emph {et~al.}(2014)\citenamefont
  {Higginbotham}, \citenamefont {Kuemmeth}, \citenamefont {Larsen},
  \citenamefont {Fitzpatrick}, \citenamefont {Yao}, \citenamefont {Yan},
  \citenamefont {Lieber},\ and\ \citenamefont
  {Marcus}}]{PhysRevLett.112.216806}%
  \BibitemOpen
  \bibfield  {author} {\bibinfo {author} {\bibfnamefont {A.~P.}\ \bibnamefont
  {Higginbotham}}, \bibinfo {author} {\bibfnamefont {F.}~\bibnamefont
  {Kuemmeth}}, \bibinfo {author} {\bibfnamefont {T.~W.}\ \bibnamefont
  {Larsen}}, \bibinfo {author} {\bibfnamefont {M.}~\bibnamefont {Fitzpatrick}},
  \bibinfo {author} {\bibfnamefont {J.}~\bibnamefont {Yao}}, \bibinfo {author}
  {\bibfnamefont {H.}~\bibnamefont {Yan}}, \bibinfo {author} {\bibfnamefont
  {C.~M.}\ \bibnamefont {Lieber}}, \ and\ \bibinfo {author} {\bibfnamefont
  {C.~M.}\ \bibnamefont {Marcus}},\ }\href {\doibase
  10.1103/PhysRevLett.112.216806} {\bibfield  {journal} {\bibinfo  {journal}
  {Phys. Rev. Lett.}\ }\textbf {\bibinfo {volume} {112}},\ \bibinfo {pages}
  {216806} (\bibinfo {year} {2014})}\BibitemShut {NoStop}%
\bibitem [{\citenamefont {Brauns}\ \emph {et~al.}(2016)\citenamefont {Brauns},
  \citenamefont {Ridderbos}, \citenamefont {Li}, \citenamefont {Bakkers},\ and\
  \citenamefont {Zwanenburg}}]{PhysRevB.93.121408}%
  \BibitemOpen
  \bibfield  {author} {\bibinfo {author} {\bibfnamefont {M.}~\bibnamefont
  {Brauns}}, \bibinfo {author} {\bibfnamefont {J.}~\bibnamefont {Ridderbos}},
  \bibinfo {author} {\bibfnamefont {A.}~\bibnamefont {Li}}, \bibinfo {author}
  {\bibfnamefont {E.~P. A.~M.}\ \bibnamefont {Bakkers}}, \ and\ \bibinfo
  {author} {\bibfnamefont {F.~A.}\ \bibnamefont {Zwanenburg}},\ }\href
  {\doibase 10.1103/PhysRevB.93.121408} {\bibfield  {journal} {\bibinfo
  {journal} {Phys. Rev. B}\ }\textbf {\bibinfo {volume} {93}},\ \bibinfo
  {pages} {121408} (\bibinfo {year} {2016})}\BibitemShut {NoStop}%
\bibitem [{\citenamefont {Gao}\ \emph {et~al.}(2020)\citenamefont {Gao},
  \citenamefont {Wang}, \citenamefont {Watzinger}, \citenamefont {Hu},
  \citenamefont {Rančić}, \citenamefont {Zhang}, \citenamefont {Wang},
  \citenamefont {Yao}, \citenamefont {Wang}, \citenamefont {Kukučka},
  \citenamefont {Vukušić}, \citenamefont {Kloeffel}, \citenamefont {Loss},
  \citenamefont {Liu}, \citenamefont {Katsaros},\ and\ \citenamefont
  {Zhang}}]{doi:10.1002/adma.201906523}%
  \BibitemOpen
  \bibfield  {author} {\bibinfo {author} {\bibfnamefont {F.}~\bibnamefont
  {Gao}}, \bibinfo {author} {\bibfnamefont {J.-H.}\ \bibnamefont {Wang}},
  \bibinfo {author} {\bibfnamefont {H.}~\bibnamefont {Watzinger}}, \bibinfo
  {author} {\bibfnamefont {H.}~\bibnamefont {Hu}}, \bibinfo {author}
  {\bibfnamefont {M.~J.}\ \bibnamefont {Rančić}}, \bibinfo {author}
  {\bibfnamefont {J.-Y.}\ \bibnamefont {Zhang}}, \bibinfo {author}
  {\bibfnamefont {T.}~\bibnamefont {Wang}}, \bibinfo {author} {\bibfnamefont
  {Y.}~\bibnamefont {Yao}}, \bibinfo {author} {\bibfnamefont {G.-L.}\
  \bibnamefont {Wang}}, \bibinfo {author} {\bibfnamefont {J.}~\bibnamefont
  {Kukučka}}, \bibinfo {author} {\bibfnamefont {L.}~\bibnamefont {Vukušić}},
  \bibinfo {author} {\bibfnamefont {C.}~\bibnamefont {Kloeffel}}, \bibinfo
  {author} {\bibfnamefont {D.}~\bibnamefont {Loss}}, \bibinfo {author}
  {\bibfnamefont {F.}~\bibnamefont {Liu}}, \bibinfo {author} {\bibfnamefont
  {G.}~\bibnamefont {Katsaros}}, \ and\ \bibinfo {author} {\bibfnamefont
  {J.-J.}\ \bibnamefont {Zhang}},\ }\href {\doibase 10.1002/adma.201906523}
  {\bibfield  {journal} {\bibinfo  {journal} {Advanced Materials}\ }\textbf
  {\bibinfo {volume} {32}},\ \bibinfo {pages} {1906523} (\bibinfo {year}
  {2020})}\BibitemShut {NoStop}%
\bibitem [{\citenamefont {Wang}\ \emph {et~al.}(2017)\citenamefont {Wang},
  \citenamefont {Deacon}, \citenamefont {Yao}, \citenamefont {Lieber},\ and\
  \citenamefont {Ishibashi}}]{wang2017electrical}%
  \BibitemOpen
  \bibfield  {author} {\bibinfo {author} {\bibfnamefont {R.}~\bibnamefont
  {Wang}}, \bibinfo {author} {\bibfnamefont {R.}~\bibnamefont {Deacon}},
  \bibinfo {author} {\bibfnamefont {J.}~\bibnamefont {Yao}}, \bibinfo {author}
  {\bibfnamefont {C.}~\bibnamefont {Lieber}}, \ and\ \bibinfo {author}
  {\bibfnamefont {K.}~\bibnamefont {Ishibashi}},\ }\href {\doibase
  10.1088/1361-6641/aa7ce6} {\bibfield  {journal} {\bibinfo  {journal}
  {Semiconductor Science and Technology}\ }\textbf {\bibinfo {volume} {32}},\
  \bibinfo {pages} {094002} (\bibinfo {year} {2017})}\BibitemShut {NoStop}%
\bibitem [{\citenamefont {Froning}\ \emph
  {et~al.}(2021{\natexlab{b}})\citenamefont {Froning}, \citenamefont
  {Ran\ifmmode \check{c}\else \v{c}\fi{}i\ifmmode~\acute{c}\else \'{c}\fi{}},
  \citenamefont {Het\'enyi}, \citenamefont {Bosco}, \citenamefont {Rehmann},
  \citenamefont {Li}, \citenamefont {Bakkers}, \citenamefont {Zwanenburg},
  \citenamefont {Loss}, \citenamefont {Zumb\"uhl},\ and\ \citenamefont
  {Braakman}}]{froning2020strong}%
  \BibitemOpen
  \bibfield  {author} {\bibinfo {author} {\bibfnamefont {F.~N.~M.}\
  \bibnamefont {Froning}}, \bibinfo {author} {\bibfnamefont {M.~J.}\
  \bibnamefont {Ran\ifmmode \check{c}\else \v{c}\fi{}i\ifmmode~\acute{c}\else
  \'{c}\fi{}}}, \bibinfo {author} {\bibfnamefont {B.}~\bibnamefont
  {Het\'enyi}}, \bibinfo {author} {\bibfnamefont {S.}~\bibnamefont {Bosco}},
  \bibinfo {author} {\bibfnamefont {M.~K.}\ \bibnamefont {Rehmann}}, \bibinfo
  {author} {\bibfnamefont {A.}~\bibnamefont {Li}}, \bibinfo {author}
  {\bibfnamefont {E.~P. A.~M.}\ \bibnamefont {Bakkers}}, \bibinfo {author}
  {\bibfnamefont {F.~A.}\ \bibnamefont {Zwanenburg}}, \bibinfo {author}
  {\bibfnamefont {D.}~\bibnamefont {Loss}}, \bibinfo {author} {\bibfnamefont
  {D.~M.}\ \bibnamefont {Zumb\"uhl}}, \ and\ \bibinfo {author} {\bibfnamefont
  {F.~R.}\ \bibnamefont {Braakman}},\ }\href {\doibase
  10.1103/PhysRevResearch.3.013081} {\bibfield  {journal} {\bibinfo  {journal}
  {Phys. Rev. Research}\ }\textbf {\bibinfo {volume} {3}},\ \bibinfo {pages}
  {013081} (\bibinfo {year} {2021}{\natexlab{b}})}\BibitemShut {NoStop}%
\bibitem [{\citenamefont {Maier}\ \emph {et~al.}(2013)\citenamefont {Maier},
  \citenamefont {Kloeffel},\ and\ \citenamefont {Loss}}]{PhysRevB.87.161305}%
  \BibitemOpen
  \bibfield  {author} {\bibinfo {author} {\bibfnamefont {F.}~\bibnamefont
  {Maier}}, \bibinfo {author} {\bibfnamefont {C.}~\bibnamefont {Kloeffel}}, \
  and\ \bibinfo {author} {\bibfnamefont {D.}~\bibnamefont {Loss}},\ }\href
  {\doibase 10.1103/PhysRevB.87.161305} {\bibfield  {journal} {\bibinfo
  {journal} {Phys. Rev. B}\ }\textbf {\bibinfo {volume} {87}},\ \bibinfo
  {pages} {161305} (\bibinfo {year} {2013})}\BibitemShut {NoStop}%
\bibitem [{\citenamefont {Landig}\ \emph {et~al.}(2018)\citenamefont {Landig},
  \citenamefont {Koski}, \citenamefont {Scarlino}, \citenamefont {Mendes},
  \citenamefont {Blais}, \citenamefont {Reichl}, \citenamefont {Wegscheider},
  \citenamefont {Wallraff}, \citenamefont {Ensslin},\ and\ \citenamefont
  {Ihn}}]{landig2018coherent}%
  \BibitemOpen
  \bibfield  {author} {\bibinfo {author} {\bibfnamefont {A.~J.}\ \bibnamefont
  {Landig}}, \bibinfo {author} {\bibfnamefont {J.~V.}\ \bibnamefont {Koski}},
  \bibinfo {author} {\bibfnamefont {P.}~\bibnamefont {Scarlino}}, \bibinfo
  {author} {\bibfnamefont {U.}~\bibnamefont {Mendes}}, \bibinfo {author}
  {\bibfnamefont {A.}~\bibnamefont {Blais}}, \bibinfo {author} {\bibfnamefont
  {C.}~\bibnamefont {Reichl}}, \bibinfo {author} {\bibfnamefont
  {W.}~\bibnamefont {Wegscheider}}, \bibinfo {author} {\bibfnamefont
  {A.}~\bibnamefont {Wallraff}}, \bibinfo {author} {\bibfnamefont
  {K.}~\bibnamefont {Ensslin}}, \ and\ \bibinfo {author} {\bibfnamefont
  {T.}~\bibnamefont {Ihn}},\ }\href {\doibase 10.1038/s41586-018-0365-y}
  {\bibfield  {journal} {\bibinfo  {journal} {Nature}\ }\textbf {\bibinfo
  {volume} {560}},\ \bibinfo {pages} {179} (\bibinfo {year}
  {2018})}\BibitemShut {NoStop}%
\bibitem [{\citenamefont {Mi}\ \emph {et~al.}(2018)\citenamefont {Mi},
  \citenamefont {Benito}, \citenamefont {Putz}, \citenamefont {Zajac},
  \citenamefont {Taylor}, \citenamefont {Burkard},\ and\ \citenamefont
  {Petta}}]{mi2018coherent}%
  \BibitemOpen
  \bibfield  {author} {\bibinfo {author} {\bibfnamefont {X.}~\bibnamefont
  {Mi}}, \bibinfo {author} {\bibfnamefont {M.}~\bibnamefont {Benito}}, \bibinfo
  {author} {\bibfnamefont {S.}~\bibnamefont {Putz}}, \bibinfo {author}
  {\bibfnamefont {D.~M.}\ \bibnamefont {Zajac}}, \bibinfo {author}
  {\bibfnamefont {J.~M.}\ \bibnamefont {Taylor}}, \bibinfo {author}
  {\bibfnamefont {G.}~\bibnamefont {Burkard}}, \ and\ \bibinfo {author}
  {\bibfnamefont {J.~R.}\ \bibnamefont {Petta}},\ }\href {\doibase
  10.1038/nature25769} {\bibfield  {journal} {\bibinfo  {journal} {Nature}\
  }\textbf {\bibinfo {volume} {555}},\ \bibinfo {pages} {599} (\bibinfo {year}
  {2018})}\BibitemShut {NoStop}%
\bibitem [{\citenamefont {Kloeffel}\ \emph {et~al.}(2013)\citenamefont
  {Kloeffel}, \citenamefont {Trif}, \citenamefont {Stano},\ and\ \citenamefont
  {Loss}}]{DRkloeffel2}%
  \BibitemOpen
  \bibfield  {author} {\bibinfo {author} {\bibfnamefont {C.}~\bibnamefont
  {Kloeffel}}, \bibinfo {author} {\bibfnamefont {M.}~\bibnamefont {Trif}},
  \bibinfo {author} {\bibfnamefont {P.}~\bibnamefont {Stano}}, \ and\ \bibinfo
  {author} {\bibfnamefont {D.}~\bibnamefont {Loss}},\ }\href {\doibase
  10.1103/PhysRevB.88.241405} {\bibfield  {journal} {\bibinfo  {journal} {Phys.
  Rev. B}\ }\textbf {\bibinfo {volume} {88}},\ \bibinfo {pages} {241405}
  (\bibinfo {year} {2013})}\BibitemShut {NoStop}%
\bibitem [{\citenamefont {Sun}\ \emph {et~al.}(2018)\citenamefont {Sun},
  \citenamefont {Deacon}, \citenamefont {Wang}, \citenamefont {Yao},
  \citenamefont {Lieber},\ and\ \citenamefont {Ishibashi}}]{sun2018helical}%
  \BibitemOpen
  \bibfield  {author} {\bibinfo {author} {\bibfnamefont {J.}~\bibnamefont
  {Sun}}, \bibinfo {author} {\bibfnamefont {R.~S.}\ \bibnamefont {Deacon}},
  \bibinfo {author} {\bibfnamefont {R.}~\bibnamefont {Wang}}, \bibinfo {author}
  {\bibfnamefont {J.}~\bibnamefont {Yao}}, \bibinfo {author} {\bibfnamefont
  {C.~M.}\ \bibnamefont {Lieber}}, \ and\ \bibinfo {author} {\bibfnamefont
  {K.}~\bibnamefont {Ishibashi}},\ }\href {\doibase
  10.1021/acs.nanolett.8b01799} {\bibfield  {journal} {\bibinfo  {journal}
  {Nano letters}\ }\textbf {\bibinfo {volume} {18}},\ \bibinfo {pages} {6144}
  (\bibinfo {year} {2018})}\BibitemShut {NoStop}%
\bibitem [{\citenamefont {Oreg}\ \emph {et~al.}(2010)\citenamefont {Oreg},
  \citenamefont {Refael},\ and\ \citenamefont {von
  Oppen}}]{PhysRevLett.105.177002}%
  \BibitemOpen
  \bibfield  {author} {\bibinfo {author} {\bibfnamefont {Y.}~\bibnamefont
  {Oreg}}, \bibinfo {author} {\bibfnamefont {G.}~\bibnamefont {Refael}}, \ and\
  \bibinfo {author} {\bibfnamefont {F.}~\bibnamefont {von Oppen}},\ }\href
  {\doibase 10.1103/PhysRevLett.105.177002} {\bibfield  {journal} {\bibinfo
  {journal} {Phys. Rev. Lett.}\ }\textbf {\bibinfo {volume} {105}},\ \bibinfo
  {pages} {177002} (\bibinfo {year} {2010})}\BibitemShut {NoStop}%
\bibitem [{\citenamefont {Maier}\ \emph {et~al.}(2014)\citenamefont {Maier},
  \citenamefont {Klinovaja},\ and\ \citenamefont {Loss}}]{PhysRevB.90.195421}%
  \BibitemOpen
  \bibfield  {author} {\bibinfo {author} {\bibfnamefont {F.}~\bibnamefont
  {Maier}}, \bibinfo {author} {\bibfnamefont {J.}~\bibnamefont {Klinovaja}}, \
  and\ \bibinfo {author} {\bibfnamefont {D.}~\bibnamefont {Loss}},\ }\href
  {\doibase 10.1103/PhysRevB.90.195421} {\bibfield  {journal} {\bibinfo
  {journal} {Phys. Rev. B}\ }\textbf {\bibinfo {volume} {90}},\ \bibinfo
  {pages} {195421} (\bibinfo {year} {2014})}\BibitemShut {NoStop}%
\bibitem [{\citenamefont {Yoneda}\ \emph {et~al.}(2018)\citenamefont {Yoneda},
  \citenamefont {Takeda}, \citenamefont {Otsuka}, \citenamefont {Nakajima},
  \citenamefont {Delbecq}, \citenamefont {Allison}, \citenamefont {Honda},
  \citenamefont {Kodera}, \citenamefont {Oda}, \citenamefont {Hoshi},
  \citenamefont {Usami}, \citenamefont {Itoh},\ and\ \citenamefont
  {Tarucha}}]{Yonedaquantumdotspinqubit2018}%
  \BibitemOpen
  \bibfield  {author} {\bibinfo {author} {\bibfnamefont {J.}~\bibnamefont
  {Yoneda}}, \bibinfo {author} {\bibfnamefont {K.}~\bibnamefont {Takeda}},
  \bibinfo {author} {\bibfnamefont {T.}~\bibnamefont {Otsuka}}, \bibinfo
  {author} {\bibfnamefont {T.}~\bibnamefont {Nakajima}}, \bibinfo {author}
  {\bibfnamefont {M.~R.}\ \bibnamefont {Delbecq}}, \bibinfo {author}
  {\bibfnamefont {G.}~\bibnamefont {Allison}}, \bibinfo {author} {\bibfnamefont
  {T.}~\bibnamefont {Honda}}, \bibinfo {author} {\bibfnamefont
  {T.}~\bibnamefont {Kodera}}, \bibinfo {author} {\bibfnamefont
  {S.}~\bibnamefont {Oda}}, \bibinfo {author} {\bibfnamefont {Y.}~\bibnamefont
  {Hoshi}}, \bibinfo {author} {\bibfnamefont {N.}~\bibnamefont {Usami}},
  \bibinfo {author} {\bibfnamefont {K.~M.}\ \bibnamefont {Itoh}}, \ and\
  \bibinfo {author} {\bibfnamefont {S.}~\bibnamefont {Tarucha}},\ }\href
  {\doibase 10.1038/s41565-017-0014-x} {\bibfield  {journal} {\bibinfo
  {journal} {Nat. Nanotechnol.}\ }\textbf {\bibinfo {volume} {13}},\ \bibinfo
  {pages} {102} (\bibinfo {year} {2018})}\BibitemShut {NoStop}%
\bibitem [{\citenamefont {Culcer}\ \emph {et~al.}(2009)\citenamefont {Culcer},
  \citenamefont {Hu},\ and\ \citenamefont {Das~Sarma}}]{culcer2009dephasing}%
  \BibitemOpen
  \bibfield  {author} {\bibinfo {author} {\bibfnamefont {D.}~\bibnamefont
  {Culcer}}, \bibinfo {author} {\bibfnamefont {X.}~\bibnamefont {Hu}}, \ and\
  \bibinfo {author} {\bibfnamefont {S.}~\bibnamefont {Das~Sarma}},\ }\href
  {\doibase 10.1063/1.3194778} {\bibfield  {journal} {\bibinfo  {journal}
  {Applied Physics Letters}\ }\textbf {\bibinfo {volume} {95}},\ \bibinfo
  {pages} {073102} (\bibinfo {year} {2009})}\BibitemShut {NoStop}%
\bibitem [{\citenamefont {Bermeister}\ \emph {et~al.}(2014)\citenamefont
  {Bermeister}, \citenamefont {Keith},\ and\ \citenamefont
  {Culcer}}]{bermeister2014charge}%
  \BibitemOpen
  \bibfield  {author} {\bibinfo {author} {\bibfnamefont {A.}~\bibnamefont
  {Bermeister}}, \bibinfo {author} {\bibfnamefont {D.}~\bibnamefont {Keith}}, \
  and\ \bibinfo {author} {\bibfnamefont {D.}~\bibnamefont {Culcer}},\ }\href
  {\doibase 10.1063/1.4901162} {\bibfield  {journal} {\bibinfo  {journal}
  {Applied Physics Letters}\ }\textbf {\bibinfo {volume} {105}},\ \bibinfo
  {pages} {192102} (\bibinfo {year} {2014})}\BibitemShut {NoStop}%
\bibitem [{\citenamefont {Wang}\ \emph {et~al.}(2019)\citenamefont {Wang},
  \citenamefont {Marcellina}, \citenamefont {Hamilton}, \citenamefont {Rogge},
  \citenamefont {Salfi},\ and\ \citenamefont {Culcer}}]{wang2019suppressing}%
  \BibitemOpen
  \bibfield  {author} {\bibinfo {author} {\bibfnamefont {Z.}~\bibnamefont
  {Wang}}, \bibinfo {author} {\bibfnamefont {E.}~\bibnamefont {Marcellina}},
  \bibinfo {author} {\bibfnamefont {A.}~\bibnamefont {Hamilton}}, \bibinfo
  {author} {\bibfnamefont {S.}~\bibnamefont {Rogge}}, \bibinfo {author}
  {\bibfnamefont {J.}~\bibnamefont {Salfi}}, \ and\ \bibinfo {author}
  {\bibfnamefont {D.}~\bibnamefont {Culcer}},\ }\href@noop {} {\bibfield
  {journal} {\bibinfo  {journal} {arXiv preprint arXiv:1911.11143}\ } (\bibinfo
  {year} {2019})}\BibitemShut {NoStop}%
\bibitem [{\citenamefont {Salfi}\ \emph {et~al.}(2016)\citenamefont {Salfi},
  \citenamefont {Mol}, \citenamefont {Culcer},\ and\ \citenamefont
  {Rogge}}]{PhysRevLett.116.246801}%
  \BibitemOpen
  \bibfield  {author} {\bibinfo {author} {\bibfnamefont {J.}~\bibnamefont
  {Salfi}}, \bibinfo {author} {\bibfnamefont {J.~A.}\ \bibnamefont {Mol}},
  \bibinfo {author} {\bibfnamefont {D.}~\bibnamefont {Culcer}}, \ and\ \bibinfo
  {author} {\bibfnamefont {S.}~\bibnamefont {Rogge}},\ }\href {\doibase
  10.1103/PhysRevLett.116.246801} {\bibfield  {journal} {\bibinfo  {journal}
  {Phys. Rev. Lett.}\ }\textbf {\bibinfo {volume} {116}},\ \bibinfo {pages}
  {246801} (\bibinfo {year} {2016})}\BibitemShut {NoStop}%
\bibitem [{\citenamefont {Benito}\ \emph {et~al.}(2019)\citenamefont {Benito},
  \citenamefont {Croot}, \citenamefont {Adelsberger}, \citenamefont {Putz},
  \citenamefont {Mi}, \citenamefont {Petta},\ and\ \citenamefont
  {Burkard}}]{PhysRevB.100.125430}%
  \BibitemOpen
  \bibfield  {author} {\bibinfo {author} {\bibfnamefont {M.}~\bibnamefont
  {Benito}}, \bibinfo {author} {\bibfnamefont {X.}~\bibnamefont {Croot}},
  \bibinfo {author} {\bibfnamefont {C.}~\bibnamefont {Adelsberger}}, \bibinfo
  {author} {\bibfnamefont {S.}~\bibnamefont {Putz}}, \bibinfo {author}
  {\bibfnamefont {X.}~\bibnamefont {Mi}}, \bibinfo {author} {\bibfnamefont
  {J.~R.}\ \bibnamefont {Petta}}, \ and\ \bibinfo {author} {\bibfnamefont
  {G.}~\bibnamefont {Burkard}},\ }\href {\doibase 10.1103/PhysRevB.100.125430}
  {\bibfield  {journal} {\bibinfo  {journal} {Phys. Rev. B}\ }\textbf {\bibinfo
  {volume} {100}},\ \bibinfo {pages} {125430} (\bibinfo {year}
  {2019})}\BibitemShut {NoStop}%
\bibitem [{\citenamefont {Tanttu}\ \emph {et~al.}(2019)\citenamefont {Tanttu},
  \citenamefont {Hensen}, \citenamefont {Chan}, \citenamefont {Yang},
  \citenamefont {Huang}, \citenamefont {Fogarty}, \citenamefont {Hudson},
  \citenamefont {Itoh}, \citenamefont {Culcer}, \citenamefont {Laucht},
  \citenamefont {Morello},\ and\ \citenamefont
  {Dzurak}}]{TanttuControllingspinorbitinteractions2018}%
  \BibitemOpen
  \bibfield  {author} {\bibinfo {author} {\bibfnamefont {T.}~\bibnamefont
  {Tanttu}}, \bibinfo {author} {\bibfnamefont {B.}~\bibnamefont {Hensen}},
  \bibinfo {author} {\bibfnamefont {K.~W.}\ \bibnamefont {Chan}}, \bibinfo
  {author} {\bibfnamefont {C.~H.}\ \bibnamefont {Yang}}, \bibinfo {author}
  {\bibfnamefont {W.~W.}\ \bibnamefont {Huang}}, \bibinfo {author}
  {\bibfnamefont {M.}~\bibnamefont {Fogarty}}, \bibinfo {author} {\bibfnamefont
  {F.}~\bibnamefont {Hudson}}, \bibinfo {author} {\bibfnamefont
  {K.}~\bibnamefont {Itoh}}, \bibinfo {author} {\bibfnamefont {D.}~\bibnamefont
  {Culcer}}, \bibinfo {author} {\bibfnamefont {A.}~\bibnamefont {Laucht}},
  \bibinfo {author} {\bibfnamefont {A.}~\bibnamefont {Morello}}, \ and\
  \bibinfo {author} {\bibfnamefont {A.~S.}\ \bibnamefont {Dzurak}},\ }\href
  {\doibase 10.1103/PhysRevX.9.021028} {\bibfield  {journal} {\bibinfo
  {journal} {Phys. Rev. X}\ }\textbf {\bibinfo {volume} {9}},\ \bibinfo {pages}
  {021028} (\bibinfo {year} {2019})}\BibitemShut {NoStop}%
\bibitem [{\citenamefont {Venitucci}\ \emph {et~al.}(2018)\citenamefont
  {Venitucci}, \citenamefont {Bourdet}, \citenamefont {Pouzada},\ and\
  \citenamefont {Niquet}}]{VenitucciElectricalmanipulationsemiconductor2018}%
  \BibitemOpen
  \bibfield  {author} {\bibinfo {author} {\bibfnamefont {B.}~\bibnamefont
  {Venitucci}}, \bibinfo {author} {\bibfnamefont {L.}~\bibnamefont {Bourdet}},
  \bibinfo {author} {\bibfnamefont {D.}~\bibnamefont {Pouzada}}, \ and\
  \bibinfo {author} {\bibfnamefont {Y.-M.}\ \bibnamefont {Niquet}},\ }\href
  {\doibase 10.1103/PhysRevB.98.155319} {\bibfield  {journal} {\bibinfo
  {journal} {Phys. Rev. B}\ }\textbf {\bibinfo {volume} {98}},\ \bibinfo
  {pages} {155319} (\bibinfo {year} {2018})}\BibitemShut {NoStop}%
\bibitem [{\citenamefont {Hook}(2012)}]{6330653}%
  \BibitemOpen
  \bibfield  {author} {\bibinfo {author} {\bibfnamefont {T.~B.}\ \bibnamefont
  {Hook}},\ }in\ \href {\doibase 10.1109/CICC.2012.6330653} {\emph {\bibinfo
  {booktitle} {Proceedings of the IEEE 2012 Custom Integrated Circuits
  Conference}}}\ (\bibinfo {year} {2012})\ pp.\ \bibinfo {pages}
  {1--7}\BibitemShut {NoStop}%
\bibitem [{\citenamefont {Kuhlmann}\ \emph {et~al.}(2018)\citenamefont
  {Kuhlmann}, \citenamefont {Deshpande}, \citenamefont {Camenzind},
  \citenamefont {Zumb{\"u}hl},\ and\ \citenamefont
  {Fuhrer}}]{kuhlmann2018ambipolar}%
  \BibitemOpen
  \bibfield  {author} {\bibinfo {author} {\bibfnamefont {A.~V.}\ \bibnamefont
  {Kuhlmann}}, \bibinfo {author} {\bibfnamefont {V.}~\bibnamefont {Deshpande}},
  \bibinfo {author} {\bibfnamefont {L.~C.}\ \bibnamefont {Camenzind}}, \bibinfo
  {author} {\bibfnamefont {D.~M.}\ \bibnamefont {Zumb{\"u}hl}}, \ and\ \bibinfo
  {author} {\bibfnamefont {A.}~\bibnamefont {Fuhrer}},\ }\href {\doibase
  10.1063/1.5048097} {\bibfield  {journal} {\bibinfo  {journal} {Applied
  Physics Letters}\ }\textbf {\bibinfo {volume} {113}},\ \bibinfo {pages}
  {122107} (\bibinfo {year} {2018})}\BibitemShut {NoStop}%
\bibitem [{\citenamefont {Geyer}\ \emph {et~al.}(2021)\citenamefont {Geyer},
  \citenamefont {Camenzind}, \citenamefont {Czornomaz}, \citenamefont
  {Deshpande}, \citenamefont {Fuhrer}, \citenamefont {Warburton}, \citenamefont
  {Zumb{\"u}hl},\ and\ \citenamefont {Kuhlmann}}]{geyer2020silicon}%
  \BibitemOpen
  \bibfield  {author} {\bibinfo {author} {\bibfnamefont {S.}~\bibnamefont
  {Geyer}}, \bibinfo {author} {\bibfnamefont {L.~C.}\ \bibnamefont
  {Camenzind}}, \bibinfo {author} {\bibfnamefont {L.}~\bibnamefont
  {Czornomaz}}, \bibinfo {author} {\bibfnamefont {V.}~\bibnamefont
  {Deshpande}}, \bibinfo {author} {\bibfnamefont {A.}~\bibnamefont {Fuhrer}},
  \bibinfo {author} {\bibfnamefont {R.~J.}\ \bibnamefont {Warburton}}, \bibinfo
  {author} {\bibfnamefont {D.~M.}\ \bibnamefont {Zumb{\"u}hl}}, \ and\ \bibinfo
  {author} {\bibfnamefont {A.~V.}\ \bibnamefont {Kuhlmann}},\ }\href {\doibase
  10.1063/5.0036520} {\bibfield  {journal} {\bibinfo  {journal} {Appl. Phys.
  Lett.}\ }\textbf {\bibinfo {volume} {118}},\ \bibinfo {pages} {104004}
  (\bibinfo {year} {2021})}\BibitemShut {NoStop}%
\bibitem [{\citenamefont {Camenzind}\ \emph {et~al.}(2021)\citenamefont
  {Camenzind}, \citenamefont {Geyer}, \citenamefont {Fuhrer}, \citenamefont
  {Warburton}, \citenamefont {Zumb{\"u}hl},\ and\ \citenamefont
  {Kuhlmann}}]{siliconFin}%
  \BibitemOpen
  \bibfield  {author} {\bibinfo {author} {\bibfnamefont {L.~C.}\ \bibnamefont
  {Camenzind}}, \bibinfo {author} {\bibfnamefont {S.}~\bibnamefont {Geyer}},
  \bibinfo {author} {\bibfnamefont {A.}~\bibnamefont {Fuhrer}}, \bibinfo
  {author} {\bibfnamefont {R.~J.}\ \bibnamefont {Warburton}}, \bibinfo {author}
  {\bibfnamefont {D.~M.}\ \bibnamefont {Zumb{\"u}hl}}, \ and\ \bibinfo {author}
  {\bibfnamefont {A.~V.}\ \bibnamefont {Kuhlmann}},\ }\href@noop {} {\bibfield
  {journal} {\bibinfo  {journal} {arXiv preprint arXiv:2103.07369}\ } (\bibinfo
  {year} {2021})}\BibitemShut {NoStop}%
\bibitem [{\citenamefont {Yang}\ \emph {et~al.}(2020)\citenamefont {Yang},
  \citenamefont {Leon}, \citenamefont {Hwang}, \citenamefont {Saraiva},
  \citenamefont {Tanttu}, \citenamefont {Huang}, \citenamefont {Lemyre},
  \citenamefont {Chan}, \citenamefont {Tan}, \citenamefont {Hudson},
  \citenamefont {Itoh}, \citenamefont {Morello}, \citenamefont
  {Pioro-Ladrière}, \citenamefont {Laucht},\ and\ \citenamefont
  {Dzurak}}]{yang2020operation}%
  \BibitemOpen
  \bibfield  {author} {\bibinfo {author} {\bibfnamefont {C.~H.}\ \bibnamefont
  {Yang}}, \bibinfo {author} {\bibfnamefont {R.}~\bibnamefont {Leon}}, \bibinfo
  {author} {\bibfnamefont {J.}~\bibnamefont {Hwang}}, \bibinfo {author}
  {\bibfnamefont {A.}~\bibnamefont {Saraiva}}, \bibinfo {author} {\bibfnamefont
  {T.}~\bibnamefont {Tanttu}}, \bibinfo {author} {\bibfnamefont
  {W.}~\bibnamefont {Huang}}, \bibinfo {author} {\bibfnamefont {J.~C.}\
  \bibnamefont {Lemyre}}, \bibinfo {author} {\bibfnamefont {K.~W.}\
  \bibnamefont {Chan}}, \bibinfo {author} {\bibfnamefont {K.}~\bibnamefont
  {Tan}}, \bibinfo {author} {\bibfnamefont {F.~E.}\ \bibnamefont {Hudson}},
  \bibinfo {author} {\bibfnamefont {K.~M.}\ \bibnamefont {Itoh}}, \bibinfo
  {author} {\bibfnamefont {A.}~\bibnamefont {Morello}}, \bibinfo {author}
  {\bibfnamefont {M.}~\bibnamefont {Pioro-Ladrière}}, \bibinfo {author}
  {\bibfnamefont {A.}~\bibnamefont {Laucht}}, \ and\ \bibinfo {author}
  {\bibfnamefont {A.~S.}\ \bibnamefont {Dzurak}},\ }\href {\doibase
  10.1038/s41586-020-2171-6} {\bibfield  {journal} {\bibinfo  {journal}
  {Nature}\ }\textbf {\bibinfo {volume} {580}},\ \bibinfo {pages} {350}
  (\bibinfo {year} {2020})}\BibitemShut {NoStop}%
\bibitem [{\citenamefont {Petit}\ \emph {et~al.}(2020)\citenamefont {Petit},
  \citenamefont {Russ}, \citenamefont {Eenink}, \citenamefont {Lawrie},
  \citenamefont {Clarke}, \citenamefont {Vandersypen},\ and\ \citenamefont
  {Veldhorst}}]{petit2020high}%
  \BibitemOpen
  \bibfield  {author} {\bibinfo {author} {\bibfnamefont {L.}~\bibnamefont
  {Petit}}, \bibinfo {author} {\bibfnamefont {M.}~\bibnamefont {Russ}},
  \bibinfo {author} {\bibfnamefont {H.}~\bibnamefont {Eenink}}, \bibinfo
  {author} {\bibfnamefont {W.}~\bibnamefont {Lawrie}}, \bibinfo {author}
  {\bibfnamefont {J.}~\bibnamefont {Clarke}}, \bibinfo {author} {\bibfnamefont
  {L.}~\bibnamefont {Vandersypen}}, \ and\ \bibinfo {author} {\bibfnamefont
  {M.}~\bibnamefont {Veldhorst}},\ }\href@noop {} {\bibfield  {journal}
  {\bibinfo  {journal} {arXiv preprint arXiv:2007.09034}\ } (\bibinfo {year}
  {2020})}\BibitemShut {NoStop}%
\bibitem [{\citenamefont {Veldhorst}\ \emph {et~al.}(2015)\citenamefont
  {Veldhorst}, \citenamefont {Yang}, \citenamefont {Hwang}, \citenamefont
  {Huang}, \citenamefont {Dehollain}, \citenamefont {Muhonen}, \citenamefont
  {Simmons}, \citenamefont {Laucht}, \citenamefont {Hudson}, \citenamefont
  {Itoh}, \citenamefont {Morello},\ and\ \citenamefont
  {Dzurak}}]{Veldhorsttwoqubitlogicgate2015}%
  \BibitemOpen
  \bibfield  {author} {\bibinfo {author} {\bibfnamefont {M.}~\bibnamefont
  {Veldhorst}}, \bibinfo {author} {\bibfnamefont {C.~H.}\ \bibnamefont {Yang}},
  \bibinfo {author} {\bibfnamefont {J.~C.~C.}\ \bibnamefont {Hwang}}, \bibinfo
  {author} {\bibfnamefont {W.}~\bibnamefont {Huang}}, \bibinfo {author}
  {\bibfnamefont {J.~P.}\ \bibnamefont {Dehollain}}, \bibinfo {author}
  {\bibfnamefont {J.~T.}\ \bibnamefont {Muhonen}}, \bibinfo {author}
  {\bibfnamefont {S.}~\bibnamefont {Simmons}}, \bibinfo {author} {\bibfnamefont
  {A.}~\bibnamefont {Laucht}}, \bibinfo {author} {\bibfnamefont {F.~E.}\
  \bibnamefont {Hudson}}, \bibinfo {author} {\bibfnamefont {K.~M.}\
  \bibnamefont {Itoh}}, \bibinfo {author} {\bibfnamefont {A.}~\bibnamefont
  {Morello}}, \ and\ \bibinfo {author} {\bibfnamefont {A.~S.}\ \bibnamefont
  {Dzurak}},\ }\href {\doibase 10.1038/nature15263} {\bibfield  {journal}
  {\bibinfo  {journal} {Nature}\ }\textbf {\bibinfo {volume} {526}},\ \bibinfo
  {pages} {410} (\bibinfo {year} {2015})}\BibitemShut {NoStop}%
\bibitem [{\citenamefont {Watson}\ \emph {et~al.}(2018)\citenamefont {Watson},
  \citenamefont {Philips}, \citenamefont {Kawakami}, \citenamefont {Ward},
  \citenamefont {Scarlino}, \citenamefont {Veldhorst}, \citenamefont {Savage},
  \citenamefont {Lagally}, \citenamefont {Friesen}, \citenamefont
  {Coppersmith}, \citenamefont {Eriksson},\ and\ \citenamefont
  {Vandersypen}}]{Watsonprogrammabletwoqubitquantum2018}%
  \BibitemOpen
  \bibfield  {author} {\bibinfo {author} {\bibfnamefont {T.~F.}\ \bibnamefont
  {Watson}}, \bibinfo {author} {\bibfnamefont {S.~G.~J.}\ \bibnamefont
  {Philips}}, \bibinfo {author} {\bibfnamefont {E.}~\bibnamefont {Kawakami}},
  \bibinfo {author} {\bibfnamefont {D.~R.}\ \bibnamefont {Ward}}, \bibinfo
  {author} {\bibfnamefont {P.}~\bibnamefont {Scarlino}}, \bibinfo {author}
  {\bibfnamefont {M.}~\bibnamefont {Veldhorst}}, \bibinfo {author}
  {\bibfnamefont {D.~E.}\ \bibnamefont {Savage}}, \bibinfo {author}
  {\bibfnamefont {M.~G.}\ \bibnamefont {Lagally}}, \bibinfo {author}
  {\bibfnamefont {M.}~\bibnamefont {Friesen}}, \bibinfo {author} {\bibfnamefont
  {S.~N.}\ \bibnamefont {Coppersmith}}, \bibinfo {author} {\bibfnamefont
  {M.~A.}\ \bibnamefont {Eriksson}}, \ and\ \bibinfo {author} {\bibfnamefont
  {L.~M.~K.}\ \bibnamefont {Vandersypen}},\ }\href {\doibase
  10.1038/nature25766} {\bibfield  {journal} {\bibinfo  {journal} {Nature}\
  }\textbf {\bibinfo {volume} {555}},\ \bibinfo {pages} {633} (\bibinfo {year}
  {2018})}\BibitemShut {NoStop}%
\bibitem [{\citenamefont {Zajac}\ \emph {et~al.}(2018)\citenamefont {Zajac},
  \citenamefont {Sigillito}, \citenamefont {Russ}, \citenamefont {Borjans},
  \citenamefont {Taylor}, \citenamefont {Burkard},\ and\ \citenamefont
  {Petta}}]{ZajacResonantlydrivenCNOT2018}%
  \BibitemOpen
  \bibfield  {author} {\bibinfo {author} {\bibfnamefont {D.~M.}\ \bibnamefont
  {Zajac}}, \bibinfo {author} {\bibfnamefont {A.~J.}\ \bibnamefont
  {Sigillito}}, \bibinfo {author} {\bibfnamefont {M.}~\bibnamefont {Russ}},
  \bibinfo {author} {\bibfnamefont {F.}~\bibnamefont {Borjans}}, \bibinfo
  {author} {\bibfnamefont {J.~M.}\ \bibnamefont {Taylor}}, \bibinfo {author}
  {\bibfnamefont {G.}~\bibnamefont {Burkard}}, \ and\ \bibinfo {author}
  {\bibfnamefont {J.~R.}\ \bibnamefont {Petta}},\ }\href {\doibase
  10.1126/science.aao5965} {\bibfield  {journal} {\bibinfo  {journal}
  {Science}\ }\textbf {\bibinfo {volume} {359}},\ \bibinfo {pages} {439}
  (\bibinfo {year} {2018})}\BibitemShut {NoStop}%
\bibitem [{\citenamefont {Xue}\ \emph {et~al.}(2019)\citenamefont {Xue},
  \citenamefont {Watson}, \citenamefont {Helsen}, \citenamefont {Ward},
  \citenamefont {Savage}, \citenamefont {Lagally}, \citenamefont {Coppersmith},
  \citenamefont {Eriksson}, \citenamefont {Wehner},\ and\ \citenamefont
  {Vandersypen}}]{PhysRevX.9.021011}%
  \BibitemOpen
  \bibfield  {author} {\bibinfo {author} {\bibfnamefont {X.}~\bibnamefont
  {Xue}}, \bibinfo {author} {\bibfnamefont {T.~F.}\ \bibnamefont {Watson}},
  \bibinfo {author} {\bibfnamefont {J.}~\bibnamefont {Helsen}}, \bibinfo
  {author} {\bibfnamefont {D.~R.}\ \bibnamefont {Ward}}, \bibinfo {author}
  {\bibfnamefont {D.~E.}\ \bibnamefont {Savage}}, \bibinfo {author}
  {\bibfnamefont {M.~G.}\ \bibnamefont {Lagally}}, \bibinfo {author}
  {\bibfnamefont {S.~N.}\ \bibnamefont {Coppersmith}}, \bibinfo {author}
  {\bibfnamefont {M.~A.}\ \bibnamefont {Eriksson}}, \bibinfo {author}
  {\bibfnamefont {S.}~\bibnamefont {Wehner}}, \ and\ \bibinfo {author}
  {\bibfnamefont {L.~M.~K.}\ \bibnamefont {Vandersypen}},\ }\href {\doibase
  10.1103/PhysRevX.9.021011} {\bibfield  {journal} {\bibinfo  {journal} {Phys.
  Rev. X}\ }\textbf {\bibinfo {volume} {9}},\ \bibinfo {pages} {021011}
  (\bibinfo {year} {2019})}\BibitemShut {NoStop}%
\bibitem [{\citenamefont {Huang}\ \emph {et~al.}(2019)\citenamefont {Huang},
  \citenamefont {Yang}, \citenamefont {Chan}, \citenamefont {Tanttu},
  \citenamefont {Hensen}, \citenamefont {Leon}, \citenamefont {Fogarty},
  \citenamefont {Hwang}, \citenamefont {Hudson}, \citenamefont {Itoh},
  \citenamefont {Morello}, \citenamefont {Laucht},\ and\ \citenamefont
  {Dzurak}}]{huang2019fidelity}%
  \BibitemOpen
  \bibfield  {author} {\bibinfo {author} {\bibfnamefont {W.}~\bibnamefont
  {Huang}}, \bibinfo {author} {\bibfnamefont {C.}~\bibnamefont {Yang}},
  \bibinfo {author} {\bibfnamefont {K.}~\bibnamefont {Chan}}, \bibinfo {author}
  {\bibfnamefont {T.}~\bibnamefont {Tanttu}}, \bibinfo {author} {\bibfnamefont
  {B.}~\bibnamefont {Hensen}}, \bibinfo {author} {\bibfnamefont
  {R.}~\bibnamefont {Leon}}, \bibinfo {author} {\bibfnamefont {M.}~\bibnamefont
  {Fogarty}}, \bibinfo {author} {\bibfnamefont {J.}~\bibnamefont {Hwang}},
  \bibinfo {author} {\bibfnamefont {F.}~\bibnamefont {Hudson}}, \bibinfo
  {author} {\bibfnamefont {K.~M.}\ \bibnamefont {Itoh}}, \bibinfo {author}
  {\bibfnamefont {A.}~\bibnamefont {Morello}}, \bibinfo {author} {\bibfnamefont
  {A.}~\bibnamefont {Laucht}}, \ and\ \bibinfo {author} {\bibfnamefont {A.~S.}\
  \bibnamefont {Dzurak}},\ }\href {\doibase 10.1038/s41586-019-1197-0}
  {\bibfield  {journal} {\bibinfo  {journal} {Nature}\ }\textbf {\bibinfo
  {volume} {569}},\ \bibinfo {pages} {532} (\bibinfo {year}
  {2019})}\BibitemShut {NoStop}%
\bibitem [{\citenamefont {Takeda}\ \emph {et~al.}(2020)\citenamefont {Takeda},
  \citenamefont {Noiri}, \citenamefont {Yoneda}, \citenamefont {Nakajima},\
  and\ \citenamefont {Tarucha}}]{TakedaResonantlyDrivenSingletTriplet2020}%
  \BibitemOpen
  \bibfield  {author} {\bibinfo {author} {\bibfnamefont {K.}~\bibnamefont
  {Takeda}}, \bibinfo {author} {\bibfnamefont {A.}~\bibnamefont {Noiri}},
  \bibinfo {author} {\bibfnamefont {J.}~\bibnamefont {Yoneda}}, \bibinfo
  {author} {\bibfnamefont {T.}~\bibnamefont {Nakajima}}, \ and\ \bibinfo
  {author} {\bibfnamefont {S.}~\bibnamefont {Tarucha}},\ }\href {\doibase
  10.1103/PhysRevLett.124.117701} {\bibfield  {journal} {\bibinfo  {journal}
  {Phys. Rev. Lett.}\ }\textbf {\bibinfo {volume} {124}},\ \bibinfo {pages}
  {117701} (\bibinfo {year} {2020})}\BibitemShut {NoStop}%
\bibitem [{\citenamefont {Lundberg}\ \emph {et~al.}(2020)\citenamefont
  {Lundberg}, \citenamefont {Li}, \citenamefont {Hutin}, \citenamefont
  {Bertrand}, \citenamefont {Ibberson}, \citenamefont {Lee}, \citenamefont
  {Niegemann}, \citenamefont {Urdampilleta}, \citenamefont {Stelmashenko},
  \citenamefont {Meunier}, \citenamefont {Robinson}, \citenamefont {Ibberson},
  \citenamefont {Vinet}, \citenamefont {Niquet},\ and\ \citenamefont
  {Gonzalez-Zalba}}]{PhysRevX.10.041010}%
  \BibitemOpen
  \bibfield  {author} {\bibinfo {author} {\bibfnamefont {T.}~\bibnamefont
  {Lundberg}}, \bibinfo {author} {\bibfnamefont {J.}~\bibnamefont {Li}},
  \bibinfo {author} {\bibfnamefont {L.}~\bibnamefont {Hutin}}, \bibinfo
  {author} {\bibfnamefont {B.}~\bibnamefont {Bertrand}}, \bibinfo {author}
  {\bibfnamefont {D.~J.}\ \bibnamefont {Ibberson}}, \bibinfo {author}
  {\bibfnamefont {C.-M.}\ \bibnamefont {Lee}}, \bibinfo {author} {\bibfnamefont
  {D.~J.}\ \bibnamefont {Niegemann}}, \bibinfo {author} {\bibfnamefont
  {M.}~\bibnamefont {Urdampilleta}}, \bibinfo {author} {\bibfnamefont
  {N.}~\bibnamefont {Stelmashenko}}, \bibinfo {author} {\bibfnamefont
  {T.}~\bibnamefont {Meunier}}, \bibinfo {author} {\bibfnamefont {J.~W.~A.}\
  \bibnamefont {Robinson}}, \bibinfo {author} {\bibfnamefont {L.}~\bibnamefont
  {Ibberson}}, \bibinfo {author} {\bibfnamefont {M.}~\bibnamefont {Vinet}},
  \bibinfo {author} {\bibfnamefont {Y.-M.}\ \bibnamefont {Niquet}}, \ and\
  \bibinfo {author} {\bibfnamefont {M.~F.}\ \bibnamefont {Gonzalez-Zalba}},\
  }\href {\doibase 10.1103/PhysRevX.10.041010} {\bibfield  {journal} {\bibinfo
  {journal} {Phys. Rev. X}\ }\textbf {\bibinfo {volume} {10}},\ \bibinfo
  {pages} {041010} (\bibinfo {year} {2020})}\BibitemShut {NoStop}%
\bibitem [{\citenamefont {Zheng}\ \emph {et~al.}(2019)\citenamefont {Zheng},
  \citenamefont {Samkharadze}, \citenamefont {Noordam}, \citenamefont {Kalhor},
  \citenamefont {Brousse}, \citenamefont {Sammak}, \citenamefont {Scappucci},\
  and\ \citenamefont {Vandersypen}}]{ZhengRapidgatebasedspin2019}%
  \BibitemOpen
  \bibfield  {author} {\bibinfo {author} {\bibfnamefont {G.}~\bibnamefont
  {Zheng}}, \bibinfo {author} {\bibfnamefont {N.}~\bibnamefont {Samkharadze}},
  \bibinfo {author} {\bibfnamefont {M.~L.}\ \bibnamefont {Noordam}}, \bibinfo
  {author} {\bibfnamefont {N.}~\bibnamefont {Kalhor}}, \bibinfo {author}
  {\bibfnamefont {D.}~\bibnamefont {Brousse}}, \bibinfo {author} {\bibfnamefont
  {A.}~\bibnamefont {Sammak}}, \bibinfo {author} {\bibfnamefont
  {G.}~\bibnamefont {Scappucci}}, \ and\ \bibinfo {author} {\bibfnamefont
  {L.~M.~K.}\ \bibnamefont {Vandersypen}},\ }\href {\doibase
  10.1038/s41565-019-0488-9} {\bibfield  {journal} {\bibinfo  {journal} {Nat.
  Nanotechnol.}\ }\textbf {\bibinfo {volume} {14}},\ \bibinfo {pages} {742}
  (\bibinfo {year} {2019})}\BibitemShut {NoStop}%
\bibitem [{\citenamefont {West}\ \emph {et~al.}(2019)\citenamefont {West},
  \citenamefont {Hensen}, \citenamefont {Jouan}, \citenamefont {Tanttu},
  \citenamefont {Yang}, \citenamefont {Rossi}, \citenamefont
  {{Gonzalez-Zalba}}, \citenamefont {Hudson}, \citenamefont {Morello},
  \citenamefont {Reilly},\ and\ \citenamefont
  {Dzurak}}]{WestGatebasedsingleshotreadout2019}%
  \BibitemOpen
  \bibfield  {author} {\bibinfo {author} {\bibfnamefont {A.}~\bibnamefont
  {West}}, \bibinfo {author} {\bibfnamefont {B.}~\bibnamefont {Hensen}},
  \bibinfo {author} {\bibfnamefont {A.}~\bibnamefont {Jouan}}, \bibinfo
  {author} {\bibfnamefont {T.}~\bibnamefont {Tanttu}}, \bibinfo {author}
  {\bibfnamefont {C.-H.}\ \bibnamefont {Yang}}, \bibinfo {author}
  {\bibfnamefont {A.}~\bibnamefont {Rossi}}, \bibinfo {author} {\bibfnamefont
  {M.~F.}\ \bibnamefont {{Gonzalez-Zalba}}}, \bibinfo {author} {\bibfnamefont
  {F.}~\bibnamefont {Hudson}}, \bibinfo {author} {\bibfnamefont
  {A.}~\bibnamefont {Morello}}, \bibinfo {author} {\bibfnamefont {D.~J.}\
  \bibnamefont {Reilly}}, \ and\ \bibinfo {author} {\bibfnamefont {A.~S.}\
  \bibnamefont {Dzurak}},\ }\href {\doibase 10.1038/s41565-019-0400-7}
  {\bibfield  {journal} {\bibinfo  {journal} {Nat. Nanotechnol.}\ }\textbf
  {\bibinfo {volume} {14}},\ \bibinfo {pages} {437} (\bibinfo {year}
  {2019})}\BibitemShut {NoStop}%
\bibitem [{\citenamefont {Yoneda}\ \emph {et~al.}(2020)\citenamefont {Yoneda},
  \citenamefont {Takeda}, \citenamefont {Noiri}, \citenamefont {Nakajima},
  \citenamefont {Li}, \citenamefont {Kamioka}, \citenamefont {Kodera},\ and\
  \citenamefont {Tarucha}}]{yoneda2020quantum}%
  \BibitemOpen
  \bibfield  {author} {\bibinfo {author} {\bibfnamefont {J.}~\bibnamefont
  {Yoneda}}, \bibinfo {author} {\bibfnamefont {K.}~\bibnamefont {Takeda}},
  \bibinfo {author} {\bibfnamefont {A.}~\bibnamefont {Noiri}}, \bibinfo
  {author} {\bibfnamefont {T.}~\bibnamefont {Nakajima}}, \bibinfo {author}
  {\bibfnamefont {S.}~\bibnamefont {Li}}, \bibinfo {author} {\bibfnamefont
  {J.}~\bibnamefont {Kamioka}}, \bibinfo {author} {\bibfnamefont
  {T.}~\bibnamefont {Kodera}}, \ and\ \bibinfo {author} {\bibfnamefont
  {S.}~\bibnamefont {Tarucha}},\ }\href {\doibase 10.1038/s41467-020-14818-8}
  {\bibfield  {journal} {\bibinfo  {journal} {Nature communications}\ }\textbf
  {\bibinfo {volume} {11}},\ \bibinfo {pages} {1} (\bibinfo {year}
  {2020})}\BibitemShut {NoStop}%
\bibitem [{\citenamefont {Xue}\ \emph {et~al.}(2020)\citenamefont {Xue},
  \citenamefont {D'Anjou}, \citenamefont {Watson}, \citenamefont {Ward},
  \citenamefont {Savage}, \citenamefont {Lagally}, \citenamefont {Friesen},
  \citenamefont {Coppersmith}, \citenamefont {Eriksson}, \citenamefont
  {Coish},\ and\ \citenamefont {Vandersypen}}]{PhysRevX.10.021006}%
  \BibitemOpen
  \bibfield  {author} {\bibinfo {author} {\bibfnamefont {X.}~\bibnamefont
  {Xue}}, \bibinfo {author} {\bibfnamefont {B.}~\bibnamefont {D'Anjou}},
  \bibinfo {author} {\bibfnamefont {T.~F.}\ \bibnamefont {Watson}}, \bibinfo
  {author} {\bibfnamefont {D.~R.}\ \bibnamefont {Ward}}, \bibinfo {author}
  {\bibfnamefont {D.~E.}\ \bibnamefont {Savage}}, \bibinfo {author}
  {\bibfnamefont {M.~G.}\ \bibnamefont {Lagally}}, \bibinfo {author}
  {\bibfnamefont {M.}~\bibnamefont {Friesen}}, \bibinfo {author} {\bibfnamefont
  {S.~N.}\ \bibnamefont {Coppersmith}}, \bibinfo {author} {\bibfnamefont
  {M.~A.}\ \bibnamefont {Eriksson}}, \bibinfo {author} {\bibfnamefont {W.~A.}\
  \bibnamefont {Coish}}, \ and\ \bibinfo {author} {\bibfnamefont {L.~M.~K.}\
  \bibnamefont {Vandersypen}},\ }\href {\doibase 10.1103/PhysRevX.10.021006}
  {\bibfield  {journal} {\bibinfo  {journal} {Phys. Rev. X}\ }\textbf {\bibinfo
  {volume} {10}},\ \bibinfo {pages} {021006} (\bibinfo {year}
  {2020})}\BibitemShut {NoStop}%
\bibitem [{\citenamefont {Seedhouse}\ \emph {et~al.}(2021)\citenamefont
  {Seedhouse}, \citenamefont {Tanttu}, \citenamefont {Leon}, \citenamefont
  {Zhao}, \citenamefont {Tan}, \citenamefont {Hensen}, \citenamefont {Hudson},
  \citenamefont {Itoh}, \citenamefont {Yoneda}, \citenamefont {Yang},
  \citenamefont {Morello}, \citenamefont {Laucht}, \citenamefont {Coppersmith},
  \citenamefont {Saraiva},\ and\ \citenamefont
  {Dzurak}}]{SeedhouseParityreadoutsilicon2020}%
  \BibitemOpen
  \bibfield  {author} {\bibinfo {author} {\bibfnamefont {A.~E.}\ \bibnamefont
  {Seedhouse}}, \bibinfo {author} {\bibfnamefont {T.}~\bibnamefont {Tanttu}},
  \bibinfo {author} {\bibfnamefont {R.~C.}\ \bibnamefont {Leon}}, \bibinfo
  {author} {\bibfnamefont {R.}~\bibnamefont {Zhao}}, \bibinfo {author}
  {\bibfnamefont {K.~Y.}\ \bibnamefont {Tan}}, \bibinfo {author} {\bibfnamefont
  {B.}~\bibnamefont {Hensen}}, \bibinfo {author} {\bibfnamefont {F.~E.}\
  \bibnamefont {Hudson}}, \bibinfo {author} {\bibfnamefont {K.~M.}\
  \bibnamefont {Itoh}}, \bibinfo {author} {\bibfnamefont {J.}~\bibnamefont
  {Yoneda}}, \bibinfo {author} {\bibfnamefont {C.~H.}\ \bibnamefont {Yang}},
  \bibinfo {author} {\bibfnamefont {A.}~\bibnamefont {Morello}}, \bibinfo
  {author} {\bibfnamefont {A.}~\bibnamefont {Laucht}}, \bibinfo {author}
  {\bibfnamefont {S.~N.}\ \bibnamefont {Coppersmith}}, \bibinfo {author}
  {\bibfnamefont {A.}~\bibnamefont {Saraiva}}, \ and\ \bibinfo {author}
  {\bibfnamefont {A.~S.}\ \bibnamefont {Dzurak}},\ }\href {\doibase
  10.1103/PRXQuantum.2.010303} {\bibfield  {journal} {\bibinfo  {journal} {PRX
  Quantum}\ }\textbf {\bibinfo {volume} {2}},\ \bibinfo {pages} {010303}
  (\bibinfo {year} {2021})}\BibitemShut {NoStop}%
\bibitem [{\citenamefont {Khaetskii}\ \emph {et~al.}(2002)\citenamefont
  {Khaetskii}, \citenamefont {Loss},\ and\ \citenamefont
  {Glazman}}]{PhysRevLett.88.186802}%
  \BibitemOpen
  \bibfield  {author} {\bibinfo {author} {\bibfnamefont {A.~V.}\ \bibnamefont
  {Khaetskii}}, \bibinfo {author} {\bibfnamefont {D.}~\bibnamefont {Loss}}, \
  and\ \bibinfo {author} {\bibfnamefont {L.}~\bibnamefont {Glazman}},\ }\href
  {\doibase 10.1103/PhysRevLett.88.186802} {\bibfield  {journal} {\bibinfo
  {journal} {Phys. Rev. Lett.}\ }\textbf {\bibinfo {volume} {88}},\ \bibinfo
  {pages} {186802} (\bibinfo {year} {2002})}\BibitemShut {NoStop}%
\bibitem [{\citenamefont {Coish}\ and\ \citenamefont
  {Loss}(2004)}]{PhysRevB.70.195340}%
  \BibitemOpen
  \bibfield  {author} {\bibinfo {author} {\bibfnamefont {W.~A.}\ \bibnamefont
  {Coish}}\ and\ \bibinfo {author} {\bibfnamefont {D.}~\bibnamefont {Loss}},\
  }\href {\doibase 10.1103/PhysRevB.70.195340} {\bibfield  {journal} {\bibinfo
  {journal} {Phys. Rev. B}\ }\textbf {\bibinfo {volume} {70}},\ \bibinfo
  {pages} {195340} (\bibinfo {year} {2004})}\BibitemShut {NoStop}%
\bibitem [{\citenamefont {Fischer}\ \emph {et~al.}(2009)\citenamefont
  {Fischer}, \citenamefont {Trif}, \citenamefont {Coish},\ and\ \citenamefont
  {Loss}}]{fischer2009spin}%
  \BibitemOpen
  \bibfield  {author} {\bibinfo {author} {\bibfnamefont {J.}~\bibnamefont
  {Fischer}}, \bibinfo {author} {\bibfnamefont {M.}~\bibnamefont {Trif}},
  \bibinfo {author} {\bibfnamefont {W.}~\bibnamefont {Coish}}, \ and\ \bibinfo
  {author} {\bibfnamefont {D.}~\bibnamefont {Loss}},\ }\href@noop {} {\bibfield
   {journal} {\bibinfo  {journal} {Solid state communications}\ }\textbf
  {\bibinfo {volume} {149}},\ \bibinfo {pages} {1443} (\bibinfo {year}
  {2009})}\BibitemShut {NoStop}%
\bibitem [{\citenamefont {Fischer}\ and\ \citenamefont
  {Loss}(2010)}]{PhysRevLett.105.266603}%
  \BibitemOpen
  \bibfield  {author} {\bibinfo {author} {\bibfnamefont {J.}~\bibnamefont
  {Fischer}}\ and\ \bibinfo {author} {\bibfnamefont {D.}~\bibnamefont {Loss}},\
  }\href {\doibase 10.1103/PhysRevLett.105.266603} {\bibfield  {journal}
  {\bibinfo  {journal} {Phys. Rev. Lett.}\ }\textbf {\bibinfo {volume} {105}},\
  \bibinfo {pages} {266603} (\bibinfo {year} {2010})}\BibitemShut {NoStop}%
\bibitem [{\citenamefont {Fischer}\ \emph {et~al.}(2008)\citenamefont
  {Fischer}, \citenamefont {Coish}, \citenamefont {Bulaev},\ and\ \citenamefont
  {Loss}}]{PhysRevB.78.155329}%
  \BibitemOpen
  \bibfield  {author} {\bibinfo {author} {\bibfnamefont {J.}~\bibnamefont
  {Fischer}}, \bibinfo {author} {\bibfnamefont {W.~A.}\ \bibnamefont {Coish}},
  \bibinfo {author} {\bibfnamefont {D.~V.}\ \bibnamefont {Bulaev}}, \ and\
  \bibinfo {author} {\bibfnamefont {D.}~\bibnamefont {Loss}},\ }\href {\doibase
  10.1103/PhysRevB.78.155329} {\bibfield  {journal} {\bibinfo  {journal} {Phys.
  Rev. B}\ }\textbf {\bibinfo {volume} {78}},\ \bibinfo {pages} {155329}
  (\bibinfo {year} {2008})}\BibitemShut {NoStop}%
\bibitem [{\citenamefont {Maier}\ and\ \citenamefont
  {Loss}(2012)}]{PhysRevB.85.195323}%
  \BibitemOpen
  \bibfield  {author} {\bibinfo {author} {\bibfnamefont {F.}~\bibnamefont
  {Maier}}\ and\ \bibinfo {author} {\bibfnamefont {D.}~\bibnamefont {Loss}},\
  }\href {\doibase 10.1103/PhysRevB.85.195323} {\bibfield  {journal} {\bibinfo
  {journal} {Phys. Rev. B}\ }\textbf {\bibinfo {volume} {85}},\ \bibinfo
  {pages} {195323} (\bibinfo {year} {2012})}\BibitemShut {NoStop}%
\bibitem [{\citenamefont {Philippopoulos}\ \emph {et~al.}(2020)\citenamefont
  {Philippopoulos}, \citenamefont {Chesi},\ and\ \citenamefont
  {Coish}}]{PhysRevB.101.115302}%
  \BibitemOpen
  \bibfield  {author} {\bibinfo {author} {\bibfnamefont {P.}~\bibnamefont
  {Philippopoulos}}, \bibinfo {author} {\bibfnamefont {S.}~\bibnamefont
  {Chesi}}, \ and\ \bibinfo {author} {\bibfnamefont {W.~A.}\ \bibnamefont
  {Coish}},\ }\href {\doibase 10.1103/PhysRevB.101.115302} {\bibfield
  {journal} {\bibinfo  {journal} {Phys. Rev. B}\ }\textbf {\bibinfo {volume}
  {101}},\ \bibinfo {pages} {115302} (\bibinfo {year} {2020})}\BibitemShut
  {NoStop}%
\bibitem [{\citenamefont {Takeda}\ \emph {et~al.}(2016)\citenamefont {Takeda},
  \citenamefont {Kamioka}, \citenamefont {Otsuka}, \citenamefont {Yoneda},
  \citenamefont {Nakajima}, \citenamefont {Delbecq}, \citenamefont {Amaha},
  \citenamefont {Allison}, \citenamefont {Kodera}, \citenamefont {Oda},\ and\
  \citenamefont {Tarucha}}]{Takedafaulttolerantaddressablespin2016}%
  \BibitemOpen
  \bibfield  {author} {\bibinfo {author} {\bibfnamefont {K.}~\bibnamefont
  {Takeda}}, \bibinfo {author} {\bibfnamefont {J.}~\bibnamefont {Kamioka}},
  \bibinfo {author} {\bibfnamefont {T.}~\bibnamefont {Otsuka}}, \bibinfo
  {author} {\bibfnamefont {J.}~\bibnamefont {Yoneda}}, \bibinfo {author}
  {\bibfnamefont {T.}~\bibnamefont {Nakajima}}, \bibinfo {author}
  {\bibfnamefont {M.~R.}\ \bibnamefont {Delbecq}}, \bibinfo {author}
  {\bibfnamefont {S.}~\bibnamefont {Amaha}}, \bibinfo {author} {\bibfnamefont
  {G.}~\bibnamefont {Allison}}, \bibinfo {author} {\bibfnamefont
  {T.}~\bibnamefont {Kodera}}, \bibinfo {author} {\bibfnamefont
  {S.}~\bibnamefont {Oda}}, \ and\ \bibinfo {author} {\bibfnamefont
  {S.}~\bibnamefont {Tarucha}},\ }\href {\doibase 10.1126/sciadv.1600694}
  {\bibfield  {journal} {\bibinfo  {journal} {Sci. Adv.}\ }\textbf {\bibinfo
  {volume} {2}},\ \bibinfo {pages} {e1600694} (\bibinfo {year}
  {2016})}\BibitemShut {NoStop}%
\bibitem [{\citenamefont {Zhao}\ \emph {et~al.}(2019)\citenamefont {Zhao},
  \citenamefont {Tanttu}, \citenamefont {Tan}, \citenamefont {Hensen},
  \citenamefont {Chan}, \citenamefont {Hwang}, \citenamefont {Leon},
  \citenamefont {Yang}, \citenamefont {Gilbert}, \citenamefont {Hudson},
  \citenamefont {Itoh}, \citenamefont {Kiselev}, \citenamefont {Ladd},
  \citenamefont {Morello}, \citenamefont {Laucht},\ and\ \citenamefont
  {Dzurak}}]{ZhaoSinglespinqubitsisotopically2019}%
  \BibitemOpen
  \bibfield  {author} {\bibinfo {author} {\bibfnamefont {R.}~\bibnamefont
  {Zhao}}, \bibinfo {author} {\bibfnamefont {T.}~\bibnamefont {Tanttu}},
  \bibinfo {author} {\bibfnamefont {K.~Y.}\ \bibnamefont {Tan}}, \bibinfo
  {author} {\bibfnamefont {B.}~\bibnamefont {Hensen}}, \bibinfo {author}
  {\bibfnamefont {K.~W.}\ \bibnamefont {Chan}}, \bibinfo {author}
  {\bibfnamefont {J.~C.~C.}\ \bibnamefont {Hwang}}, \bibinfo {author}
  {\bibfnamefont {R.~C.~C.}\ \bibnamefont {Leon}}, \bibinfo {author}
  {\bibfnamefont {C.~H.}\ \bibnamefont {Yang}}, \bibinfo {author}
  {\bibfnamefont {W.}~\bibnamefont {Gilbert}}, \bibinfo {author} {\bibfnamefont
  {F.~E.}\ \bibnamefont {Hudson}}, \bibinfo {author} {\bibfnamefont {K.~M.}\
  \bibnamefont {Itoh}}, \bibinfo {author} {\bibfnamefont {A.~A.}\ \bibnamefont
  {Kiselev}}, \bibinfo {author} {\bibfnamefont {T.~D.}\ \bibnamefont {Ladd}},
  \bibinfo {author} {\bibfnamefont {A.}~\bibnamefont {Morello}}, \bibinfo
  {author} {\bibfnamefont {A.}~\bibnamefont {Laucht}}, \ and\ \bibinfo {author}
  {\bibfnamefont {A.~S.}\ \bibnamefont {Dzurak}},\ }\href {\doibase
  10.1038/s41467-019-13416-7} {\bibfield  {journal} {\bibinfo  {journal} {Nat.
  Commun.}\ }\textbf {\bibinfo {volume} {10}},\ \bibinfo {pages} {1} (\bibinfo
  {year} {2019})}\BibitemShut {NoStop}%
\bibitem [{\citenamefont {Voisin}\ \emph {et~al.}(2016)\citenamefont {Voisin},
  \citenamefont {Maurand}, \citenamefont {Barraud}, \citenamefont {Vinet},
  \citenamefont {Jehl}, \citenamefont {Sanquer}, \citenamefont {Renard},\ and\
  \citenamefont {De~Franceschi}}]{voisin2016electrical}%
  \BibitemOpen
  \bibfield  {author} {\bibinfo {author} {\bibfnamefont {B.}~\bibnamefont
  {Voisin}}, \bibinfo {author} {\bibfnamefont {R.}~\bibnamefont {Maurand}},
  \bibinfo {author} {\bibfnamefont {S.}~\bibnamefont {Barraud}}, \bibinfo
  {author} {\bibfnamefont {M.}~\bibnamefont {Vinet}}, \bibinfo {author}
  {\bibfnamefont {X.}~\bibnamefont {Jehl}}, \bibinfo {author} {\bibfnamefont
  {M.}~\bibnamefont {Sanquer}}, \bibinfo {author} {\bibfnamefont
  {J.}~\bibnamefont {Renard}}, \ and\ \bibinfo {author} {\bibfnamefont
  {S.}~\bibnamefont {De~Franceschi}},\ }\href {\doibase
  10.1021/acs.nanolett.5b02920} {\bibfield  {journal} {\bibinfo  {journal}
  {Nano letters}\ }\textbf {\bibinfo {volume} {16}},\ \bibinfo {pages} {88}
  (\bibinfo {year} {2016})}\BibitemShut {NoStop}%
\bibitem [{\citenamefont {Crippa}\ \emph {et~al.}(2018)\citenamefont {Crippa},
  \citenamefont {Maurand}, \citenamefont {Bourdet}, \citenamefont
  {Kotekar-Patil}, \citenamefont {Amisse}, \citenamefont {Jehl}, \citenamefont
  {Sanquer}, \citenamefont {Lavi\'eville}, \citenamefont {Bohuslavskyi},
  \citenamefont {Hutin}, \citenamefont {Barraud}, \citenamefont {Vinet},
  \citenamefont {Niquet},\ and\ \citenamefont
  {De~Franceschi}}]{PhysRevLett.120.137702}%
  \BibitemOpen
  \bibfield  {author} {\bibinfo {author} {\bibfnamefont {A.}~\bibnamefont
  {Crippa}}, \bibinfo {author} {\bibfnamefont {R.}~\bibnamefont {Maurand}},
  \bibinfo {author} {\bibfnamefont {L.}~\bibnamefont {Bourdet}}, \bibinfo
  {author} {\bibfnamefont {D.}~\bibnamefont {Kotekar-Patil}}, \bibinfo {author}
  {\bibfnamefont {A.}~\bibnamefont {Amisse}}, \bibinfo {author} {\bibfnamefont
  {X.}~\bibnamefont {Jehl}}, \bibinfo {author} {\bibfnamefont {M.}~\bibnamefont
  {Sanquer}}, \bibinfo {author} {\bibfnamefont {R.}~\bibnamefont
  {Lavi\'eville}}, \bibinfo {author} {\bibfnamefont {H.}~\bibnamefont
  {Bohuslavskyi}}, \bibinfo {author} {\bibfnamefont {L.}~\bibnamefont {Hutin}},
  \bibinfo {author} {\bibfnamefont {S.}~\bibnamefont {Barraud}}, \bibinfo
  {author} {\bibfnamefont {M.}~\bibnamefont {Vinet}}, \bibinfo {author}
  {\bibfnamefont {Y.-M.}\ \bibnamefont {Niquet}}, \ and\ \bibinfo {author}
  {\bibfnamefont {S.}~\bibnamefont {De~Franceschi}},\ }\href {\doibase
  10.1103/PhysRevLett.120.137702} {\bibfield  {journal} {\bibinfo  {journal}
  {Phys. Rev. Lett.}\ }\textbf {\bibinfo {volume} {120}},\ \bibinfo {pages}
  {137702} (\bibinfo {year} {2018})}\BibitemShut {NoStop}%
\bibitem [{\citenamefont {Reed}\ \emph {et~al.}(1986)\citenamefont {Reed},
  \citenamefont {Bate}, \citenamefont {Bradshaw}, \citenamefont {Duncan},
  \citenamefont {Frensley}, \citenamefont {Lee},\ and\ \citenamefont
  {Shih}}]{reed1986spatial}%
  \BibitemOpen
  \bibfield  {author} {\bibinfo {author} {\bibfnamefont {M.}~\bibnamefont
  {Reed}}, \bibinfo {author} {\bibfnamefont {R.}~\bibnamefont {Bate}}, \bibinfo
  {author} {\bibfnamefont {K.}~\bibnamefont {Bradshaw}}, \bibinfo {author}
  {\bibfnamefont {W.}~\bibnamefont {Duncan}}, \bibinfo {author} {\bibfnamefont
  {W.}~\bibnamefont {Frensley}}, \bibinfo {author} {\bibfnamefont
  {J.}~\bibnamefont {Lee}}, \ and\ \bibinfo {author} {\bibfnamefont
  {H.}~\bibnamefont {Shih}},\ }\href {\doibase 10.1116/1.583331} {\bibfield
  {journal} {\bibinfo  {journal} {Journal of Vacuum Science \& Technology B:
  Microelectronics Processing and Phenomena}\ }\textbf {\bibinfo {volume}
  {4}},\ \bibinfo {pages} {358} (\bibinfo {year} {1986})}\BibitemShut {NoStop}%
\bibitem [{\citenamefont {Warburton}(2002)}]{doi:10.1080/00107510210123845}%
  \BibitemOpen
  \bibfield  {author} {\bibinfo {author} {\bibfnamefont {R.~J.}\ \bibnamefont
  {Warburton}},\ }\href {\doibase 10.1080/00107510210123845} {\bibfield
  {journal} {\bibinfo  {journal} {Contemporary Physics}\ }\textbf {\bibinfo
  {volume} {43}},\ \bibinfo {pages} {351} (\bibinfo {year} {2002})}\BibitemShut
  {NoStop}%
\bibitem [{\citenamefont {Venitucci}\ and\ \citenamefont
  {Niquet}(2019)}]{PhysRevB.99.115317}%
  \BibitemOpen
  \bibfield  {author} {\bibinfo {author} {\bibfnamefont {B.}~\bibnamefont
  {Venitucci}}\ and\ \bibinfo {author} {\bibfnamefont {Y.-M.}\ \bibnamefont
  {Niquet}},\ }\href {\doibase 10.1103/PhysRevB.99.115317} {\bibfield
  {journal} {\bibinfo  {journal} {Phys. Rev. B}\ }\textbf {\bibinfo {volume}
  {99}},\ \bibinfo {pages} {115317} (\bibinfo {year} {2019})}\BibitemShut
  {NoStop}%
\bibitem [{\citenamefont {Xiong}\ \emph {et~al.}(2021)\citenamefont {Xiong},
  \citenamefont {Guan}, \citenamefont {Luo},\ and\ \citenamefont
  {Li}}]{xiong2020strong}%
  \BibitemOpen
  \bibfield  {author} {\bibinfo {author} {\bibfnamefont {J.-X.}\ \bibnamefont
  {Xiong}}, \bibinfo {author} {\bibfnamefont {S.}~\bibnamefont {Guan}},
  \bibinfo {author} {\bibfnamefont {J.-W.}\ \bibnamefont {Luo}}, \ and\
  \bibinfo {author} {\bibfnamefont {S.-S.}\ \bibnamefont {Li}},\ }\href
  {\doibase 10.1103/PhysRevB.103.085309} {\bibfield  {journal} {\bibinfo
  {journal} {Phys. Rev. B}\ }\textbf {\bibinfo {volume} {103}},\ \bibinfo
  {pages} {085309} (\bibinfo {year} {2021})}\BibitemShut {NoStop}%
\bibitem [{\citenamefont {Het\'enyi}\ \emph {et~al.}(2020)\citenamefont
  {Het\'enyi}, \citenamefont {Kloeffel},\ and\ \citenamefont
  {Loss}}]{PhysRevResearch.2.033036}%
  \BibitemOpen
  \bibfield  {author} {\bibinfo {author} {\bibfnamefont {B.}~\bibnamefont
  {Het\'enyi}}, \bibinfo {author} {\bibfnamefont {C.}~\bibnamefont {Kloeffel}},
  \ and\ \bibinfo {author} {\bibfnamefont {D.}~\bibnamefont {Loss}},\ }\href
  {\doibase 10.1103/PhysRevResearch.2.033036} {\bibfield  {journal} {\bibinfo
  {journal} {Phys. Rev. Research}\ }\textbf {\bibinfo {volume} {2}},\ \bibinfo
  {pages} {033036} (\bibinfo {year} {2020})}\BibitemShut {NoStop}%
\bibitem [{\citenamefont {Luttinger}(1956)}]{LuttingerModel}%
  \BibitemOpen
  \bibfield  {author} {\bibinfo {author} {\bibfnamefont {J.~M.}\ \bibnamefont
  {Luttinger}},\ }\href {\doibase 10.1103/PhysRev.102.1030} {\bibfield
  {journal} {\bibinfo  {journal} {Phys. Rev.}\ }\textbf {\bibinfo {volume}
  {102}},\ \bibinfo {pages} {1030} (\bibinfo {year} {1956})}\BibitemShut
  {NoStop}%
\bibitem [{Note1()}]{Note1}%
  \BibitemOpen
  \bibinfo {note} {If $\protect \textbf {E}=0$ and the cross-section is
  inversion symmetric, then $V(x,y)\equiv V_\protect \text
  {HW}(x,y)+V_E(x,y)=\protect \mathcal {I} V(x,y)\protect \mathcal {I}^\dagger
  = V(-x,-y)$, where $\protect \mathcal {I}=e^{-i\pi F_z}$ is the inversion
  operator in the $(x,y)$ plane; $F_z$ is the component parallel to the wire of
  the total angular momentum. It follows that the ground subspace of
  $H_\protect \text {LK}(p_z=0)+V(x,y)$ can be labelled by $| i, \uparrow
  \downarrow \rangle $ and is composed of a degenerate Kramers doublet
  $|\uparrow \downarrow \rangle $ that is also an eigenstate of $\protect
  \mathcal {I}$ to eigenvalue $i$. Let us consider the correction to the LK
  Hamiltonian $H_1 p_z$ to linear order in $p_z$, such that the states $| i,
  \uparrow \downarrow \rangle $ are still approximate eigenstates. Because
  $H_1$ anticommutes with $\protect \mathcal {I}$, i.e. $\protect \mathcal {I}
  H_1 \protect \mathcal {I}^\dagger =-H_1$, $H_1$ is an block off-diagonal
  matrix in the basis of the eigenstates of $\protect \mathcal {I}$.
  Consequently, in the groundstate subspace there cannot be spin-orbit
  interactions linear in $p_z$ and because time-reversal symmetry prohibits
  different masses for the Kramers partners, the first possible spin-orbit
  interactions are $\propto p_z^3$.}\BibitemShut {Stop}%
\bibitem [{\citenamefont {McCartin}(2003)}]{mccartin2011laplacian}%
  \BibitemOpen
  \bibfield  {author} {\bibinfo {author} {\bibfnamefont {B.~J.}\ \bibnamefont
  {McCartin}},\ }\href {\doibase 10.1137/S003614450238720} {\bibfield
  {journal} {\bibinfo  {journal} {Siam Review}\ }\textbf {\bibinfo {volume}
  {45}},\ \bibinfo {pages} {267} (\bibinfo {year} {2003})}\BibitemShut
  {NoStop}%
\bibitem [{\citenamefont {Bravyi}\ \emph {et~al.}(2011)\citenamefont {Bravyi},
  \citenamefont {DiVincenzo},\ and\ \citenamefont
  {Loss}}]{bravyi2011schrieffer}%
  \BibitemOpen
  \bibfield  {author} {\bibinfo {author} {\bibfnamefont {S.}~\bibnamefont
  {Bravyi}}, \bibinfo {author} {\bibfnamefont {D.~P.}\ \bibnamefont
  {DiVincenzo}}, \ and\ \bibinfo {author} {\bibfnamefont {D.}~\bibnamefont
  {Loss}},\ }\href {\doibase 10.1016/j.aop.2011.06.004} {\bibfield  {journal}
  {\bibinfo  {journal} {Annals of physics}\ }\textbf {\bibinfo {volume}
  {326}},\ \bibinfo {pages} {2793} (\bibinfo {year} {2011})}\BibitemShut
  {NoStop}%
\bibitem [{Note2()}]{Note2}%
  \BibitemOpen
  \bibinfo {note} {The results presented in Sec.~\ref {Sec:Homogeneous_E} are
  qualitatively valid even in the presence of asymmetries that lead to a finite
  $E_x$. In particular, we find that the value of the spin-orbit velocity at
  $E_y=0$ is increased by $E_x$ and consequently the spin-orbit switch is
  pushed to higher values of $E_y$.}\BibitemShut {Stop}%
\bibitem [{\citenamefont {Lee}\ \emph {et~al.}(2005)\citenamefont {Lee},
  \citenamefont {Fitzgerald}, \citenamefont {Bulsara}, \citenamefont {Currie},\
  and\ \citenamefont {Lochtefeld}}]{lee2005strained}%
  \BibitemOpen
  \bibfield  {author} {\bibinfo {author} {\bibfnamefont {M.~L.}\ \bibnamefont
  {Lee}}, \bibinfo {author} {\bibfnamefont {E.~A.}\ \bibnamefont {Fitzgerald}},
  \bibinfo {author} {\bibfnamefont {M.~T.}\ \bibnamefont {Bulsara}}, \bibinfo
  {author} {\bibfnamefont {M.~T.}\ \bibnamefont {Currie}}, \ and\ \bibinfo
  {author} {\bibfnamefont {A.}~\bibnamefont {Lochtefeld}},\ }\href {\doibase
  10.1063/1.1819976} {\bibfield  {journal} {\bibinfo  {journal} {Journal of
  Applied Physics}\ }\textbf {\bibinfo {volume} {97}},\ \bibinfo {pages} {1}
  (\bibinfo {year} {2005})}\BibitemShut {NoStop}%
\bibitem [{\citenamefont {Liles}\ \emph {et~al.}(2020)\citenamefont {Liles},
  \citenamefont {Martins}, \citenamefont {Miserev}, \citenamefont {Kiselev},
  \citenamefont {Thorvaldson}, \citenamefont {Rendell}, \citenamefont {Jin},
  \citenamefont {Hudson}, \citenamefont {Veldhorst}, \citenamefont {Itoh},
  \citenamefont {Sushkov}, \citenamefont {Ladd}, \citenamefont {Dzurak},\ and\
  \citenamefont {Hamilton}}]{liles2020electrical}%
  \BibitemOpen
  \bibfield  {author} {\bibinfo {author} {\bibfnamefont {S.~D.}\ \bibnamefont
  {Liles}}, \bibinfo {author} {\bibfnamefont {F.}~\bibnamefont {Martins}},
  \bibinfo {author} {\bibfnamefont {D.~S.}\ \bibnamefont {Miserev}}, \bibinfo
  {author} {\bibfnamefont {A.~A.}\ \bibnamefont {Kiselev}}, \bibinfo {author}
  {\bibfnamefont {I.~D.}\ \bibnamefont {Thorvaldson}}, \bibinfo {author}
  {\bibfnamefont {M.~J.}\ \bibnamefont {Rendell}}, \bibinfo {author}
  {\bibfnamefont {I.~K.}\ \bibnamefont {Jin}}, \bibinfo {author} {\bibfnamefont
  {F.~E.}\ \bibnamefont {Hudson}}, \bibinfo {author} {\bibfnamefont
  {M.}~\bibnamefont {Veldhorst}}, \bibinfo {author} {\bibfnamefont {K.~M.}\
  \bibnamefont {Itoh}}, \bibinfo {author} {\bibfnamefont {O.~P.}\ \bibnamefont
  {Sushkov}}, \bibinfo {author} {\bibfnamefont {T.~D.}\ \bibnamefont {Ladd}},
  \bibinfo {author} {\bibfnamefont {A.~S.}\ \bibnamefont {Dzurak}}, \ and\
  \bibinfo {author} {\bibfnamefont {A.~R.}\ \bibnamefont {Hamilton}},\
  }\href@noop {} {\bibfield  {journal} {\bibinfo  {journal} {arXiv preprint
  arXiv:2012.04985}\ } (\bibinfo {year} {2020})}\BibitemShut {NoStop}%
\bibitem [{\citenamefont {Niquet}\ \emph {et~al.}(2012)\citenamefont {Niquet},
  \citenamefont {Delerue},\ and\ \citenamefont
  {Krzeminski}}]{niquet2012effects}%
  \BibitemOpen
  \bibfield  {author} {\bibinfo {author} {\bibfnamefont {Y.-M.}\ \bibnamefont
  {Niquet}}, \bibinfo {author} {\bibfnamefont {C.}~\bibnamefont {Delerue}}, \
  and\ \bibinfo {author} {\bibfnamefont {C.}~\bibnamefont {Krzeminski}},\
  }\href {\doibase 10.1021/nl3010995} {\bibfield  {journal} {\bibinfo
  {journal} {Nano letters}\ }\textbf {\bibinfo {volume} {12}},\ \bibinfo
  {pages} {3545} (\bibinfo {year} {2012})}\BibitemShut {NoStop}%
\bibitem [{\citenamefont {Stan}\ \emph {et~al.}(2010)\citenamefont {Stan},
  \citenamefont {Krylyuk}, \citenamefont {Davydov},\ and\ \citenamefont
  {Cook}}]{stan2010compressive}%
  \BibitemOpen
  \bibfield  {author} {\bibinfo {author} {\bibfnamefont {G.}~\bibnamefont
  {Stan}}, \bibinfo {author} {\bibfnamefont {S.}~\bibnamefont {Krylyuk}},
  \bibinfo {author} {\bibfnamefont {A.}~\bibnamefont {Davydov}}, \ and\
  \bibinfo {author} {\bibfnamefont {R.~F.}\ \bibnamefont {Cook}},\ }\href
  {\doibase 10.1021/nl100062n} {\bibfield  {journal} {\bibinfo  {journal} {Nano
  letters}\ }\textbf {\bibinfo {volume} {10}},\ \bibinfo {pages} {2031}
  (\bibinfo {year} {2010})}\BibitemShut {NoStop}%
\bibitem [{\citenamefont {Thorbeck}\ and\ \citenamefont
  {Zimmerman}(2015)}]{thorbeck2015formation}%
  \BibitemOpen
  \bibfield  {author} {\bibinfo {author} {\bibfnamefont {T.}~\bibnamefont
  {Thorbeck}}\ and\ \bibinfo {author} {\bibfnamefont {N.~M.}\ \bibnamefont
  {Zimmerman}},\ }\href {\doibase 10.1063/1.4928320} {\bibfield  {journal}
  {\bibinfo  {journal} {AIP Advances}\ }\textbf {\bibinfo {volume} {5}},\
  \bibinfo {pages} {087107} (\bibinfo {year} {2015})}\BibitemShut {NoStop}%
\bibitem [{\citenamefont {Bir}\ and\ \citenamefont
  {Pikus}(1974)}]{bir1974symmetry}%
  \BibitemOpen
  \bibfield  {author} {\bibinfo {author} {\bibfnamefont {G.~L.}\ \bibnamefont
  {Bir}}\ and\ \bibinfo {author} {\bibfnamefont {G.~E.}\ \bibnamefont
  {Pikus}},\ }\href@noop {} {\emph {\bibinfo {title} {Symmetry and
  strain-induced effects in semiconductors}}},\ Vol.\ \bibinfo {volume} {484}\
  (\bibinfo  {publisher} {Wiley New York},\ \bibinfo {year} {1974})\BibitemShut
  {NoStop}%
\bibitem [{\citenamefont {Del~Vecchio}\ \emph {et~al.}(2020)\citenamefont
  {Del~Vecchio}, \citenamefont {Lodari}, \citenamefont {Sammak}, \citenamefont
  {Scappucci},\ and\ \citenamefont {Moutanabbir}}]{del2020vanishing}%
  \BibitemOpen
  \bibfield  {author} {\bibinfo {author} {\bibfnamefont {P.}~\bibnamefont
  {Del~Vecchio}}, \bibinfo {author} {\bibfnamefont {M.}~\bibnamefont {Lodari}},
  \bibinfo {author} {\bibfnamefont {A.}~\bibnamefont {Sammak}}, \bibinfo
  {author} {\bibfnamefont {G.}~\bibnamefont {Scappucci}}, \ and\ \bibinfo
  {author} {\bibfnamefont {O.}~\bibnamefont {Moutanabbir}},\ }\href {\doibase
  10.1103/PhysRevB.102.115304} {\bibfield  {journal} {\bibinfo  {journal}
  {Phys. Rev. B}\ }\textbf {\bibinfo {volume} {102}},\ \bibinfo {pages}
  {115304} (\bibinfo {year} {2020})}\BibitemShut {NoStop}%
\bibitem [{Note3()}]{Note3}%
  \BibitemOpen
  \bibinfo {note} {A more precise condition for the validity of the nanowire
  Hamiltonian in Eq.~\protect \textup {\hbox {\mathsurround \z@ \protect
  \normalfont (\ignorespaces \ref {eq:wire-Ham}\unskip \@@italiccorr )}} is
  that the harmonic frequency $\hbar \omega _z=\hbar ^2/(m^*l^2)$ is much
  smaller than the energy gap $\Delta E$ between the ground-state and the first
  excited state of the nanowire. For the devices considered and in the range of
  parameters shown the minimal energy gap is $\Delta E\sim 0.05 \epsilon _c$ at
  the switching field, and so we obtain the condition $1\gg \hbar \omega
  _z/\Delta E\sim {3L^2}/({0.05\times 16 \pi ^2 \gamma _1 m^* l^2})\sim
  {L^2}/({\gamma _1 l^2})$, where we used the effective mass in Eq.~\protect
  \textup {\hbox {\mathsurround \z@ \protect \normalfont (\ignorespaces \ref
  {eq:Effective_masses_zero_field}\unskip \@@italiccorr )}}. Because of the
  prefactor $1/\gamma _1\sim 0.22$, the nanowire Hamiltonian works reasonably
  well when $L\approx l_{so}\approx l$.}\BibitemShut {Stop}%
\bibitem [{\citenamefont {Levitov}\ and\ \citenamefont
  {Rashba}(2003)}]{PhysRevB.67.115324}%
  \BibitemOpen
  \bibfield  {author} {\bibinfo {author} {\bibfnamefont {L.~S.}\ \bibnamefont
  {Levitov}}\ and\ \bibinfo {author} {\bibfnamefont {E.~I.}\ \bibnamefont
  {Rashba}},\ }\href {\doibase 10.1103/PhysRevB.67.115324} {\bibfield
  {journal} {\bibinfo  {journal} {Phys. Rev. B}\ }\textbf {\bibinfo {volume}
  {67}},\ \bibinfo {pages} {115324} (\bibinfo {year} {2003})}\BibitemShut
  {NoStop}%
\bibitem [{\citenamefont {Golovach}\ \emph {et~al.}(2006)\citenamefont
  {Golovach}, \citenamefont {Borhani},\ and\ \citenamefont
  {Loss}}]{PhysRevB.74.165319}%
  \BibitemOpen
  \bibfield  {author} {\bibinfo {author} {\bibfnamefont {V.~N.}\ \bibnamefont
  {Golovach}}, \bibinfo {author} {\bibfnamefont {M.}~\bibnamefont {Borhani}}, \
  and\ \bibinfo {author} {\bibfnamefont {D.}~\bibnamefont {Loss}},\ }\href
  {\doibase 10.1103/PhysRevB.74.165319} {\bibfield  {journal} {\bibinfo
  {journal} {Phys. Rev. B}\ }\textbf {\bibinfo {volume} {74}},\ \bibinfo
  {pages} {165319} (\bibinfo {year} {2006})}\BibitemShut {NoStop}%
\bibitem [{\citenamefont {Golovach}\ \emph {et~al.}(2004)\citenamefont
  {Golovach}, \citenamefont {Khaetskii},\ and\ \citenamefont
  {Loss}}]{PhysRevLett.93.016601}%
  \BibitemOpen
  \bibfield  {author} {\bibinfo {author} {\bibfnamefont {V.~N.}\ \bibnamefont
  {Golovach}}, \bibinfo {author} {\bibfnamefont {A.}~\bibnamefont {Khaetskii}},
  \ and\ \bibinfo {author} {\bibfnamefont {D.}~\bibnamefont {Loss}},\ }\href
  {\doibase 10.1103/PhysRevLett.93.016601} {\bibfield  {journal} {\bibinfo
  {journal} {Phys. Rev. Lett.}\ }\textbf {\bibinfo {volume} {93}},\ \bibinfo
  {pages} {016601} (\bibinfo {year} {2004})}\BibitemShut {NoStop}%
\bibitem [{\citenamefont {Makhlin}\ \emph {et~al.}(2004)\citenamefont
  {Makhlin}, \citenamefont {Sch\"on},\ and\ \citenamefont
  {Shnirman}}]{MAKHLIN2004315}%
  \BibitemOpen
  \bibfield  {author} {\bibinfo {author} {\bibfnamefont {Y.}~\bibnamefont
  {Makhlin}}, \bibinfo {author} {\bibfnamefont {G.}~\bibnamefont {Sch\"on}}, \
  and\ \bibinfo {author} {\bibfnamefont {A.}~\bibnamefont {Shnirman}},\ }\href
  {\doibase https://doi.org/10.1016/j.chemphys.2003.09.025} {\bibfield
  {journal} {\bibinfo  {journal} {Chemical Physics}\ }\textbf {\bibinfo
  {volume} {296}},\ \bibinfo {pages} {315 } (\bibinfo {year}
  {2004})}\BibitemShut {NoStop}%
\bibitem [{\citenamefont {Petersson}\ \emph {et~al.}(2010)\citenamefont
  {Petersson}, \citenamefont {Petta}, \citenamefont {Lu},\ and\ \citenamefont
  {Gossard}}]{PhysRevLett.105.246804}%
  \BibitemOpen
  \bibfield  {author} {\bibinfo {author} {\bibfnamefont {K.~D.}\ \bibnamefont
  {Petersson}}, \bibinfo {author} {\bibfnamefont {J.~R.}\ \bibnamefont
  {Petta}}, \bibinfo {author} {\bibfnamefont {H.}~\bibnamefont {Lu}}, \ and\
  \bibinfo {author} {\bibfnamefont {A.~C.}\ \bibnamefont {Gossard}},\ }\href
  {\doibase 10.1103/PhysRevLett.105.246804} {\bibfield  {journal} {\bibinfo
  {journal} {Phys. Rev. Lett.}\ }\textbf {\bibinfo {volume} {105}},\ \bibinfo
  {pages} {246804} (\bibinfo {year} {2010})}\BibitemShut {NoStop}%
\bibitem [{\citenamefont {Zumb{\"u}hl}()}]{DZ_private}%
  \BibitemOpen
  \bibfield  {author} {\bibinfo {author} {\bibfnamefont {D.~M.}\ \bibnamefont
  {Zumb{\"u}hl}},\ }\href@noop {} {}\bibinfo {howpublished} {Private
  communication}\BibitemShut {NoStop}%
\bibitem [{COM()}]{COMSOL}%
  \BibitemOpen
  \href {www.comsol.com} {}\bibinfo {howpublished} {COMSOL
  Multiphysics{\textregistered} v. 5.5. COMSOL AB, Stockholm,
  Sweden}\BibitemShut {NoStop}%
\end{thebibliography}%
\end{document}